\documentclass[]{aa}  
\usepackage{amsfonts}
\usepackage{amssymb}
\usepackage[abs]{overpic}
\usepackage{amsmath}
\usepackage{array}
\usepackage{makecell}
\usepackage{siunitx}
\usepackage[]{graphicx}
\usepackage{subfig}
\usepackage{xcolor}
\usepackage[varg]{txfonts}
\usepackage{footnote}
\usepackage{lscape}
\usepackage{ragged2e}
\usepackage{natbib}
\usepackage{rotating}
\usepackage{multirow}
\usepackage{hhline}
\usepackage{makecell}
\usepackage{tabularx}
\usepackage[perpage,para,symbol,hang,flushmargin]{footmisc}
\usepackage{booktabs}
\usepackage{caption}
\usepackage{threeparttable}
\usepackage{multirow}
\usepackage[abs]{overpic}
\usepackage{enumerate}
\usepackage{textpos}   
\usepackage{color}
\usepackage{txfonts}
\usepackage[T1]{fontenc}
\usepackage{textcomp}
\newcommand{\magarc}{mag arcsec$^{\mathrm{-2}}$}
\newcommand{\mulim}{$\mu_{lim}$}

\def\raw{{\tt{raw.fits}}}
\def\flt{{\tt{flt.fits}}}
\def\ima{{\tt{ima.fits}}}
\def\calwf3{{\tt{calwf3}}}

\begin{document}

\title{The missing light of the Hubble Ultra Deep Field\thanks{The \textsf{ABYSS} HUDF mosaics are available in electronic form
at the CDS via anonymous ftp to cdsarc.u-strasbg.fr (130.79.128.5), via http://cdsweb.u-strasbg.fr/cgi-bin/qcat?J/A+A/ and the project official webpage http://www.iac.es/proyecto/abyss/}}
\authorrunning{Borlaff et al.}

\author{Alejandro Borlaff\inst{1,2}, Ignacio Trujillo\inst{1,2}, Javier Román\inst{1,2}, John E. Beckman\inst{1,2,3}, M.~Carmen Eliche-Moral\inst{1},\\ Ra\'{u}l Infante-S{\'a}inz\inst{1,2}, Alejandro Lumbreras-Calle\inst{1,2}, Rodrigo Takuro Sato Martín de Almagro\inst{4},\\ Carlos G\'omez-Guijarro\inst{5},  Mar\'{i}a Cebri\'{a}n\inst{1,2}, Antonio Dorta\inst{1,2}, Nicol\'{a}s Cardiel\inst{6},\\ Mohammad Akhlaghi\inst{7}, Cristina Martínez-Lombilla\inst{1,2}} 

\institute{
Instituto de Astrof\'{i}sica de Canarias, C/ V\'{i}a L\'actea, E-38200 La Laguna, Tenerife, Spain
\\\email{asborlaff@iac.es}
\and
Facultad de F\'{i}sica, Universidad de La Laguna, Avda. Astrof\'{i}sico Fco. S\'{a}nchez s/n, 38200, La Laguna, Tenerife, Spain
\and
Consejo Superior de Investigaciones Cient\'{i}ficas, Spain
\and
Instituto de Ciencias Matem\'{a}ticas, Consejo Superior de Investigaciones Científicas, Madrid, Spain
\and
Cosmic Dawn Center (DAWN), Niels Bohr Institute, University of Copenhagen, Copenhagen, Denmark
\and
Departamento de Astrof\'{\i}sica y CC.~de la Atm\'osfera, Universidad Complutense de Madrid, E-28040 Madrid, Spain
\and
Univ. Lyon, Univ. Lyon1, ENS de Lyon, CNRS, Centre de Recherche Astrophysique de Lyon UMR5574, 69230 Saint-Genis-Laval, France
}

 
  \abstract
   {The Hubble Ultra Deep field (HUDF) is the deepest region ever observed with the \emph{Hubble Space Telescope}. With the main objective of unveiling the nature of galaxies up to $z\sim7-8$, the observing and reduction strategy have focused on the properties of small and unresolved objects, rather than the outskirts of the largest objects, which are usually over-subtracted.}
   {We aim to create a new set of WFC3 IR mosaics of the HUDF using novel techniques to preserve the properties of the low surface brightness regions.}
   {We created {\textsf{ABYSS}}: a pipeline that optimises the estimate and modelling of low-level systematic effects to obtain a robust background subtraction. We have improved four key points in the reduction: 1) creation of new absolute sky flat fields, 2) extended persistence models, 3) dedicated sky background subtraction and 4) robust co-adding.}
   {The new mosaics successfully recover the low surface brightness structure removed on the previous HUDF published reductions. The amount of light recovered with a mean surface brightness dimmer than $\overline{\mu}=26$ \magarc\ is equivalent to a $m=19$ mag source when compared to the XDF and a $m=20$ mag compared to the HUDF12.}
   {We present a set of techniques to reduce ultra-deep images ($\mu>32.5$ \magarc, $3\sigma$ in $10\times10$ arcsec boxes), that successfully allow to detect the low surface brightness structure of extended sources on ultra deep surveys. The developed procedures are applicable to HST, JWST, EUCLID and many other space and ground-based observatories. We made the final {\textsf{ABYSS}} WFC3 IR HUDF mosaics publicly available at http://www.iac.es/proyecto/abyss/.}


   \maketitle
%

\section{Introduction}
\label{Sec:Intro}

The Hubble Ultra Deep field is a 11 arcmin$^2$ region of the sky located in the southern hemisphere ($\alpha$=3h 32m 39.0s, $\delta=\ang{-27;47;29.1}$, J2000), in the Fornax Constellation. Included inside the Chandra Deep Field South \citep{Giacconi2002} and the GOODS-South Field \citep{Giavalisco2003}, it was observed by \citet{Beckwith2006} during 10$^6$s of the Hubble Space Telescope Director's Discretionary Time with the Advanced Camera for Surveys (ACS), becoming the deepest image of the sky ever obtained. The authors divided the exposure time between the available filters (F435W, F606W, F775W and F850LP) with the main objective of creating a robust sample of galaxies between $4<z<7$ by using the Lyman break dropout method \citep{1992AJ....104..941S,1996AJ....112..352S,1996ApJ...462L..17S}.

Since then, an increasingly number of follow up projects with different telescopes has continued the observations of this field, including observation at radio wavelengths (VLA, \citealt[][]{Rujopakarn2016}; ALMA, \citealt[][]{Dunlop2016,Aravena2016,Aravena2016a,Walter2016}), in the infrared \citep[Spitzer IRAC,][]{Labbe2015}, optical and near-infrared (near IR) spectroscopy \citep[VLT MUSE,][]{Bacon2017}, ultraviolet \citep[WFC3 UVIS,][]{Teplitz2013}, and X-rays (Chandra, \citealt[][]{Xue2011,Luo2016}; XMM-Newton, \citealt[][]{Comastri2011}). In addition to these, the replacement in 2009 of the Wide Field and Planetary Camera 2 by the Wide Field Camera 3 (WFC3) during the Hubble Space Telescope Servicing Mission 4 (STS-125) allowed astronomers to continue the exploration of this field with a deep, high-resolution survey in near IR. With the main objective of finding the earliest sources in the Universe, HUDF09 \citep{Oesch2009,Bouwens2009} and the Hubble Ultra Deep Field 2012 \citep[HUDF12 hereafter,][]{Koekemoer2012} have increased the number of filters and exposure time in the HUDF. In addition, the eXtreme deep field \citep[XDF hereafter,][]{Illingworth2013}, reprocessed all the HST ACS and WFC3 IR available data in the HUDF. The addition of four new bands in WFC3 IR (F105W, F125W, F140W and F160W) has permitted the detection of galaxies out to $z\sim9-10$. Additionally, the extraordinary depth of the HUDF12 allows the study of galaxy stellar halos with surface brightness profiles down to $\mu_{lim} \sim 31$ \magarc\ \citep{Buitrago2017}.

Detecting extended sources in the low-surface brightness regime is an extremely challenging task. Systematic effects such as sky background, persistence, or the PSF may dominate the light distribution of the science images. Aggressive sky background subtraction may be a tempting solution to get rid of the diffuse light, whether it is caused by real astronomical sources or by artificial sky background gradients. Nevertheless, such approach have a major setback. It removes the possibility of using the sciences mosaics to study the outskirts of the largest objects, distorting the photometry of the structures up to $2-3$ \magarc\ brighter than the limiting magnitude. Due to this, is common to find over-subtracted zones with negative fluxes around large galaxies in the majority of the surveys \citep[i.e, see the Hyper Suprime-Cam survey\footnote{Hyper Suprime-Cam Subaru Data Release 1: https://hsc-release.mtk.nao.ac.jp/doc/index.php/known-problems-in-dr1/},][]{Aihara2018}. It is for this reason that careful sky-subtraction is a crucial step to preserve the properties of the low-surface brightness features of extended sources. This effect is clearly visible around the largest objects of the XDF mosaics \citep{Illingworth2013}. We must note that the main objective of the XDF project was not the study of the stellar haloes of the nearest galaxies of the HUDF ($z<1$) but to identify unresolved sources across a redshift range from $z\sim4$ to $z\sim12$ with aperture photometry. Nonetheless, the over subtraction of the diffuse outskirts of nearby galaxies (which cover a large fraction of the total field-of-view) can significantly affect the photometry of the background high-$z$ objects.  

In order to overcome the increasing challenge of studying the low surface brightness Universe, a number of recent observational studies has shown the way to reach and surpass the frontier of \mulim $\sim 30$ \magarc\ \citep[see][]{Trujillo2016}. The variety of systematic problems that significantly affect the background level of the images depends on the required imaging depth, such as fringing \citep{Wong2010}, ghosts \citep{Yang2002}, gain differences between chip amplifiers, scattered light \citep{Fowler2017} and even the effect of the point-spread function \citep[PSF,][]{Sandin2014, Sandin2015}. As a consequence of this, the use of robust statistical tools for the study of the outskirts of galaxies is mandatory for these scientific objectives. 

The low surface brightness Universe is one of the key fields for cosmology and to unveil the origin and evolution of galaxies. In particular, the processes that give rise to the galaxy discs are not completely clear. Apparent exponential discs (Type-I) on first inspection may on closer examination suffer from a number of deviations such as truncations and down-bending profiles (Type-II) or the opposite phenomenon: anti-truncations \citep[Type-III discs,][]{Erwin2005,Pohlen2006,Erwin2007,Gutierrez2011}. There are various scenarios which entail a variety of possible mechanisms probing the variety of galactic discs, such as different types of mergers and gravitational interactions \citep{Laurikainen2001,Penarrubia2006,Younger2007,Kazantzidis2009,Borlaff2014}, internal and stellar formation related processes \citep{Roskar2007, Herpich2015,Herpich2015a,Elmegreen2016,Struck2016,Struck2017} or the presence of different components on the structure of disc galaxies \citep{Comeron2012, Comeron2014}. Until now, there have been few studies that have tried to study the detailed shape of galaxy discs along significant cosmological times \citep{Azzollini2008,Azzollini2008a,Trujillo2013,Maltby2014,Borlaff2017,Borlaff2018}. The main reason for this is that PSF-effects and cosmological dimming make difficult to study the outskirts of galaxies at increasing redshifts, where these effects become critical \citep{Sandin2014, Sandin2015, Trujillo2016, Borlaff2017}. Moreover, while near IR imaging allows us to explore higher $z$ ranges, it includes additional problems such as the high sky background variability or persistence effects, which may contaminate the science images. Consequently, there is an increasing need for very deep observations and surveys as redshift increases.  

The outskirts of some nearest galaxies show traces of their formation mechanisms, such as tidal tails, haloes, plumes and satellites. Simulation-based studies predict that a hypothetical survey with a limiting magnitude fainter than $\sim 30$ \magarc\ would detect up to a dozen of accretion features around Milky Way-type galaxies \citep{Johnston2008,Cooper2009}. Apparently isolated galaxies show large tidal tails, warped discs and other asymmetric features in sufficiently deep images, as well as large numbers of satellites
\citep{Schweizer1990,Martinez-Delgado2008, Martinez-Delgado2008b,Chonis2011}. In fact, volume-limited samples of nearby galaxies detect that almost 14\% of the galaxies present diffuse features compatible with minor merger events at a limiting magnitude of 28 \magarc\ \citep{Morales2018}. An increasing number of low surface brightness explorations have been performed on individual galaxies and deep fields, but there is an increasing need for telescope-dedicated large-scale low surface brightness surveys \citep[see The Australian Space Eye and The MESSIER surveyor]{Horton2016,Valls-Gabaud2016}. Direct imaging of the stellar haloes and their structure is one of the major test for $\Lambda$CDM scenario of galaxy formation \citep{Abadi2005,Bullock2005}. Although the study of the streams of the stellar haloes has been tested for a limited sample of galaxies in the Local Group using star counts \citep{Ibata2009,McConnachie2009,Tanaka2011,Ibata2013,Peacock2014}, this method is limited to a maximum distance of 16 Mpc with HST \citep{Zackrisson2012}. 

It is for these reasons that the main objective for this paper is to explore the capabilities of HST performing a specific reduction that intends to optimise the limiting depth around extended objects and pave the way for the exploration of the outskirts of galaxies, discs, satellites, tidal streams and their stellar haloes. The proposed techniques are directly applicable to its successor in the infrared -- the James Webb Space Telescope (JWST) -- and similar missions such as EUCLID, for deep integrated photometry of extended sources beyond the Local Universe.

The paper is structured as follows. The full methodology used for the reduction is described in detail in Sect.\,\ref{Sec:Methods}. The results are presented and discussed in Sect.\,\ref{Sec:Results}. The final conclusions can be found in Sect.\,\ref{Sec:Conclusions}. We assume a concordance cosmology \citep[$\Omega_{\mathrm{M}} = 0.3,\Omega_{\mathrm{\Lambda}}=0.7, H_{0} =70 $ km s$^{-1}$ Mpc$^{-1}$, see][]{Spergel2006}. All magnitudes are in the AB system \citep{Oke1971} unless otherwise noted.



\section{Methods}
\label{Sec:Methods}

\begin{figure*}[]
\centering
\vspace{0.25cm}
\includegraphics[width=0.9\textwidth]{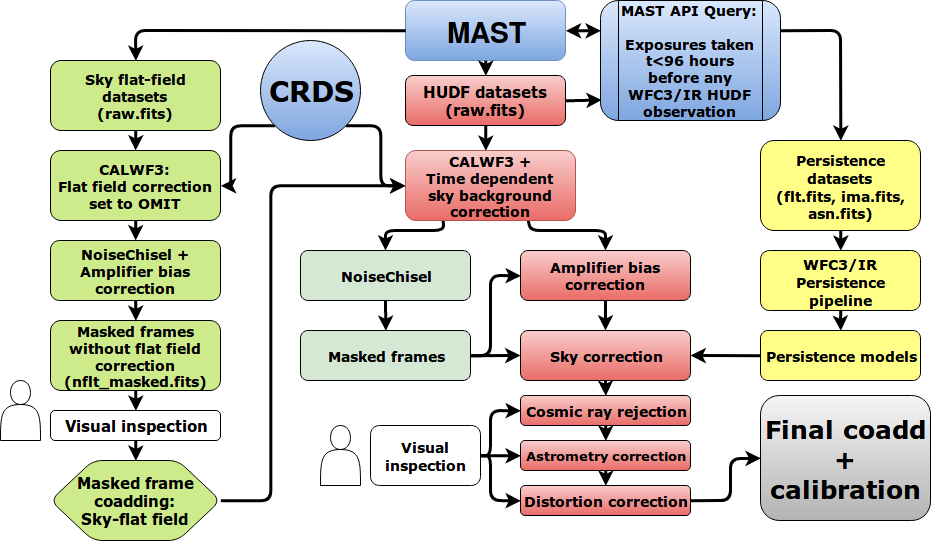}
\vspace{0.25cm}
\caption[]{Flowchart for the {\textsf{ABYSS}} HUDF reduction pipeline. Each major step in the pipeline discussed in the text is shown. The full process can be divided in three major branches: \emph{Green steps:} sky flat field creation. \emph{Yellow steps:} modelling of the extended persistence arrays. \emph{Red steps:} main reduction of the HUDF exposures. \emph{Blue steps:} required queries to the MAST or the CRDS database.}
\vspace{0.25cm}
\label{fig:flowchart}
\end{figure*}

In this section we provide an outline of the reduction process that we have followed to create a version of the HUDF WFC3 IR mosaics optimised for the study of the low surface brightness Universe (the {\textsf{ABYSS}}\footnote{{\textsf{ABYSS}}: a low surface brightness dedicated reduction for the HUDF WFC3 IR mosaics: http://www.iac.es/proyecto/abyss/} pipeline, hereafter). The full process (represented in the flowchart of Fig.\,\ref{fig:flowchart}) can be divided into three main branches:
\begin{itemize}

    \item Creation of sky flat fields for the four filters. This process is fully described in Sect.\,\ref{Subsec:flatfield}. 
    \item Creation of a catalogue of all WFC3 IR datasets that may affect our mosaics (including calibration exposures) to generate a set of improved persistence models for each exposure of the HUDF. We detail this process in Sect.\,\ref{Subsec:persistence}. 
    \item Download and reduction of all the WFC3 IR datasets that include observations using the F105W, F125W, F140W and F160W filters on the HUDF.
\end{itemize}

For each individual exposure, we conduct the following steps. We refer the reader to the corresponding sections for a detailed description: 

\begin{enumerate}
\item Preliminary calibration (bias, dark, flat field), including a time variation sky background correction (see Sects.\,\ref{Subsec:rawdata} and \ref{Subsec:timesky}).
\item Create masked frames for each exposure (Sects.\,\ref{Subsec:masking} and \ref{Subsec:dataquality}).
\item Perform amplifier level correction (Sect.\,\ref{Subsec:EqualizerAMP}).
\item Perform sky correction for each readout, using the individual masks and robust statistical estimators (Sect.\,\ref{Subsec:skycor}).  
\item Perform cosmic rays rejection and alignment of each exposure. (Sect.\,\ref{Subsec:CRAstro}). 
\item Transformation of the corrected frames into a geometric distortion corrected frame, combination and calibration of the final mosaic (Sect.\,\ref{Subsec:coadd}).
\end{enumerate}

\subsection{Initial data and preliminary calibration}
\label{Subsec:rawdata}

For the present work, we started downloading all the individual exposures of the WFC3 (symbolised here by \raw) within $r<3$ arcmin radius of the original HUDF centre coordinates ($\alpha$=3h 32m 39.0s, $\delta=\ang{-27;47;29.1}$, J2000). In order to do that, we make use of The Barbara A. Mikulski Archive for Space Telescopes (MAST)\footnote{MAST is managed by Space Telescope Science Institute (STScI) and is publicly available at https://mast.stsci.edu/portal/Mashup/Clients/Mast/Portal.html}. MAST allows us to perform queries with multiple constraints, such as sky-coordinates ($\alpha,\delta$), telescope, instrument, detector, filter, proposal ID and observation date. Given the large number of files to download, we used the MAST Python API\footnote{MAST Python API: https://mast.stsci.edu/api/v0/}. The MAST API allows us for programmatic queries and implement them easily into any Python code. We included MAST API subroutines into our pipeline to automatically download required files from MAST when needed.

Subsequently, we made use of the HST Calibration Reference Data System\footnote{HST Calibration Reference Data System: https://hst-crds.stsci.edu/} \citep[CRDS,][]{Greenfield2016} to download the best reference files available for each individual exposure of the HUDF. The CRDS is publicly available and can be installed as part of {\textsf{AstroConda}}\footnote{{\textsf{AstroConda}} is a free {\textsf{Conda}} channel maintained by the Space Telescope Science Institute (STScI). It provides tools and utilities required to process and analyse data from the Hubble Space Telescope (HST), James Webb Space Telescope (JWST), and others: http://AstroConda.readthedocs.io/}. For HST, CRDS has a command line tool that assigns and automatically downloads the best reference files to the FITS headers of the \raw\ files. We processed all our \raw\ files using the Space Telescope Science Data Analysis System (STSDAS) task \calwf3. \calwf3 is available as part of the {\texttt{stsci\_python}} package on {\textsf{AstroConda}}, and it corrects for instrumental effects and generates calibrated frames. In addition, \calwf3\ can process multiple readouts for the same exposure to create improved data products. 

For WFC3 IR observations, it is possible to sample the signal multiple times as an exposure accumulates, before the end of the exposure. This allows to 1) to record the signal of a pixel before it saturates, 2) to perform a better cosmic-ray rejection and 3) to reduce the net effective read noise. This observation mode is named {\tt{MULTIACCUM}}, and is the default observation mode for WFC3 IR. Each \raw\ file contains the information of each readout (up to 16) of the chip during a certain exposure. The individual processing steps that \calwf3\ performs for each readout are: 
\begin{enumerate}
    \item Flagging of known bad pixels in the data quality (DQ) array.
    \item Identification of pixels in the initial read that contain detectable source signal.
    \item Subtraction of bias drifts using the reference pixels.
    \item Subtraction of the zeroth (first) read.
    \item Estimation of the noise model for each pixel and record in the error (ERR) array.
    \item Photometric non-linearity and saturated pixels correction.
    \item Subtraction of dark image. 
    \item Calculation of the photometric header keyword values for flux conversion. 
    \item Conversion of the data from counts to count rates. 
\end{enumerate}

After this, \calwf3 uses each readout of the {\tt{MULTIACCUM}} mode to create a single image for each individual exposure. This is done by analysing the count differences between readout as a function of time (up-the-ramp fitting). This new array represents the best-fit count rate for each pixel. Finally, the pipeline performs flat fielding and gain conversion (transforming from counts to count-rates). The final result is an \flt\ file, which contains the science best-fit count-rate array (SCI), the error array (ERR), a data-quality array (which flags bad-pixels and cosmic rays), the number of samples (SAMP) and integration time (TIME) arrays. Each one of the \calwf3\ steps can be omitted or performed by switching their corresponding keywords in the headers of the \raw\ files of each exposure. For those frames that are going to be used here to calculate our dedicated sky-flat, we set the corresponding keyword ({\tt{FLATCORR}}) to ''OMIT'' prior to their processing.
\begin{figure*}[h!]
\centering
\vspace{0.25cm}
\begin{overpic}[width=0.48\textwidth]{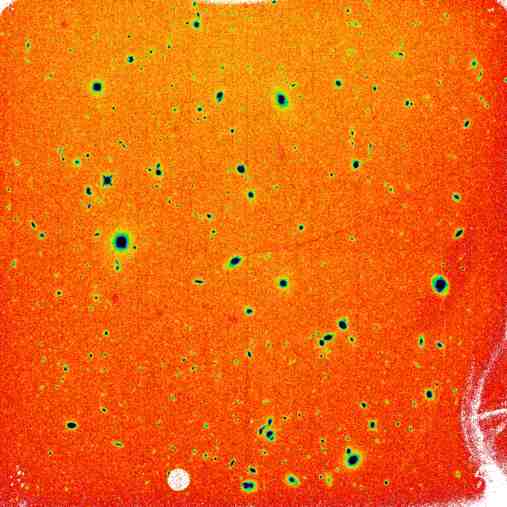}
\put(10,235){\color{black} \colorbox{white}{\textbf{\large Before flat correction}}}
\end{overpic}
\begin{overpic}[width=0.48\textwidth]{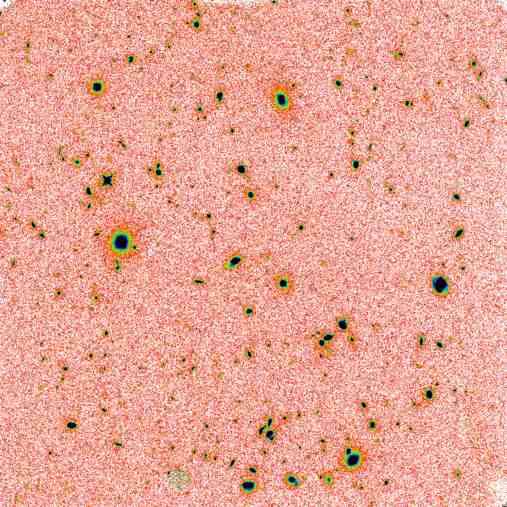}
\put(10,235){\color{black} \colorbox{white}{\textbf{\large After flat correction}}}
\end{overpic}
\begin{overpic}[width=0.48\textwidth]{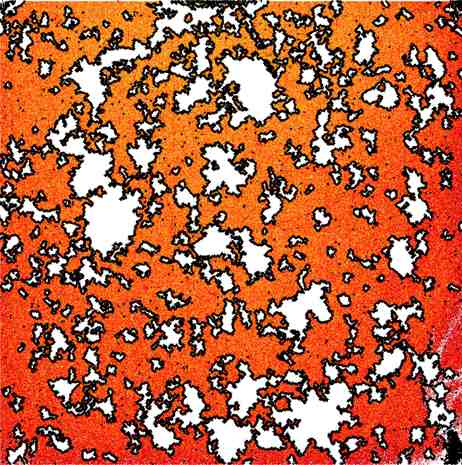}
\put(10,235){\color{black} \colorbox{white}{\textbf{\large Masked without flat correction}}}
\end{overpic}
\begin{overpic}[width=0.48\textwidth]{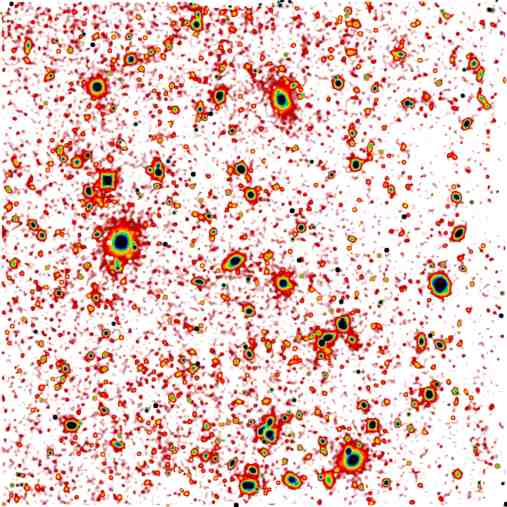}
\put(10,235){\color{black} \colorbox{white}{\textbf{\large After flat correction with 5-pixel Gaussian smooth}}}
\end{overpic}
\caption{Example of the masking process. \emph{Top left:} Pre-calibrated frame, without flat field correction. \emph{Top right:} Pre-calibrated frame, with flat field correction applied. \emph{Bottom left:} Masked frame without flat field correction. \emph{Bottom right:} Flat field corrected frame with Gaussian smoothing. Notice that {\textsf{NoiseChisel}} \citep{Akhlaghi2015} efficiently detects objects with very different shapes and sizes, masking the outskirts and diffuse regions that are barely visible even in the convolved frame.} 
\vspace{0.25cm}
\label{fig:masking_process}
\end{figure*}

\subsection{Image masking}
\label{Subsec:masking}

In order to create the sky flat field models, estimate the sky level of each exposure and obtain the final catalogue, we require an accurate masking of the individual sources in the images. For this task we used {\textsf{Gnuastro}}'s\footnote{GNU Astronomy Utilities (ascl.net/1801.009) project is part of the Free Software Foundation and is freely available at: https://www.gnu.org/software/gnuastro/} program (version 0.5) for detection and segmentation: {\textsf{NoiseChisel}} \citep{Akhlaghi2015}. {\textsf{NoiseChisel}} is a recently developed free software to detect signal in astronomical images based on erosion of pixels. This method is non-parametric, allowing to efficiently detect astronomical objects with irregular morphologies that are immersed in noise. In addition to this, {\textsf{NoiseChisel}} algorithm is based on the properties of the image noise (not only on those of the signal dominated pixels) reducing the input from the user and thus it is extremely robust against the sky level value or the properties of the sources to mask, making it an excellent choice for highly different exposures, such as the ones analysed in this paper. {\textsf{NoiseChisel}} accepts FITS files as input, and returns a multi-extension FITS file that contains: 1) the input image, 2) the segmentation map with the detection labels for each object, 3) a secondary segmentation map for the clumps inside each detection, 4) {\textsf{NoiseChisel}} estimation of the final sky value on each pixel, and 5) the standard deviation for each pixel. The final results are based on version 0.5 of {\textsf{Gnuastro}}. The masking of each individual exposure is performed as follows: 

\begin{enumerate}
    \item In order to properly detect sources in the input images with {\textsf{NoiseChisel}} they have to be flat field corrected. Otherwise, we would not have a similar sensitivity across the detector, and our results would be biased. We perform preliminary calibration of the \raw\ files, including flat field calibration. During the creation of our own sky flat fields, we have to correct the images before masking. For this step, we use the official MAST flat fields. After the sky flat fields are created, if the input image is a HUDF exposure, we use our own sky-flat fields. 
   
    \item We create the segmentation maps for each \flt\ image using {\textsf{NoiseChisel}} with default configuration (tile size equal to $50\times50$ pixels).
    
    \item If the images are being masked for the generation of the sky flat fields, we multiply back the images by its corresponding official MAST flat fields. The result of this process is a properly masked flat field uncorrected exposure.  
\end{enumerate}

In Fig.\,\ref{fig:masking_process} we show an example of the masking process of a F160W exposure, included in the HUDF field ({\tt{ib5x2elbq}}, PID 11563). The top left panel represents the \flt\ file after preliminary calibration through \calwf3, but before flat field correction (see Sect.\,\ref{Subsec:rawdata}). The top right panel shows the same \flt\ file after flat field correction. After this step, we analyse the flat field corrected \flt\ image with {\textsf{NoiseChisel}}, which produces a segmentation map for each exposure. We identify all the pixels that are included as part of a source obtaining a masked, flat field uncorrected \flt\ file, ready to be combined for the creation of a sky flat field. We show this final masked frame in the bottom-left panel of Fig.\,\ref{fig:masking_process}. In order to illustrate the accuracy of the segmentation map results, we compare in the bottom panels the {\textsf{NoiseChisel}} mask with the \flt\ image convolved by a Gaussian kernel with $\sigma = 5$ pixels. It is clearly visible that this process accurately masks objects with very different ranges of size, even if part of them (or the whole object) is not visible in the non-convolved frame. We refer the reader to Sect.\,\ref{Subsec:skycor} for a systematic benchmark analysis of the masking procedure effect on the sky background subtraction and its comparison with other methods.   

\subsection{Amplifier relative gain correction}
\label{Subsec:EqualizerAMP}

\begin{figure}[]
\centering
\vspace{0.25cm}
\includegraphics[width=0.5\textwidth]{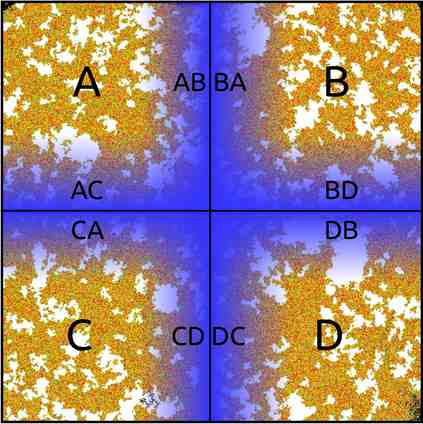}
\caption[]{Amplifier gain correction diagram. We label each one of the four $507\times507$ pixel sections of the WFC3 IR detector as A, B, C, D (see figure). We represent with a blue transparent gradient the weights (bluer represents higher weights) applied to the corresponding pixels of each region, labelled accordingly. The white regions in the background are masked pixels, either because they are part of a light source or are affected by persistence. We do not include those pixels in the amplifier gain correction estimation.} 
\vspace{0.25cm}
\label{fig:equalizerAMP}
\end{figure}

The WFC3 IR detector contains $1024\times1024$ pixels, which are divided into four quadrants of $512\times512$ pixels. There is a border of 5 non-illuminated pixels around the edges of the detector, which are used to provide constant-voltage reference values for the detector. Due to this, the light exposed area of the detector includes $1014\times1014$, divided in four $507\times507$ pixels. Although the default pipeline corrects for differences in gain between the four different sections of the WFC3 IR detector, it is common to see residual differences in the flat field corrected images and specially in the combined mosaics. The amplitude and sign of these differences between amplifiers may vary from dataset to dataset, and can produce significant effects in both the final images and the sky flat fields.

\begin{figure*}[!ht]
\centering
\vspace{0.25cm}
\includegraphics[width=0.48\textwidth]{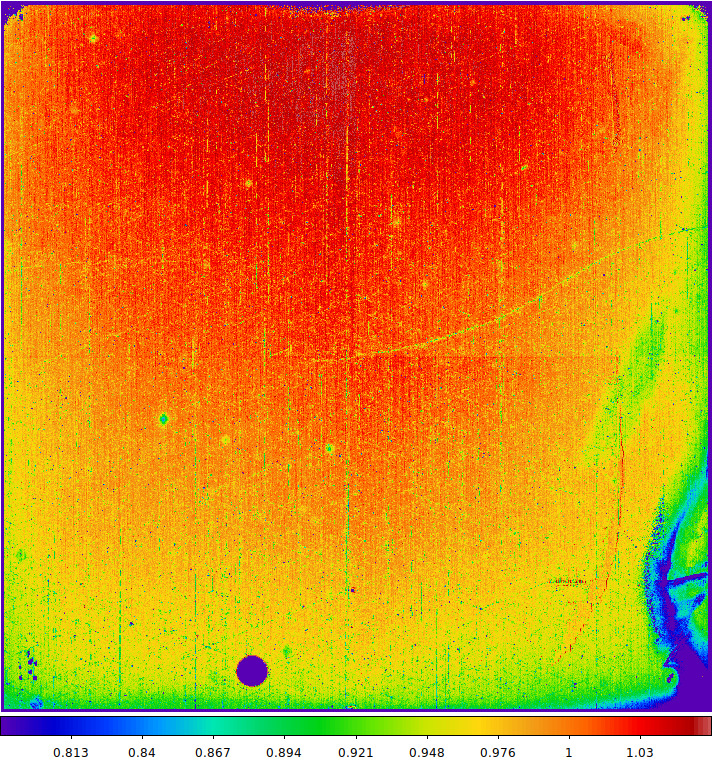}
\includegraphics[width=0.48\textwidth]{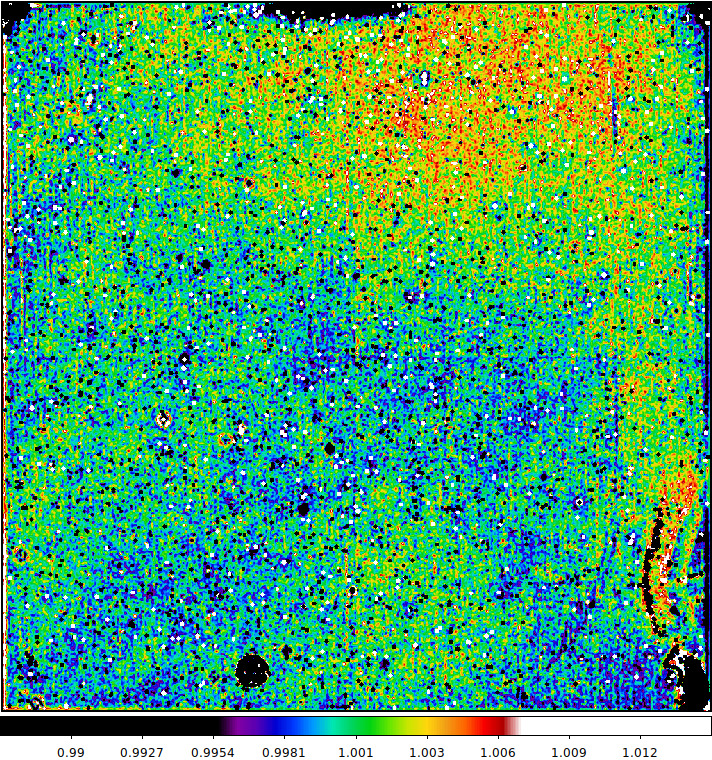}

\caption[]{Sky flat field analysis for the F160W filter. \emph{Left panel:} Absolute sky flat field measured for the June 2009 to July 2013 period, using the datasets corresponding to the proposal IDs listed in Table \ref{table:skyflat_selection}. \emph{Right panel:} Ratio of our sky flat field for the F160W to its corresponding MAST SD flat field {\tt{uc721145i}}. Notice the different colour scale, spanning only a $\sim 1\%$ for the flat field ratio panel. The right panel was smoothed with a 3 pixel wide Gaussian kernel to enhance the differences between different regions.} 
\vspace{0.25cm}
\label{fig:flat_F160W}
\end{figure*}

In order to correct for this effect we have to calculate the median differences in flux between each section of the chip and compensate them. We cannot simply use the median values for the full area covered by each section, because that would result in a wrong estimate of the median bias if there are large scale gradients across the detector. Instead, we calculate the ratios between columns or rows of pixels at equal distances to the frontier between chips. Then we estimate the weighted median of the resulting array of ratios, where the weights are equal to the inverse distance of each pixel to the frontier of the chip. By doing this, we ensure that the differences between pixels which contribute the most are the closest ones to the frontier, but we still make use of all the information available in the image. We set as reference the $A$ section of the chip (top left) and correct the remaining sections ($B$, $C$ and $D$) to it. The weighted sky-levels for each frontier between chips are labelled with two letters, being the first letter the one to which the pixels of the sample belong (see Fig.\,\ref{fig:equalizerAMP}). The correction factors are: 

\begin{equation}\label{eq:equalizerAMP}
\delta AB = <\frac{\overrightarrow{AB}}{\overleftarrow{BA}}>,
\end{equation}
\begin{equation}\label{eq:equalizerAMP}
\delta AC = <\frac{\overrightarrow{AC}}{\overleftarrow{CA}}>,
\end{equation}
\begin{equation}\label{eq:equalizerAMP}
\delta CD = <\frac{\overrightarrow{CD}}{\overleftarrow{DC}}>,
\end{equation}
\begin{equation}\label{eq:equalizerAMP}
\delta BD = <\frac{\overrightarrow{BD}}{\overleftarrow{DB}}>,
\end{equation}

\noindent where the $\overrightarrow{}$ and $\overleftarrow{}$ mean the original and axis reversed arrays, respectively, and the $< >$ represents the weighted mean by the distance of each pixel to the frontier between chips. By using these, we can calculate the $\delta AD$ correction factor as follows:
\begin{equation}\label{eq:equalizerAMP}
\delta AD1 = \delta AB \cdot \delta BD,
\end{equation}
\begin{equation}\label{eq:equalizerAMP}
\delta AD2 = \delta AC \cdot \delta CD,
\end{equation}
\begin{equation}\label{eq:equalizerAMP}
\delta AD = \frac{\delta AD1 + \delta AD2}{2},
\end{equation}

In addition, as in the case of sky-correction, we must avoid contamination by the objects in the field of view and persistence. In order to avoid such effects, we masked all the pixels flagged as part of sources by {\textsf{NoiseChisel}} (see Sect.\,\ref{Subsec:masking}). Additionally, for the HUDF field images, we also flagged all those pixels affected by persistence according to our custom-improved models (see Sect.\,\ref{Subsec:persistence}). Finally, we applied this amplifier gain correction to all the exposures used in this study before the final sky-subtraction.

\subsection{Sky flat fielding}
\label{Subsec:flatfield}

\begin{table*}
{\small 
\begin{center}
\begin{tabular}{cccc}
\toprule
Sky-flat & Period & No. of exposures & Proposal IDs\\
(1)&(2)&(3)&(4)\\
\midrule
F105W A & 55000 - 56500 & 719 & \thead{11563, 11584, 11738, 12060, 12065, 12067, 12068, 12069, 12099, 12100,\\ 12101, 12102, 12103, 12104, 12184, 12286, 12442, 12451, 12452, 12453,\\ 12454, 12455, 12456, 12457, 12458, 12459, 12460, 12461, 12496, 12498,\\ 12553, 12590, 12787, 12788, 12789, 12790, 12791, 12949, 13063} \\[1.2ex]
F105W B & 56500 - 58000 & 747 & \thead{13386, 13420, 13459, 13495, 13496, 13641, 13677, 13687, 13718, 13767,\\ 13779, 13790, 13792, 14037, 14038, 14096, 14122, 14227, 14327, 14808} \\[1.2ex]
F125W & 55000 - 56500 & 1811 & \thead{11144, 11149, 11189, 11359, 11520, 11557, 11563, 11678, 11700, 11702,\\ 12025, 12028, 12036, 12060, 12061, 12062, 12063, 12064, 12065, 12066,\\ 12067, 12068, 12069, 12099, 12100, 12101, 12102, 12103, 12265, 12286,\\ 12329, 12440, 12443, 12444, 12445, 12451, 12452, 12453, 12454, 12459,\\ 12460, 12461, 12572, 12590, 12616, 12960, 13063} \\[1.2ex]
F140W & 55000 - 56500 & 875 & \thead{11359, 11600, 11696, 12067, 12068, 12099, 12100, 12101, 12102, 12103,\\ 12166, 12177, 12190, 12203, 12217, 12328, 12330, 12452, 12458, 12459,\\ 12461, 12471, 12498, 12544, 12547, 12568, 12726, 12896, 13063} \\[1.2ex]
F160W & 55000 - 56500 & 1727 & \thead{11142, 11149, 11189, 11359, 11520, 11563, 11584, 11647, 11663, 11694,\\ 11696, 11700, 11702, 11735, 11738, 11838, 11840, 12028, 12036, 12055,\\ 12060, 12061, 12062, 12063, 12064, 12065, 12066, 12067, 12068, 12069,\\ 12072, 12075, 12099, 12100, 12101, 12102, 12104, 12167, 12194, 12195,\\ 12197, 12224, 12265, 12267, 12283, 12286, 12292, 12307, 12329, 12378,\\ 12440, 12443, 12444, 12445, 12447, 12451, 12452, 12453, 12454, 12459,\\ 12461, 12498, 12502, 12590, 12613, 12616, 12686, 12709, 12764, 12866,\\ 12990, 13063} \\[1.2ex]
\hline
\bottomrule
\end{tabular}
\caption{Summary of the exposures used for the sky flat field analysis. \emph{Columns:} 1) Flat field identifier. 2) Exposure date limits used for the selection of the datasets for each flat field. 3) Number of exposures used for each flat field. 4) Proposal IDs of the exposures used. We remark that not all the images of a given proposal ID were used in its corresponding flat field. A careful visual inspection of each one of the images was carried out by the authors, before image combination.}\vspace{-0.5cm}
\label{table:skyflat_selection}
\end{center}
}
\end{table*}

In order to measure the relative sensitivity of the pixels of a detector (flat field), the optimal process would be to observe a uniform external source of light. Although this is certainly not possible in
most cases, there are several strategies to reproduce these conditions or to compensate for any possible inhomogeneities in the illumination. In ground-based observations, the combination of several out-of focus dome images is the easiest way to estimate the flat field. Observations of the twilight sky at the start and the end of the night (twilight flats) are a reasonable alternative to avoid the possible inhomogeneities and gradients caused by the dome of the telescope. However, neither dome flats nor twilight flats are free of gradients due to non-perfect flat illumination, and the latter suffer from time variation of the sky background level. In space, twilight flats are possible using dark Earth limb
observations (see e.g., HST Cycle 17 WFC3 calibration proposal 11917) or the moonlit illuminated earth \citep{Bohlin2008}. For WFC3, high signal-to noise ground-based flat fields \citep{Bushouse2008} were generated prior to launch in the CASTLE HST simulator (hereafter CASTLE LP-flats). Nevertheless, the CASTLE simulator is not a perfect replica of the optical path and conditions of HST. Observations of the star clusters Omega-Cen and 47 Tuc (proposals CAL-11453 and CAL-11928) demonstrated that the CASTLE LP-flats were not able to correct the large scale structure in WFC3 IR channel \citep{Pirzkal2011}.

An alternative to these strategies is to take advantage of the observations previously accumulated. Sky background in the near IR is high enough, even in space, to use it as a uniform source of light. However, diffuse objects and the extended point spread function can introduce severe biases in the median images. It is because of this, that careful masking of any source in the field-of view is necessary to avoid contamination in the final calibration images. In addition to this, not all the images are suitable for this analysis, because the flux of the sky background is relatively low \citep[$\sim0.3-1.0$\,\,  e$^{-}\,$s$^{-1}$pix$^{-1}$, see][]{Pirzkal2011}. Thus, creating accurate sky-flat fields requires the combination of a large number of exposures. The standard method to create sky-flats from observations is to calculate the median of the masked science images. In ground-based observations, a masterflat per night is created by using the corresponding normalised science images of that night. The reason of this is to avoid any unwanted effects due to slight changes of focus, weather conditions, or vignetting. The high stability of space telescopes compared to the ground-based ones permits the making of sky-flat fields using images that have been taken over longer time periods. Moreover, the use of sky-flat fields have the advantage that they are measuring the relative sensitivity between pixels at the same intensity level of the images that have to be corrected. This accounts for any possible spatial differences in the linearity across the detector as a function of the input intensity.   

In \citet{Pirzkal2011}, the authors used the observations taken between 2009 and 2010 with WFC3 IR to create a second order correction to the CASTLE LP-flats (delta sky flat or SD-flats). In order to do that, they first identified a large number of datasets with exposure times longer than 300s. Secondly, they aggresively masked all the sources in the images with SExtractor \citep{Bertin1996} in the field of view. This step was repeated for several hundreds of images per filter. Finally the masks were normalised and combined to the CASTLE-LP flats, creating the new SD-flats. Although the SD-flats for the F160W included nearly a thousand masked images, some of the other filters did not have enough data to create a reliable flat field, which was the case for F105W, F110W and F140W, where only $\sim100$ datasets were used. For these filters, the signal-to-noise ratio was too low and the masking process left no available data in some regions of the detector, creating holes were the corresponding SD-flat had no data. Finally, in order to reach a reliable final SD-flat solution, the authors opted to combine the $\sim2000$ datasets from all the filters into one SD-flat (grey SD-flat, hereafter). In addition to this, they smoothed the grey SD-flat by using a $\sigma = 10$ pixel kernel before combination with the CASTLE LP-flats. The authors did not find any significant dependence on the filter or variation of the flat fields with time between 2009 and 2010. They successfully tested the improved flat field with 32 F160W datasets from the HUDF proposal 11563, removing a large scale cross-like structure in the background. These combined CASTLE LP-flats + SD-flat fields were included into the calibration database system in December 7th, 2010, until present date. 

In this paper, we follow a similar procedure as the presented in \citet{Pirzkal2011}. We generate a sky flat field per filter by combination of exposures between 55000 (June 2009) to 56500 (July 2013). Additionally, we create a second flat for the F105W filter considering observations between 56500 and 58000 (April 2017). Due to the small size of the dithering pattern, and the fact that all images were taken with very similar rotation angles, we cannot create a sky-flat field just by using the exposures from the HUDF observation programmes. To tackle this problem, we selected the deep observations as the AEGIS, COSMOS, GOODS-N, GOODS-S, and UDS fields. We also included multiple exposures from other fields that were taken in the 96 hours before the HUDF exposures. We list the amount of exposures per filter, time-period, and the proposal IDs of all the images used for each flat field in Table \ref{table:skyflat_selection}. 

In Fig.\,\ref{fig:flat_F160W} we present an example of our results. In the left panel we show our absolute sky-flat field for the F160W filter, according to the robust median of the masked datasets from Table \ref{table:skyflat_selection}. In the right panel we show the ratio between our sky flat field and its corresponding SD-flat from MAST. Notice that both panels have different colour scales, and that the right panel has been convolved with a 3 pixel wide Gaussian kernel in order to enhance the differences between flats. We do not find differences larger than $\sim1\%$, but there is a significant coherent large scale variation, in addition to residuals on the bottom-right corner of the detector (wagon-wheel). 

\subsection{Persistence correction}
\label{Subsec:persistence}

\begin{figure*}[]
\centering
\vspace{0.25cm}
\includegraphics[width=0.48\textwidth]{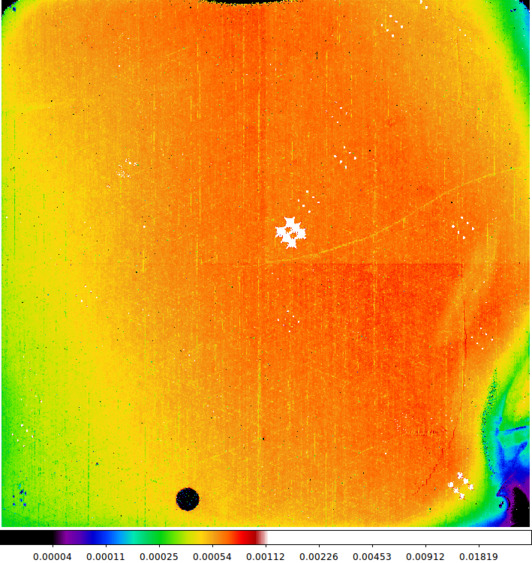}
\includegraphics[width=0.48\textwidth]{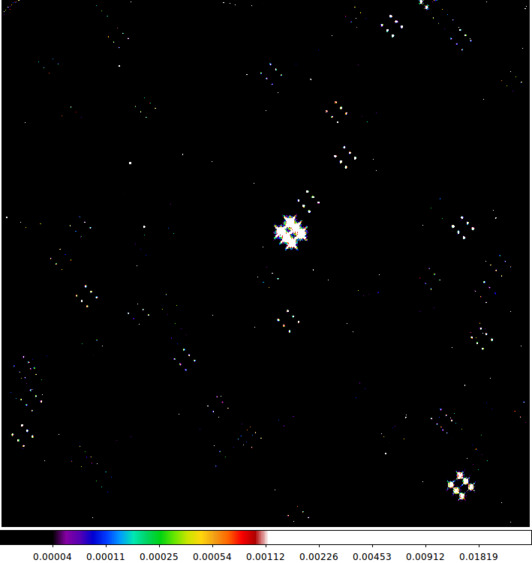}
\includegraphics[width=0.48\textwidth]{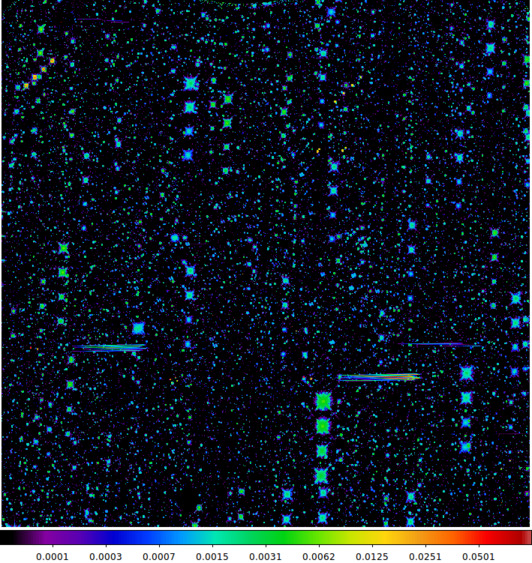}
\includegraphics[width=0.48\textwidth]{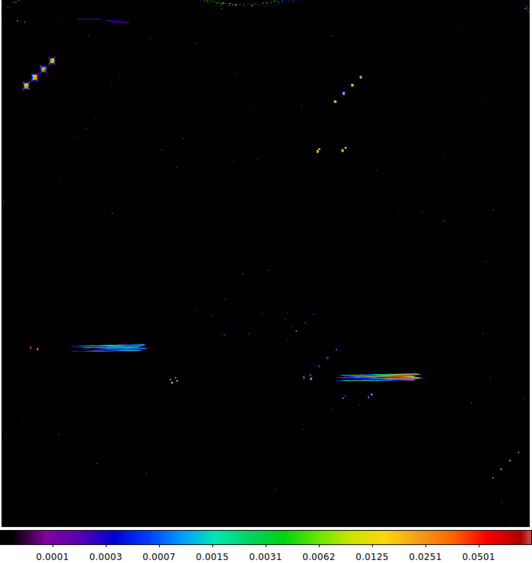}

\caption[]{Comparison between the WFC3 IR persistence models calculated taking into account 96 hours of previous observations (left column) and the persistence models available at MAST, with a limit of 16 hours of lookback time (right column). \emph{Top panels:} Persistence models for dataset {\tt{ib5x22b8q}}. \emph{Bottom panels:} Persistence models for dataset {\tt{icxt25byq}}. The images corresponding to the same dataset are at the same colour scale. Note on the top left panel a background of persistence covering the complete field of view caused by the flat calibration lamp in the previous hours to the observation of the HUDF, not detected on the persistence model available at MAST (top right panel).}
\vspace{0.25cm}
\label{fig:persistence}
\end{figure*}

\begin{figure*}[]
\centering
\vspace{0.25cm}
\includegraphics[width=0.49\textwidth, trim={0cm 4.0cm 5.1cm 1.5cm},clip]{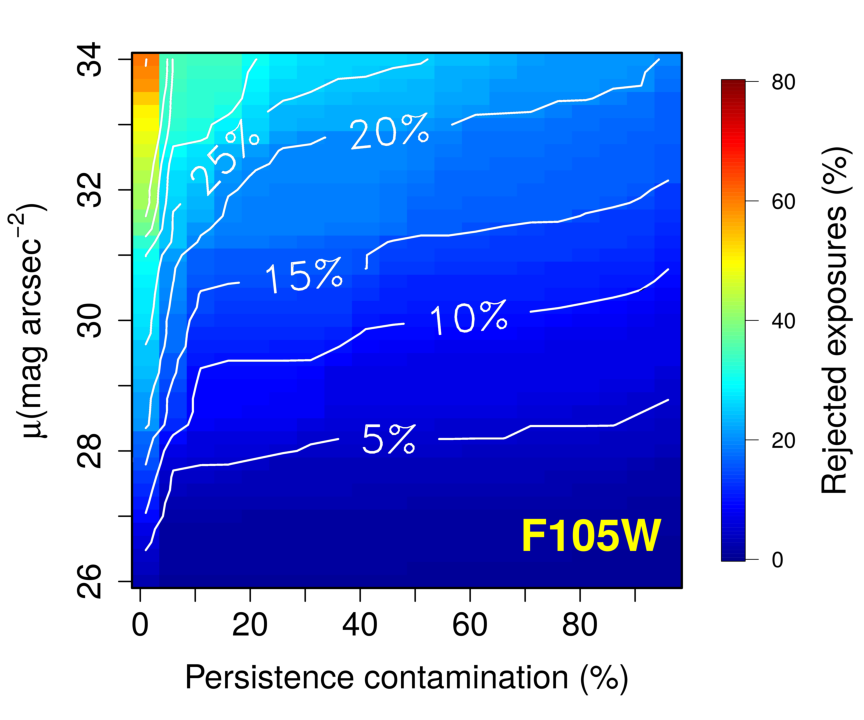}
\includegraphics[width=0.49\textwidth, trim={4.6cm 4.0cm 0.5cm 1.5cm},clip]{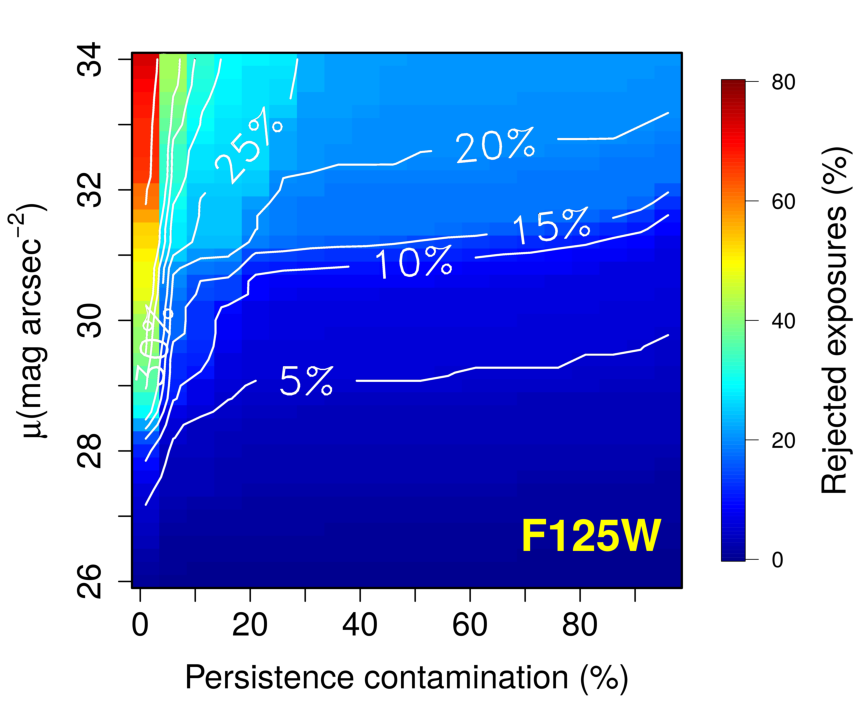}
\includegraphics[width=0.49\textwidth, trim={0cm 0.5cm 5.1cm 1.0cm},clip]{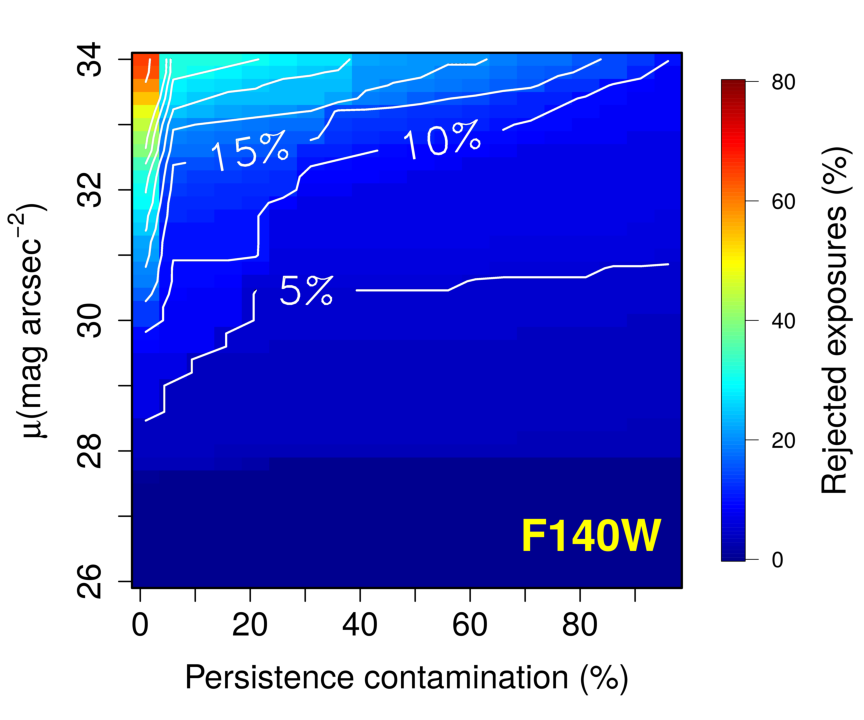}
\includegraphics[width=0.49\textwidth, trim={4.6cm 0.5cm 0.5cm 1.0cm},clip]{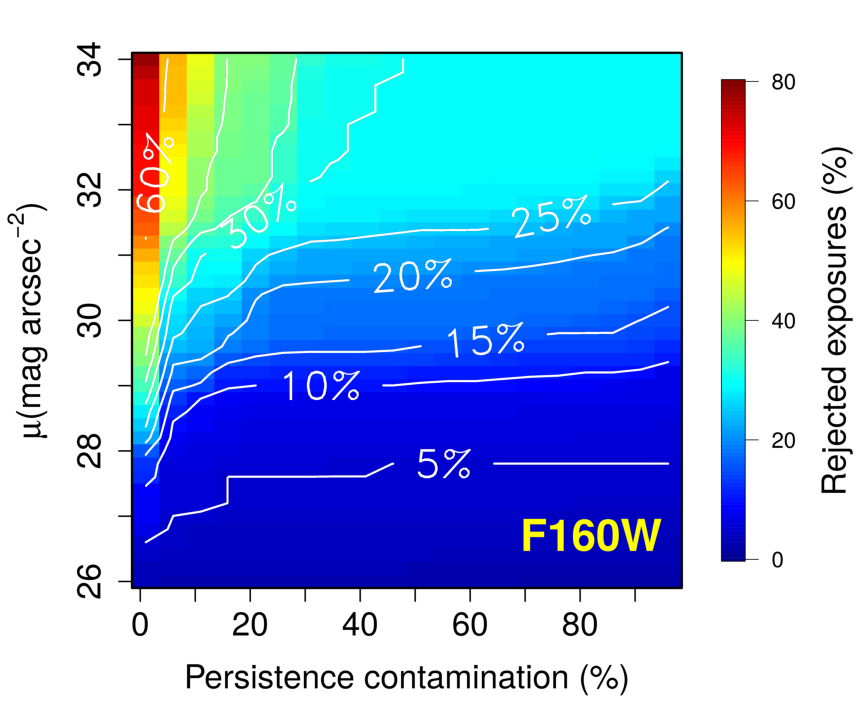}

\caption[]{Analysis of persistence contamination on the WFC3 IR HUDF exposures. In colour bins and white labelled contours, we represent the fraction of exposures that present a certain level of surface brightness contamination by persistence or higher (vertical axis) as a function of the fraction of the total amount of pixels that are affected (horizontal axis), for the F105W (top left), F125W (top right), F140W (bottom left) and F160W (bottom right) bands.}
\vspace{0.25cm}
\label{fig:persistence_hist}
\end{figure*}

A known effect that affects HgCdTe IR array detectors (as is the case of the WFC3 IR) is persistence. Persistence shows up as an afterglow on the pixels that were exposed to a bright source of light in a previous exposure. This charges arise from imperfections in the photo-diodes of IR detectors \citep{Smith2008a,Smith2008}. Neither non-destructive read-out nor resets can change significantly the rate at which this charge is released. These bright sources then re-appear in the following exposures as ghost images in the same regions of the detector. The intensity of this effect decays with time and eventually becomes negligible. Even so, persistence is an important effect to take into account as it may create false detections in science images. The effect of persistence and thus the accuracy of its correction becomes more challenging as we move towards lower surface brightness ranges.

Persistence is an intrinsic effect of the WFC3 IR detector and thus it cannot be avoided, but it can be partially corrected afterwards. STScI MAST WFC3 Persistence Project\footnote{WFC3 Persistence Project: https://archive.stsci.edu/prepds/persist/} provides the necessary tools to check if a certain exposure was affected by persistence and a complete set of models to correct the effects. The current method of persistence correction of WFC3 IR consists in modelling the number of electrons that would be created by persistence in each pixel by all the previous exposures (up to a certain time) that were taken before the one to correct \citep{Long2012}. These models are pre-calculated for all WFC3 IR exposures and publicly available through MAST. Nevertheless, these models only take into account the exposures that were taken up to 16 hours before for the creation of the persistence model (Dr. Knox Long, private communication). 

In \citet{Long2015} the authors published an improved pipeline to create persistence models for WFC3 IR. The initial models predicted persistence only based on the observed flux and the time between exposures \citep{Long2012}. Nevertheless, in \citet{Long2013a,Long2013b}, the authors carried out a set of experiments where they demonstrated that the persistence depended also with the amount of exposures taken and the time that a pixel remained filled with charge. Based on these findings, the authors developed a more accurate prediction of persistence in the IR channel of WFC3. This new pipeline was included into the set of software tools used to estimate the persistence in all WFC3 IR images and also was made publicly available\footnote{STScI Persistence repository: https://github.com/kslong/Persistence/wiki}.

In this paper, we make use of the last update to date (git commit hash b0b9cbeaf7, 26 Jan 2016) of the prediction software made by STScI in order to create a set of dedicated persistence models for each individual exposure of the HUDF. We increase the lookback time from 16 hours to 96 hours to take into account the longest time that our computer resources permit us to create the persistence models, using the default observational "A-gamma" model \citep{Long2015}. To do that, we downloaded all the exposures taken with WFC3 IR 96 hours before each exposure of the HUDF. Finally, we run the persistence pipeline for each exposure in the HUDF field. The software creates the persistence models and automatically stores them in their corresponding FITS files. 

We show two examples of the improvement of the persistence models in Fig.\,\ref{fig:persistence}. The left column shows the improved persistence models for the exposures of the HUDF {\tt{ib5x22b8q}} and {\tt{icxt25byq}}. The right column shows their corresponding MAST official persistence models taking into account the previous 16 hours. In the first case {\tt{ib5x22b8q}}, the persistence model (top-left panel) is dominated by a large gradient produced by a set of calibration observations (Cycle 17, CAL WFC3 category, 11915 - IR Internal Flat Fields, PI: Bryan Hilbert)\footnote{A detailed overview of proposal 11915 can be found in: http://www.stsci.edu/hst/phase2-public/11915.pdf}. These observations were meant to create a new set of flat fields for WFC3 IR using the internal tungsten flat field lamp to illuminate the detector. As a consequence of this, the camera was flooded with persistence just 18 hours before the observations of the HUDF with the F105W filter. Unfortunately, the results of the proposal 11915 were never published (Ben Sunnquist, WFC3 Help Desk, private communication). Many of the HUDF exposures of the F105W filter were left unusable because of this calibration program, since we are interested on the lowest surface brightness limit, and we cannot rely on the persistence correction to fully correct the images. This effect does not appear in the official MAST persistence arrays (top-right panel), due to the 16 hours limit for the lookback calculation of persistence, and thus, were not taken into account in the previous versions of the HUDF.

Nevertheless, large scale gradients can hardly affect the results for high-$z$, unresolved objects. In that case, the observers must pay more attention to flag any point-like source of light that may be caused by persistence. In the bottom panels of Fig.\,\ref{fig:persistence} we show another example ({\tt{icxt25byq}}) of the improvement of our persistence models. We identify two major causes of the persistence contamination, proposals 14074 (Opening the Window on Galaxy Assembly: Ages and Structural Parameters of Globular Clusters Towards the Galactic Bulge, PI: Roger Cohen) and the WFC3 G102 grism observations of proposal 14227 (Cycle 23 proposal 14227: The CANDELS Lyman-alpha Emission At Reionization Experiment, PI: Casey Papovich), which ended just $\sim3$ minutes before the start the HUDF exposure {\tt{icxt25byq}}. The former one (which ended 40 hours before the HUDF exposure, hence, was not taken into account for persistence) is responsible for the star-like sources across the detector, and the latter one for the two elongated sources. 

\begin{figure*}[]
\centering
\vspace{0.25cm}
\includegraphics[width=0.33\textwidth, trim={0.6cm 0.65cm 0.1cm 0.0cm},clip]{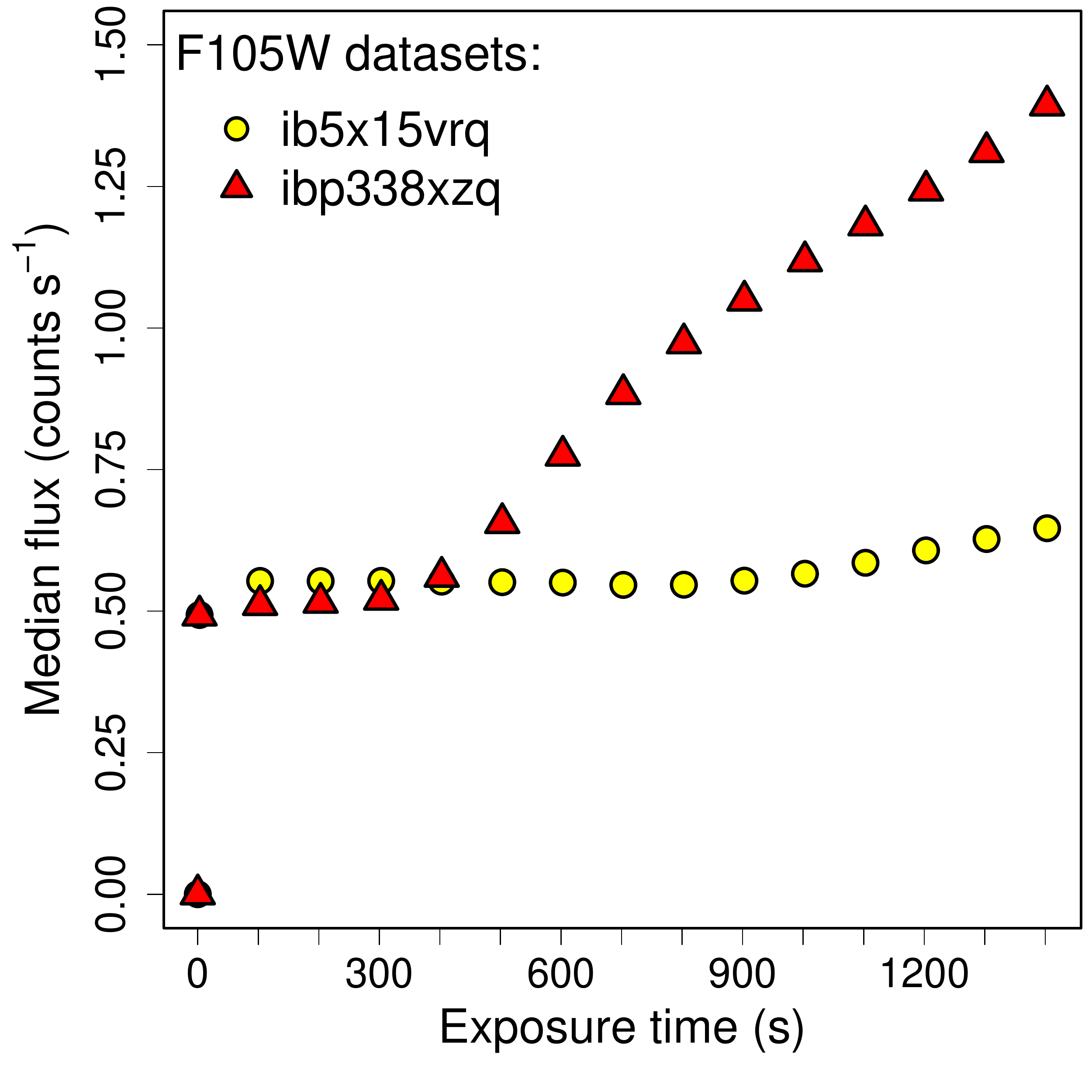}
\begin{overpic}[width=0.33\textwidth]{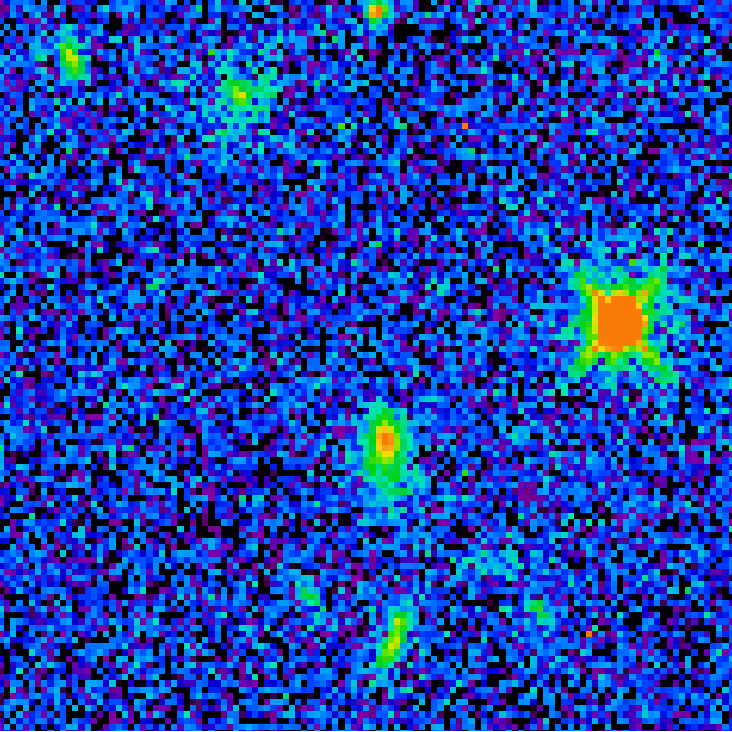}


\put(1,162){\color{black} \colorbox{white}{\textbf{\normalsize Without time sky-variation correction}}}
\end{overpic}
\begin{overpic}[width=0.33\textwidth]{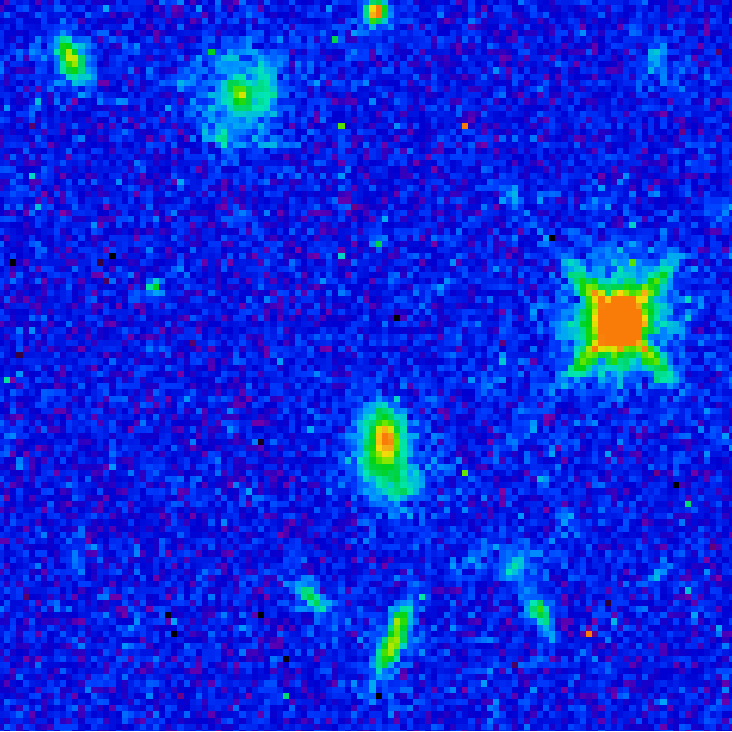}
\put(1,162){\color{black} \colorbox{white}{\textbf{\normalsize Time sky-variation corrected}}}
\end{overpic}
\caption[]{Example of the effects of time varying sky background in two different datasets of WFC3 IR in the F105W band. \emph{Left panel:} Median background flux of {\tt{ib5x15vrq}} (yellow circles) and {\tt{ibp338xzq}} (red triangles). While the dataset {\tt{ib5x15vrq}} shows a nearly constant flux on all the different readouts ($\sim 0.50$ counts s$^{-1}$), {\tt{ibp338xzq}} is clearly affected by time variation of the sky background, increasing from $\sim0.50$ counts s$^{-1}$ to $\sim1.3$ counts s$^{-1}$ after 400s from the start of the exposure. \emph{Middle panel:} Sky-corrected {\tt{ibp338xzq}} \flt\ image without time-varying sky background correction applied. \emph{Right panel:} Sky-corrected {\tt{ibp338xzq}} \flt\ image accounting for the time-varying sky background correction. Both images are set to the same colour scale.} 
\vspace{0.25cm}
\label{fig:timesky}
\end{figure*}

In Fig.\,\ref{fig:persistence_hist} we analyse the persistence contamination level of the WFC3 IR exposures of the HUDF. We measured the amount of pixels that were affected by persistence as a function of the surface brightness contamination that this effect creates. The expected surface brightness level created only by persistence is estimated using the persistence flux predicted by the models. We then determine the fraction of each image that is affected as a function of the surface brightness persistence contamination level. We found that the F160W and F105W are the filters most affected by this issue. For F160W band, $\sim17\%$ of the images present at least half of the image contaminated at surface brightness of $\mu=30$ \magarc. The filter less affected by persistence is F140W, with less than $\sim5\%$ of the images with that level of contamination on at least half of the image. Given that the persistence models are an approximation to the real level of persistence, and they do not fully correct for all the contamination, the general recommendation by STScI is to use them to correct the images and flag those pixels affected up to a certain level, that has to be chosen depending on the science target. After correcting all the images with our improved persistence models, we chose a conservative compromise between losing exposure time and possible contamination by the persistence residuals. We reject all the images that present more than 50\% of the pixels affected at surface brightness brighter than $\mu=30$ \magarc\ (this removes a $10\%$, $8\%$, $5\%$ and $17\%$ of the total amount of exposures from the F105W, F125W, F140W and F160W filter respectively). Finally, we flag any pixel in the valid images that presents a persistence surface brightness larger than $\mu=30$ \magarc.

We conclude that our persistence models are more robust and include many potential persistence sources that were not taken into account in the previous versions of the HUDF mosaics. The reason for this is that we used a four times larger lookback time for the persistence calculation than the official MAST persistence models and the last WFC3 Persistence pipeline as presented in \citet{Long2015}.

\subsection{Time-dependent sky background variation}
\label{Subsec:timesky}

During long exposures, sky background can vary noticeably, introducing a non-linear component to the count rates calculated by \calwf3. This causes non-Gaussian properties in the \flt\ frames of many exposures, with a severe impact in the depth of the final product. This problematic effect is specially common in F105W filter observations. The reason for this is the presence of atmospheric He 10,830\AA\ emission line in the upper atmosphere \citep{Brammer2014}. This emission line falls into the F105W and F110W filters and both HST WFC3 IR grisms. The effect is strongest at low Earth limb angles and under direct sunlight, and usually negligible in the Earth's shadow, but sometimes can be strong even when observing at 40 degrees above the Earth limb. In the worst case scenario, the He 10,830\AA\ line fully dominates the sky background emission \citep[see ][Figure 7.13]{Dressel2012}. Observations with long exposure times will approach closer to the Earth's limb and will be potentially more contaminated by Earth's scattered light and atmospheric emission.

The pixels of those exposures affected by time-dependent sky background will be wrongly classified by \calwf3\ as cosmic rays, thus impeding any type of alignment (see Sect.\,\ref{Subsec:CRAstro}). In order to correct for this effect, we follow a similar procedure as in the Sect. 3.2 of \citet{Koekemoer2012}. We take advantage of the flexibility of \calwf3 to stop and re-start the calibration process at any point, and we subtract the sky background of each independent readout for each exposure (typically 16 readout per exposure), before the cosmic ray rejection process (steps 7 and 8 of Sect.\ref{Subsec:rawdata}). We perform this correction in four steps:

\begin{enumerate}
    \item We run \calwf3, stopping the procedure before the "up-the-ramp" fitting and the cosmic rays identification (Sect.\,\ref{Subsec:rawdata}).
    \item We individually estimate and subtract the sky background emission from each readout of the intermediate \ima\ files. 
    \item We resume \calwf3, obtaining a first approximation to the \flt\ file. 
    \item We use this first \flt\ combined image to create an object mask, using {\textsf{NoiseChisel}} (see Sect.\,\ref{Subsec:masking})
    \item We run again steps 1-3, using this mask to have a better determination of the sky background on each readout. The output is the final \flt\ file.  
\end{enumerate}

\begin{figure*}[h!]
\centering
\vspace{0.25cm}

\begin{overpic}[width=0.33\textwidth]{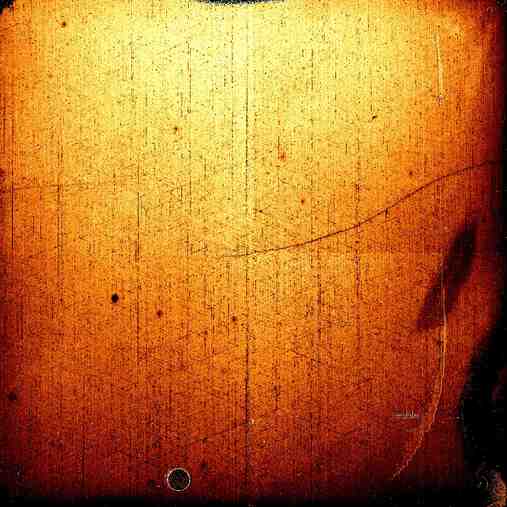}
\put(1,162){\color{black} \colorbox{white}{\textbf{\large 55000 MJD}}}
\end{overpic}
\begin{overpic}[width=0.33\textwidth]{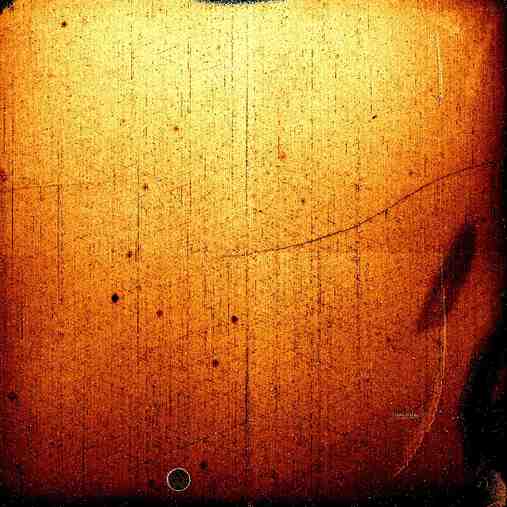}
\put(1,162){\color{black} \colorbox{white}{\textbf{\large 56500 MJD}}}
\end{overpic}
\includegraphics[width=0.33\textwidth]{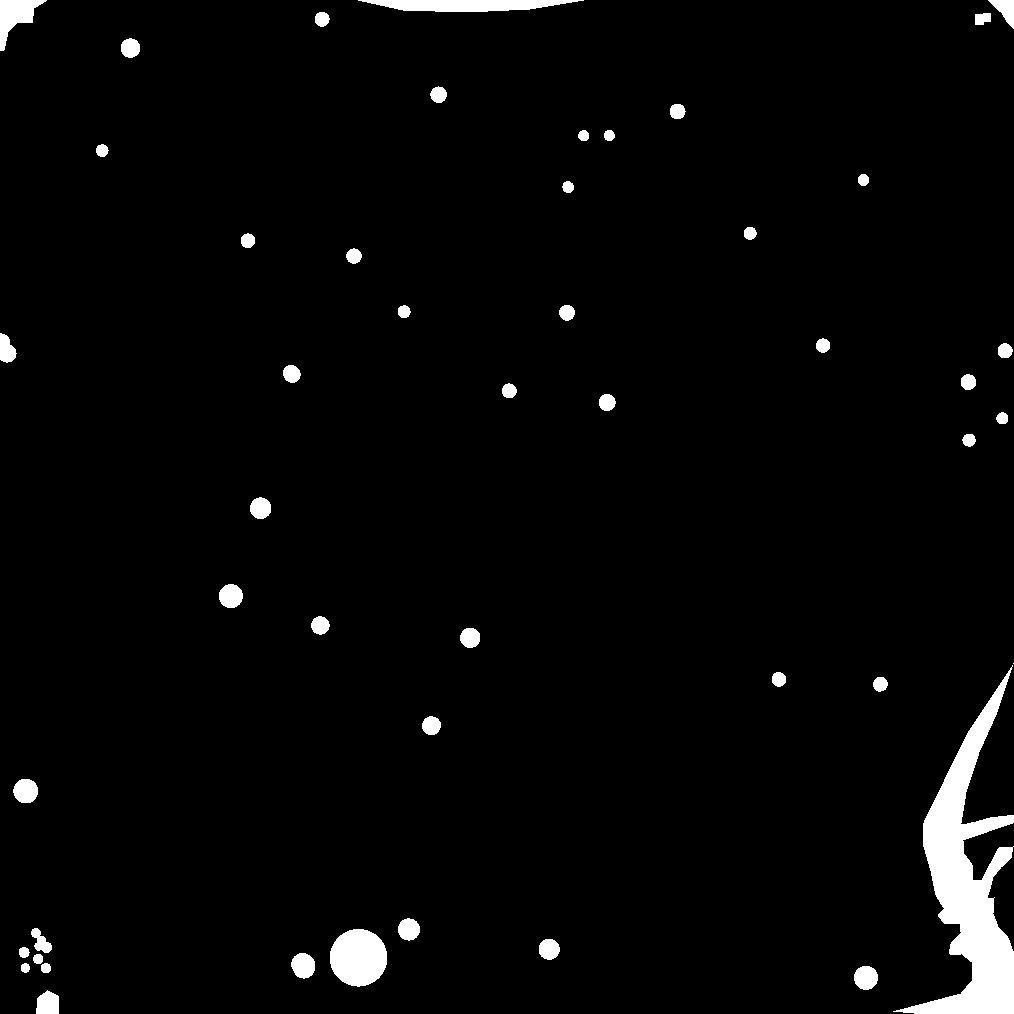}

\caption[]{Example of identification using two extreme flat fields, calculated for June 2009 (55000 MJD, \emph{left panel}) and July 2013 (56500 MJD, \emph{central panel}). \emph{Right panel:} Final data quality array. We represent in white those regions of the detector labelled as bad pixels. We note that this data quality array has to be combined with the individual data quality arrays from each individual exposure, containing the cosmic rays masks, the pixels affected by persistence, and the default time-dependent WFC3 IR blobs.} 
\vspace{-0.25cm}
\label{fig:dq}
\end{figure*}
The final sky-background level for each exposure is calculated as the median sky value of the half of the readouts with less differences between them (first or second half). This constant sky level will be subtracted in a later step. By doing this, we successfully corrected most of the exposures affected by this effect. We must note that this correction is only valid if the time variation of sky background is flat, this is, it does not include any variation of a large scale gradient across the image. We visually inspected each dataset and rejected those exposures that were clearly affected by sky background gradients.   
We show in Fig.\,\ref{fig:timesky} an example of an exposure affected by time variation of the sky background. The left panel shows the median flux for two different exposures as a function of the readout time. In an ideal case, the background count rate should be zero or at least constant, but exposure {\tt{ibp338xzq}} shows a clear increase on the sky background flux starting at $\sim 400$ s from the start of the exposure. Running the standard \calwf3 pipeline leads to the image shown in the central panel of Fig.\,\ref{fig:timesky}, where the sky background has a high-noise level which is not well represented by a Gaussian approximation. We present in the right panel the same exposure corrected by the procedure described above. We stress that both images are calculated from the same \raw\ file, and are represented with the same colour scale. 

In conclusion, this procedure allows us to successfully correct for the effect of time-variation of the sky background, which is a common issue in HST WFC3 IR exposures.

\subsection{Extended data quality array}
\label{Subsec:dataquality}

In order to avoid systematic biases due to the presence of defects in some regions of the detector, we created a manual data quality mask to flag those regions were the flat field cannot fully correct the differences in sensitivity. Besides the effect of those regions of the detector known to have lower sensitivity (wagon wheel, death star), we must pay attention to the presence of the WFC3 IR blobs. These blobs are regions that appear sporadically and cumulatively over the detector and present 5-10\% less sensitivity on the detector \citep{Pirzkal2010}. Blobs are not related to any damage on the detector, but rather to particles that stick to the mirror of the Channel Select Mechanism \citep{Calvi2014}. Their effect is strong enough to become a problem in the final mosaics. One of the most important problems is that the WFC3 IR blobs appear as a function of time. Although there is a continuous monitoring of the presence of new IR blobs \citep{Pirzkal2012}, their appearance cannot be predicted. 

To avoid including frames affected by blobs that were not flagged yet, we conservatively decided to mask manually those pixels with the largest blobs regardless of the epoch when the exposure was taken. We created a set of 16 sky-flat fields per filter (every 100 days, since 55000 to 56500 MJD) as a function of time in order to visually track the appearance of blobs (see Sect.\,\ref{Subsec:flatfield}). We show an example of the time-dependent flat fields used for the selection process in the left and central panels of Fig.\,\ref{fig:dq}. Notice that in the central panel, there are several additional IR blobs visible in the sky flat field. We carefully flagged any region with clear time variation according to our time-dependent flat fields and added them to our master data quality array, which is shown in the right panel of Fig.\,\ref{fig:dq}. We also flagged those regions with low sensitivity or none at all, which are clearly seen in the flat fields \citep[wagon wheel, death star, the unbounded pixels of the top corners and middle strip, and the seven point-like damaged regions on the low left regions, see][Chapter 5.7.7]{Dressel2012}. Finally, we flag in the individual data quality (DQ) extension of all \flt\ files all the pixels included in our extended data quality array and mask from the individual images before sky-subtraction and co-addition.

\subsection{Sky-subtraction}
\label{Subsec:skycor}

In this Section we describe the methods used to remove the sky background from the individual exposures and the final mosaics of the HUDF. We divide this task in two parts: flat sky background estimation (constant across the detector, Sect.\,\ref{Subsec:flat_skycor}) and removal of large scale residual gradients with a two-dimensional sky background (Sect.\,\ref{Subsec:gradients}).

\subsubsection{Flat sky-subtraction}
\label{Subsec:flat_skycor}
\begin{figure*}[]
\centering
\vspace{-0.2cm}
\includegraphics[width=0.44\textwidth]{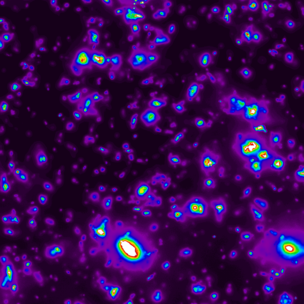}
\includegraphics[width=0.44\textwidth]{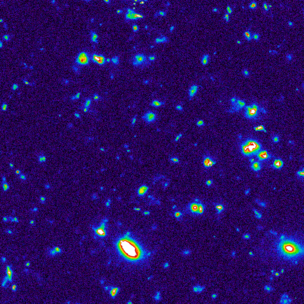}\vspace{0.025cm}
\includegraphics[width=0.44\textwidth]{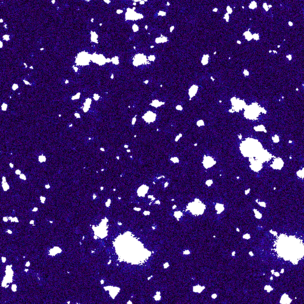}
\includegraphics[width=0.44\textwidth]{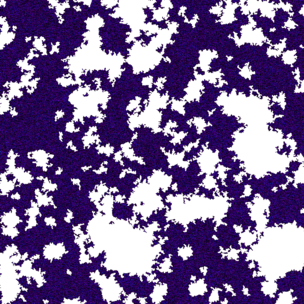}
\includegraphics[width=\textwidth]{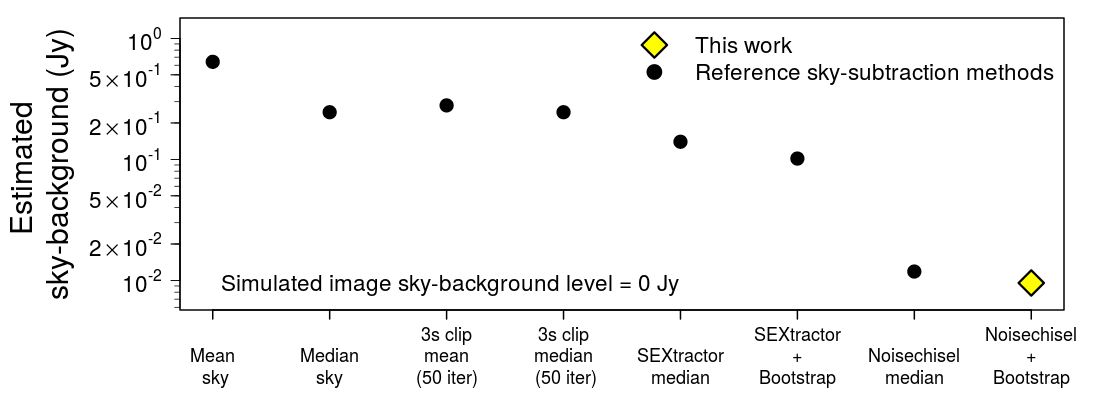}

\caption[]{Comparison of sky-subtraction methods, over a section of the Illustris simulation images. \emph{Top-left panel:} Simulated Illustris Deep Field B section.  \emph{Top-right panel:} Same image with a 1 Jy  wide Poissonian noise added. \emph{Middle-right panel:} Simulated Illustris Deep Field B with noise masked with {\textsf{SEXtractor}} (1$\sigma$ threshold level). \emph{Middle-left panel:} Same image masked with {\textsf{NoiseChisel}} (default configuration). \emph{Bottom panel:} Estimated sky background as a function of different methods. \emph{Black dots:} Reference methods. \emph{Yellow diamond:} Method applied in the present paper. Notice that all methods overestimate the true sky background level, which is equal to zero in the simulated image. See the legend and axis for details.} 
\label{fig:skymethods}
\end{figure*}

A key step in the reduction process of astronomical images is the matching of the sky background level before the final co-adding of the individual frames. Because the near IR background is brighter than in the visible and its time variation is higher, this step can be particularly complicated even for space-based observations. Furthermore, for observations of extended sources or very crowded fields – such as the case of WFC3 IR observations of the HUDF – there are many low-surface brightness features that are buried deep under the $1\sigma$ noise levels of the individual images, only detectable in the final mosaics. Extended discs, stellar halos, and diffuse objects are extremely hard to mask and they contribute to bias the distribution of the pixels selected for sky background determination towards higher values. This leads to a common tendency to over-subtract in the reduction of deep images. Moreover, a biased determination of the sky background level of the different images before co-adding can greatly affect the final mosaic by including additional noise and thus distorting low surface brightness features of the final image. In this section we describe our flat sky background determination method and compare the results with other standard methods including the method applied by default in {\textsf{Astrodrizzle}}.

We first select those pixels that will be used to calculate a flat sky background level by using the masks created with {\textsf{NoiseChisel}} (see Sect.\,\ref{Subsec:masking}). In addition, we remove (set as NaN) from the sample those pixels that are flagged as bad pixels in the DQ array of each image, including those in the extended DQ array (see Sect.\,\ref{Subsec:dataquality}). We also remove those pixels affected by persistence, according to the custom models calculated in Sect.\,\ref{Subsec:persistence}, flagging all pixels that present a persistence surface brightness brighter than $\mu = 30$ \magarc. Then, we calculate the probability density distribution of the median value of the sky background level by using random re-sampling with replacement (bootstrapping). One of the main benefits of the re-sampling methods is to avoid any assumptions of normality on the sample and hence obtain a more accurate distribution for certain statistics. In addition, the shape of the resulting probability distributions gives us information about the presence of outliers and irregularities. This method is much accurate than the sigma clipping methods applied by default by {\textsf{Astrodrizzle}} \citep{Koekemoer2002}, as the main source of bias for a masked array are not the brightest pixels in the image, but the unmasked outskirts of the largest and most diffuse objects in the field of view. Those contaminated pixels are well below the $1\sigma$ interval of the sky background probability distribution. Careful masking based on the noise-based non-parametric algorithms such as {\textsf{NoiseChisel}} combined with robust statistic methods such as bootstrapping provides much more accurate estimations of the sky level.

We present a systematic comparison of multiple sky-subtraction methods on Fig.\,\ref{fig:skymethods}. In order to make a reasonable experiment, we used the simulated images of the Illustris Proyect\footnote{The Illustris Simulation: http://www.illustris-project.org/} \citep{Vogelsberger2014,Nelson2015}. The Illustris project is a large cosmological simulation of galaxy formation, which simulates a volume of ($106.5 $Mpc)$^3$, since $z=127$ to $z=0$. The project made available a set of model deep images\footnote{The Illustris ultra deep fields are publicly available at https://archive.stsci.edu/prepds/illustris/index.html} simulating the observations of HST ACS and WFC3 detectors, JWST MIRI and WFIRST. We took a section of the same size of WFC3 exposure frames ($1014\times 1014$) of the WFC3 F160W Illustris Field B. We added Poissonian noise with a standard deviation equal to 1 Jy. This choice is arbitrary and only for testing purposes. In the bottom panel of Fig.\,\ref{fig:skymethods} we compare the following sky-subtraction methods:
\begin{enumerate}
\item Mean of the unmasked image. 
\item Median of the unmasked image. 
\item 3$\sigma$ sigma clipped (50 iterations) mean of the unmasked image. 
\item 3$\sigma$ sigma clipped (50 iterations) median of the unmasked image. 
\item Median of the {\textsf{SEXtractor}} masked image (using $1\sigma$ detection limit for sources). 
\item Bootstrapping median of the {\textsf{SEXtractor}} masked image (using $1\sigma$ detection limit for sources). 
\item Median of the {\textsf{Noisechisel}} masked image (default configuration).
\item Bootstrapping median of the {\textsf{Noisechisel}} masked image.
\end{enumerate}

We remark that the sky background of the simulated image is equal to zero. This means that the methods that measure sky background levels closer to zero can be classified as better than those that predict higher levels. We found several interesting results: 1) every single tested sky-correction method systematically overestimates the true sky level, which for this simulation equals to zero Jy \citep[see also][for a similar result based a 2D fitting modelling method]{Ji2018}. 2) {\textsf{NoiseChisel}} is extremely efficient to remove the outer and dim regions of the simulated galaxies. The difference with respect to sigma clipping (or median sky methods is larger than one order of magnitude. 3) Bootstrapping does not add a significant improvement to the estimation of the median value, although it is still less biased than a simple median (only 2.5\% less biased), even with {\textsf{NoiseChisel}} masked images. 4) As expected, sigma clipping is not a reliable method for sky-subtraction. The reason to this is that the main bias contributors to the sample of pixels are precisely those pixels which are well below the $1\sigma$ level, and thus those are not masked in the process. Finally, we choose the best subtraction method from those analysed (Bootstrapping + {\textsf{NoiseChisel}} masking) to estimate the final sky value of each exposure before co-addition.

\subsubsection{Two-dimensional sky-subtraction and diffuse light gradients}
\label{Subsec:gradients}

As we increase the depth of astronomical images, any analysis is less affected by the statistical uncertainties of the sky noise and become more dominated by systematic biases (PSF wings, sky gradients, or diffuse scatter light contamination). As a consequence of this, measuring the structure of objects with small but resolved angular sizes in highly crowded fields (such as the HUDF) is a increasingly challenging task. Even in an ideal case without any noise or residual sky background gradients, the structure of small objects is influenced by the presence of other objects in the field of view. For example, most of the objects with small angular sizes from Illustris simulation are on top of the extended wings of a larger close companion (see Fig.\,\ref{fig:skymethods}). 

Besides the scatter light caused by real sources, no astronomical image is absolutely free of artificial sky gradients, and their effects can only be partially corrected. One of the most common methods for tackling with this problem is the subtraction of a two-dimensional sky background, by using $n-$degree polynomial fits, bicubic-spline interpolation on a mesh grid (i.e, {\textsf{SEXtractor}}), or multiple subtractions of median filtered masked frames with different grid sizes \citep[see Sects.\,4.1 and 3.4 from][respectively]{Koekemoer2012,Illingworth2013}. Most, if not all of these methods have to assume some kind of minimum spatial scale for the variation of the sky background (i.e, mesh grid for {\textsf{SEXtractor}}). Any sky background variations larger than this minimum spatial scale will be fitted and subtracted from the corrected image, at least partially, regardless of their astronomical or instrumental nature. Correspondingly, smaller scales will contain an unknown fraction of artificial sky-gradient residuals and real astronomical diffuse light. Therefore, such methods cannot subtract a two-dimensional sky background that fully preserves the outskirts at all angular scales. 

As stated before, highly aggressive sky background subtraction methods allow us to obtain very flat final mosaics on small spatial scales. While this can be an asset for photometry of unresolved objects, HUDF contains objects of very different sizes (from $\sim20$ arcsec to a fraction of arcsec in diameter). Such methods present the disadvantage of removing the outer parts of the largest objects, which are the primary objective of this paper. In \citet{Illingworth2013} and \citet{Koekemoer2012}, the authors decided to calculate a non-flat sky background correction in order to remove residual background features. The chosen sky background spatial scale ranged from 100 pixels (13 arcsec, HUDF12) to 39 pixels (5 arcsec, XDF). The process consisted in the subtraction of two-dimensional sky background arrays, after masking the individual images. In the case of the XDF, the sky background of all the masked images was stacked and subtracted from the individual images. Masking the frames previous to sky background correction partially reduces the over subtraction of the source on the images. Nevertheless --while detectable on the final mosaics-- the outskirts of the largest objects extend far beyond the $1\sigma$ limit of the individual images, and thus usually get included in the two-dimensional sky background fit, biasing the final result. 

Since our primary objective is to recover the outskirts of galaxies in the HUDF at lower redshifts ($z\sim0.6 - 1$), we have to be more careful with our sky-subtraction techniques. \citet[][see Appendix A.6]{Akhlaghi2015} show that even extensively used model-based methods to calculate the background, such as \textsf{SExtractor} \citep{Bertin1996}, fail to interpolate the sky level for the corners of the detector, unless the observers perform a careful and individual setting of the mesh size ({\tt{BACK\_SIZE}}) for each image. This process may depend on the objects and their position on the detector. 
For this paper, we set the parameters of {\textsf{NoiseChisel}} checking the spatial distribution of the valid tiles (a subset or a section of the input array) for sky background estimation. We refer the reader to {\textsf{Gnuastro}} tutorial 2.3 - Detecting large extended targets\footnote{{\textsf{Gnuastro 2.3}} Tutorial - Detecting large extended targets: https://www.gnu.org/s/gnuastro/manual/html\_node/\\Detecting-large-extended-targets.html}
for a detailed tutorial on detecting the low surface brightness wings of extended sources on astronomical images and using {\textsf{NoiseChisel}} to avoid systematic biases on sky subtraction. In our configuration, {\textsf{NoiseChisel}} calculates the sky background over rectangular sections of the image called tiles, removing from the analysis those tiles which present significant differences between the mode and the median. The distance between valid tiles depends on the signal to noise ratio and the distribution of the sources on the image, being less valid tiles (or none) on more crowded regions of the image. For the individual images we set the tile size to 100 pixels, with the {\tt{smoothwidth}} parameter fixed to 5 and a mode-median quantile difference equal to 0.005. The needed number of neighbour tiles for interpolation was set to 5. We modified those parameters until no valid tiles were accepted by the program near the most extended objects. As a final step, we subtracted a final sky background gradient for each mosaic, using again a tile size of 100 pixels with {\tt{smoothwidth}} equal to 5 and the mode-median quantile difference equal to 0.01.

We conclude that as a consequence of protecting the outskirts of the largest objects in our reduction process, we cannot neglect the presence of artificial sky-gradient residuals with spatial scales of several hundreds of pixels or smaller. This is an inevitable consequence of the reduction process that has to be corrected on a case-by-case basis, taking into account the intensity, shape, angular size, and environment of each object to study. In Sect.\,\ref{Sec:Results} we provide some examples to illustrate the benefits of our dedicated mosaics and discuss the required corrections to analyse diffuse light contaminated objects.  
\subsection{Image alignment}
\label{Subsec:CRAstro}

\begin{figure}[t!]
\centering
\vspace{0.25cm}
\includegraphics[width=0.5\textwidth]{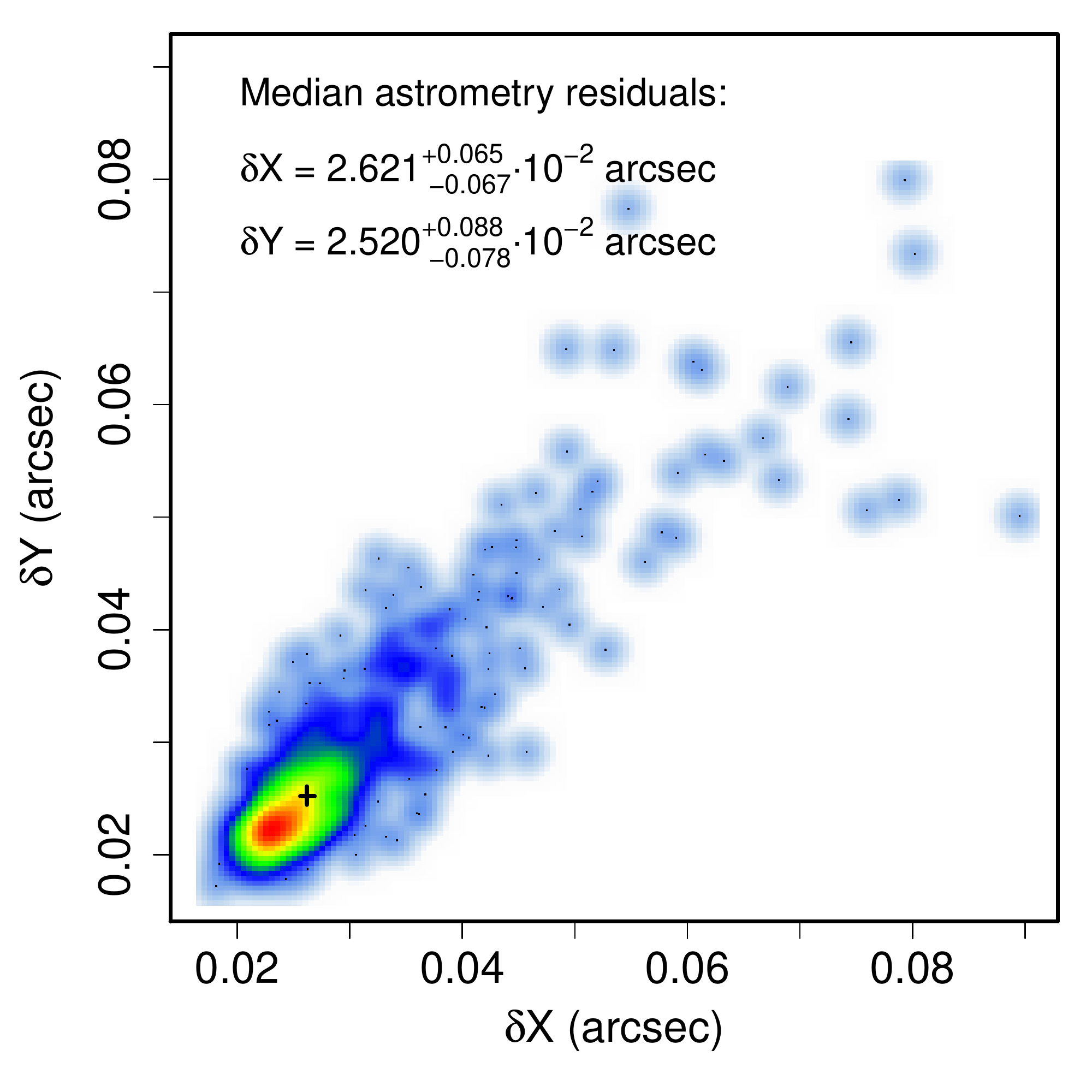}
\caption[]{Colour coded density two-dimensional distribution of the astrometry dispersion of the residuals on the HUDF dataset, according to {\textsf{Tweakreg}}. The black cross represents the median value for the astrometry residuals, in arcsec. Consult the values on the panel.} 
\vspace{0.25cm}
\label{fig:astrometry}
\end{figure}

The astrometric solution of the HST images is precise up to a fraction of arcsec \citep[typically 2-5 pixels,][]{Gonzaga2012}. As a consequence, when comparing images from different visits, it is usual to see that they are not exactly aligned. In order to exploit the full resolving capabilities of WFC3, we need to carefully re-align the images of different visits to a single reference world coordinate system solution (WCS hereafter). To perform this correction we use {\textsf{Tweakreg}}\footnote{{\textsf{Tweakreg}} is publicly available at: http://drizzlepac.readthedocs.io/en/deployment/tweakreg.html}. {\textsf{Tweakreg}} is a task part of {\textsf{Drizzlepac}} and it allows the user to align sets of images to each other and/or an external astrometric reference frame or image. In this paper, we perform this correction following the method presented on \citet{Lucas2015} and \citet{Lucas2015a}: 

\begin{enumerate}
\item We generate four reference catalogues (F105W, F125W, F140W and F160W) for the XDF final mosaics using {\textsf{SExtractor}} \citep{Bertin1996}. We iteratively increased the threshold level ({\tt{DETECT\_THRESH}} and {\tt{ANALYSIS\_THRESH}}) from 1$\sigma$ to 3$\sigma$ in order to remove spurious sources that may affect the final solution. 

\item We group the individual exposures by visits, using {\textsf{AstroDrizzle}} to correct for cosmic rays. We use {\textsf{AstroDrizzle}} to analyse each group of images from the same visit and filter and substitute the identified cosmic rays and flagged pixels with the blotted median of the group \citep[see Chapter 4.2.7 of][]{Gonzaga2012}

\item We generate one catalogue per exposure to correct using the cosmic ray corrected frames.

\item We use {\textsf{Tweakreg}} to find the necessary shift and scaling for each exposure. The program uses the cosmic rays cleaned image, their corresponding source catalog and the reference catalog to its filter, and it matches the astrometry of the individual images to the XDF mosaic.

\item We repeat the previous process, visually checking for misalignments during each iteration and modifying the sigma level of the catalogues and the searching radius of {\textsf{Tweakreg}}. 

\item Finally, we copy the astrometric solution to the \flt\ files. 
\end{enumerate}

One of the outputs from {\textsf{Tweakreg}} is the residuals of the positions for each object and its reference object according to the final astrometric solution for each dataset. In order to measure the average precision that we achieve on our images, we measure the median dispersion of the residuals on all our datasets. The results are shown in Fig.\,\ref{fig:astrometry}. The median dispersion of the residuals is $\sim 0.2$ pixels on both directions of the array, which corresponds to $\sim 0.025$ arcsec. We conclude that our images are sufficiently well-aligned to use a final pixel scale of the mosaics is equal to $0.06$ arcsec, which is the same used by HUDF12 and XDF teams.   

\subsection{Image combination: {\textsf{BootIma}}} 
\label{Subsec:coadd}

Image combination is one of the most common tasks for astronomers. Avoiding systematic biases at this step can be a challenging task. Ground and space based surveys are subdued to a large number of time dependent conditions, either external, such as the effect of cosmic rays, the variation of the sky background or stray-light contamination, that may affect in the form of gradients, or internal, such as variations on the sensitivity of the detector or the position of the different objects over the detector. All these effects produce large variations on the final mosaics, and their typical time-scale can vary between the length of the exposure time to years. The most common methods for co-adding are different types of robust median. Most (if not all) HST images rely on the {\textsf{imcombine}} task of {\textsf{AstroDrizzle}} \citep{Koekemoer2002}. 
{\textsf{AstroDrizzle}} allows the user to use different methods of image combination. The recommended method for a large number of images (more than ten) is "imedian", which corresponds to the median value for each pixel, with the exception of those regions were all pixels were flagged as bad. In that case, the algorithm returns the last value of the stack. This will prevent to have holes in the center of stars, for example. 

Nevertheless, this type of processes does not provide accurate uncertainties for the final mosaics, which are extremely useful for many scientific objectives (that is, measuring the local sky noise, performing accurate photometry, or two-dimensional decomposition). For that objective, we created {\textsf{BootIma}}. {\textsf{BootIma}} is a set of programs written in Python and {\textsf{HTCondor}}\footnote{{\textsf{HTCondor}} is an open-source high-throughput management system for computing-intensive jobs: https://research.cs.wisc.edu/htcondor/index.html} to perform image combination through Bootstrapping. We use this program for the flat field combination and to create the final mosaics. The main task of {\textsf{BootIma}} is the robust combination of a large number of images and estimation of their uncertainties. Such task is computationally expensive, and we optimised it to work in parallel processors and specially, {\textsf{HTCondor}}. We can summarise in two the main reasons for using bootstrapping to calculate the median images: 

\begin{enumerate}

\item It estimates a robust measurement of the confidence intervals, which are provided with the final mosaics and flats obtained with this task, which are extremely useful in the case of deep imaging studies. 

\item It allows us to combine images using weights. This is particularly useful for the HUDF, where multiple observing programs used different exposure times. This also means that the observing sample does not present a pure Gaussian distribution and has to be treated with non-parametric methods. 
\end{enumerate}

A previous step to image combination is the distortion correction. The images produced by WFC3 are affected by geometric distortion, caused by the tilt of the image surface with respect to the path of light \citep{Kozhurina-Platais2012}. If the distortion correction is not applied or accurate enough, the image combination will produce blurred images and distort the PSF. {\textsf{AstroDrizzle}} drizzling code corrects for this geometric distortion using the calibration {\tt{IDCTAB}} files, with a precision of 0.1 pixels. After geometric transformation, {\textsf{AstroDrizzle}} reassigns each pixel to a new and undistorted pixel grid \citep{Fruchter1998}. This process is called drizzling. The new pixel grid can take advantage of the dithering in order to reconstruct a final image with smaller pixel scale by oversampling the input data. We refer the reader to Chapter 6.3 of \citet{Gonzaga2012} for a detailed description of this process. The fraction by which each input pixel is shrunk before being drizzled onto the input image is controlled by the {\tt{final\_pixfrac}} parameter, which we set to 0.8, following the prescriptions for the HUDF12 and the XDF. We choose the Gaussian kernel to distribute the flux of each input pixel onto the new pixel grid. Finally, we obtain a new set of images, corrected for geometric distortion, with a pixel scale of 0.06 arcsec, and aligned into a common pixel grid, ready for image combination. 

{\textsf{BootIma}} performs the median image as follows: first, we register exposure time and date for all the input images. Images are re-ordered as a function of time, to simplify visual inspection. Second, we create an {\tt{hdf5}} datacube, where we store all the individual image arrays, along with their exposure time and date. Once the master {\tt{hdf5}} file is created, we calculate a robust median value along the date direction using a Python-{\textsf{HTCondor}} program built for this purpose. We use the weight images created with {\textsf{Astrodrizzle}} for each image weights during random re-sampling, as they take into account the exposure time and the fraction of counts that correspond to each pixel in the new drizzle grid for each pixel from each original pixel grid. Notice that our datacube contains "holes" (that is, NaN values due to persistence masking and bad pixels). This prevents us from using random re-sampling directly over the full set of images, because all bootstrapping simulations would not have the same amount of pixels. Each pixel of the image has a different sample size, and thus requires independent analysis. Finally, we reconstruct the final image into a single frame. 

We conclude that our method allows us to obtain precise error frames, extremely useful for many science cases while performing robust combination of the images into the final mosaics. In the FITS files that contain the {\textsf{ABYSS}} images, we included two additional FITS extensions, that contain the $+1\sigma$ and $-1\sigma$ uncertainty intervals for each pixel in the field of view.  


\section{Results and discussion}
\label{Sec:Results}

\begin{figure*}[t!]
\vspace{-0.0cm}
\centering
\begin{overpic}[width=\textwidth]{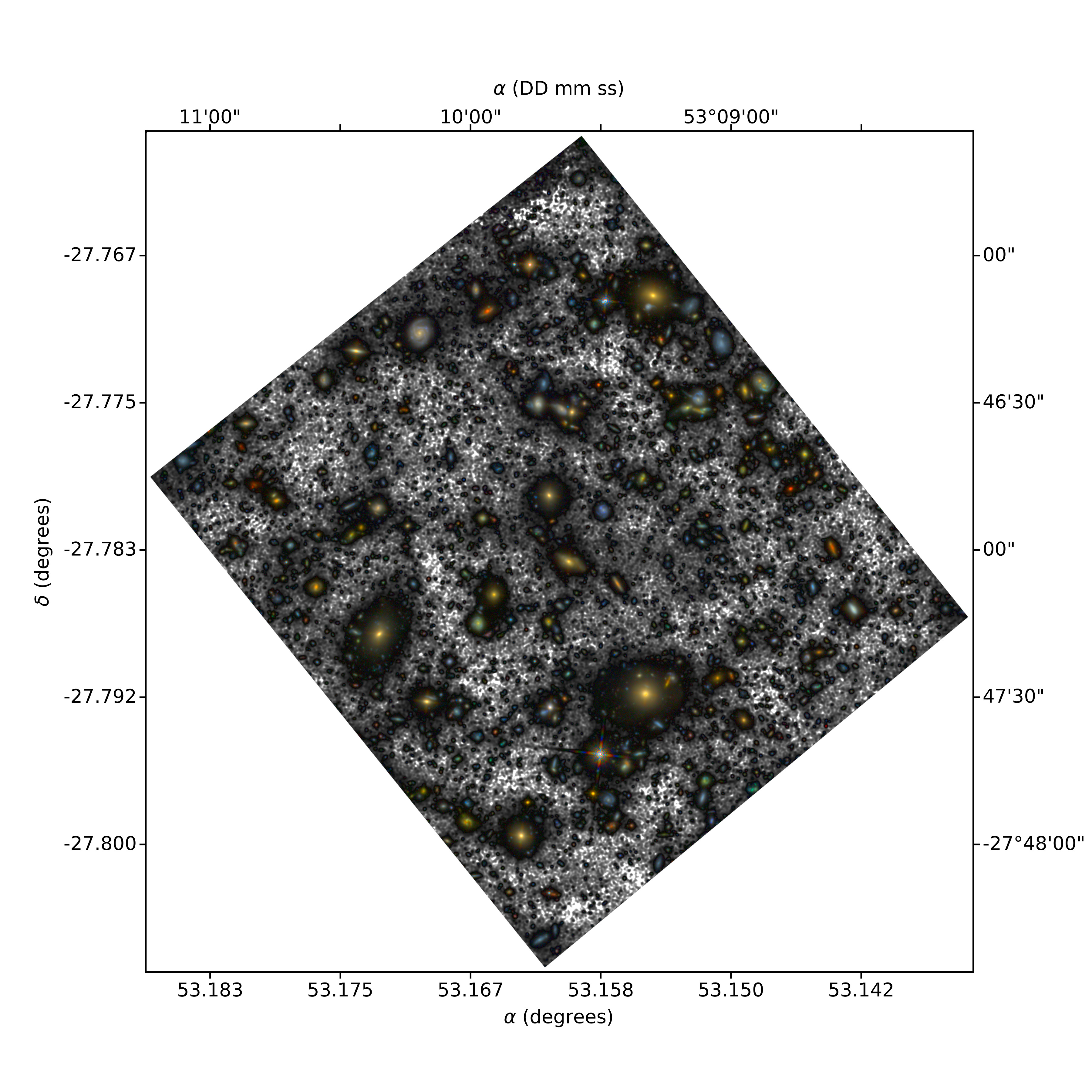}
\put(72,445){\color{black}{\Large\textbf{ABYSS WFC3 IR HUDF:}}}
\put(72,430){\color{blue}{\Large\textbf{Blue: F105W}}}
\put(72,415){\color{green}{\Large\textbf{Green: F125W+F140W}}}
\put(72,400){\color{red}{\Large\textbf{Red: F160W}}}

\put(72,95){\color{black}{\Large\textbf{ABYSS background:}}}
\put(72,80){\color{black}{\Large\textbf{Black: F105W + F125W}}}
\put(72,65){\color{black}{\Large\textbf{\,\,\,\,\,\,\,\,\,\,\,\,\,\,\,\,\,\,F140W + F160W}}}

\end{overpic}

\caption[width=\textwidth]{Luminance-RGB image showing the full depth of the mosaics of the \textsf{ABYSS} version of the HUDF WFC3 IR. The high signal-to-noise parts of the mosaics are represented with colours (\emph{Red:} F160W, \emph{Green:} mean of F125W and F140W bands, \emph{Blue:} F105W). The low signal-to-noise regions are represented in as a black and white background (black regions are brighter than white regions) according to the mean image of the four mosaics (F105W, F125W, F140W, F160W) of the mosaics of the \textsf{ABYSS} (covering HUDF, deep WFC3 IR region).}  
\vspace{0.25cm}
\label{fig:HUDF_deep_color}
\end{figure*}

\subsection{General properties}
\label{Subsec:Results}

In this section we detail the morphological and photometric properties of the newly reduced mosaics, which we present in the colour image of Fig.\,\ref{fig:HUDF_deep_color}. We compare the results from our images to the original HUDF12 \citep{Koekemoer2012} and the XDF \citep{Illingworth2013} mosaics. 

\subsubsection{Recovering the low-surface brightness structure}
\label{Subsec:mosaic_difference}

\begin{figure*}[]
\centering
\vspace{-0.25cm}
\begin{overpic}[width=0.49\textwidth, clip, trim=0 0 80 40]{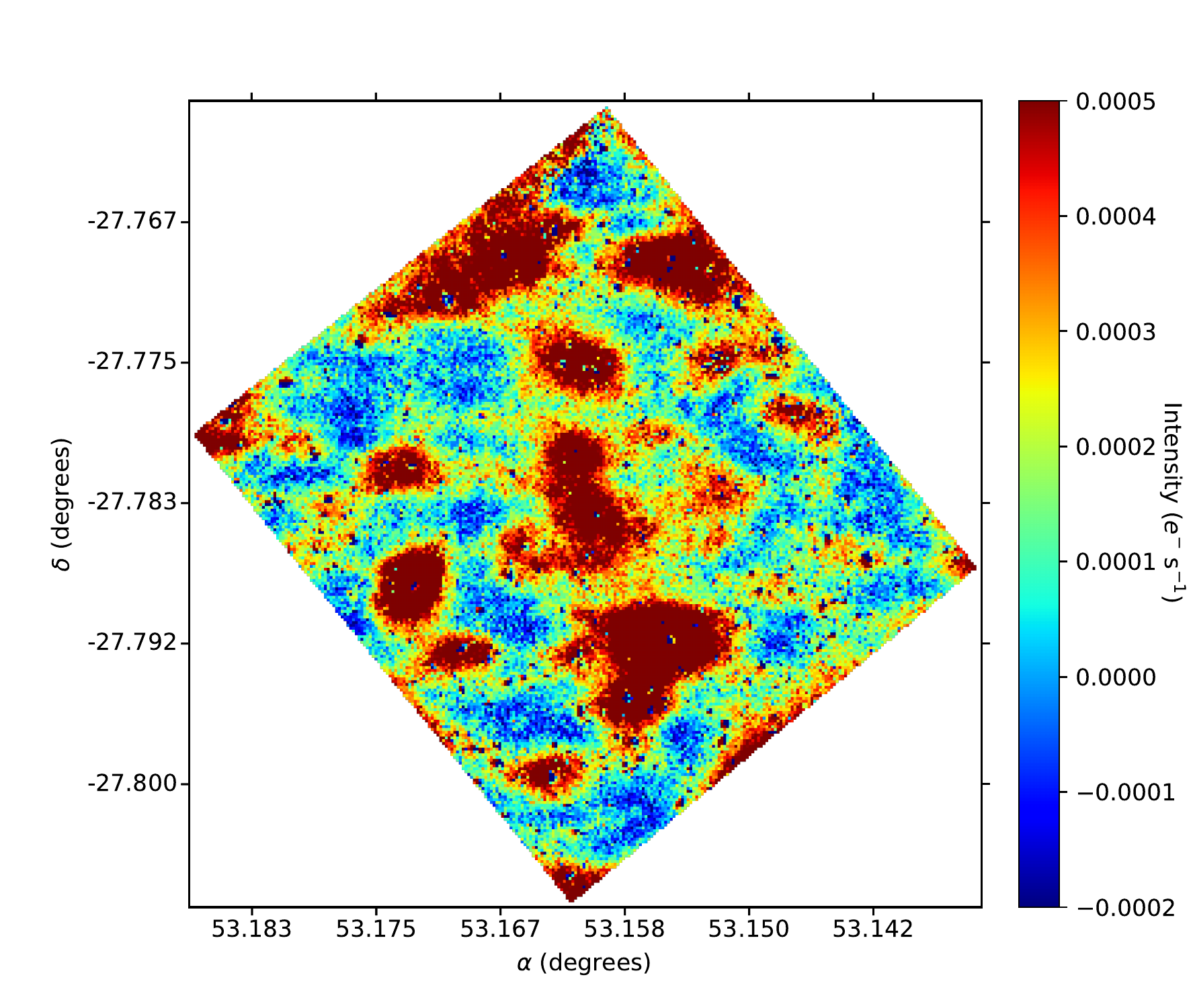}
\put(55,215){\color{black} \textbf{ABYSS-XDF}}
\put(55,205){\color{black} \textbf{F105W}}
\end{overpic}
\begin{overpic}[width=0.49\textwidth, clip, trim=80 0 0 40]{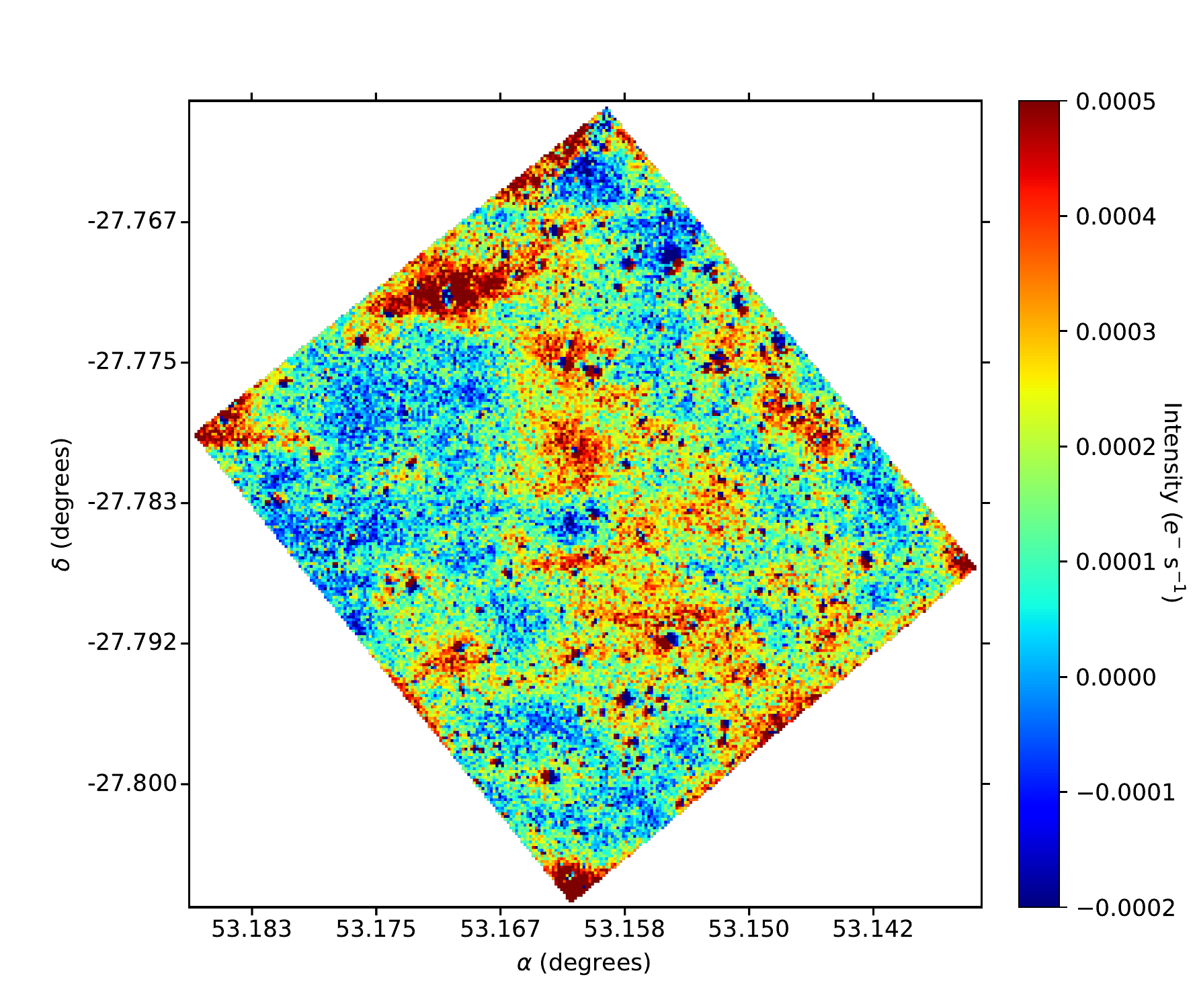}
\put(10,215){\color{black} \textbf{ABYSS-HUDF12}}
\put(10,205){\color{black} \textbf{F105W}}
\end{overpic}
\begin{overpic}[width=0.49\textwidth, clip, trim=0 0 80 40]{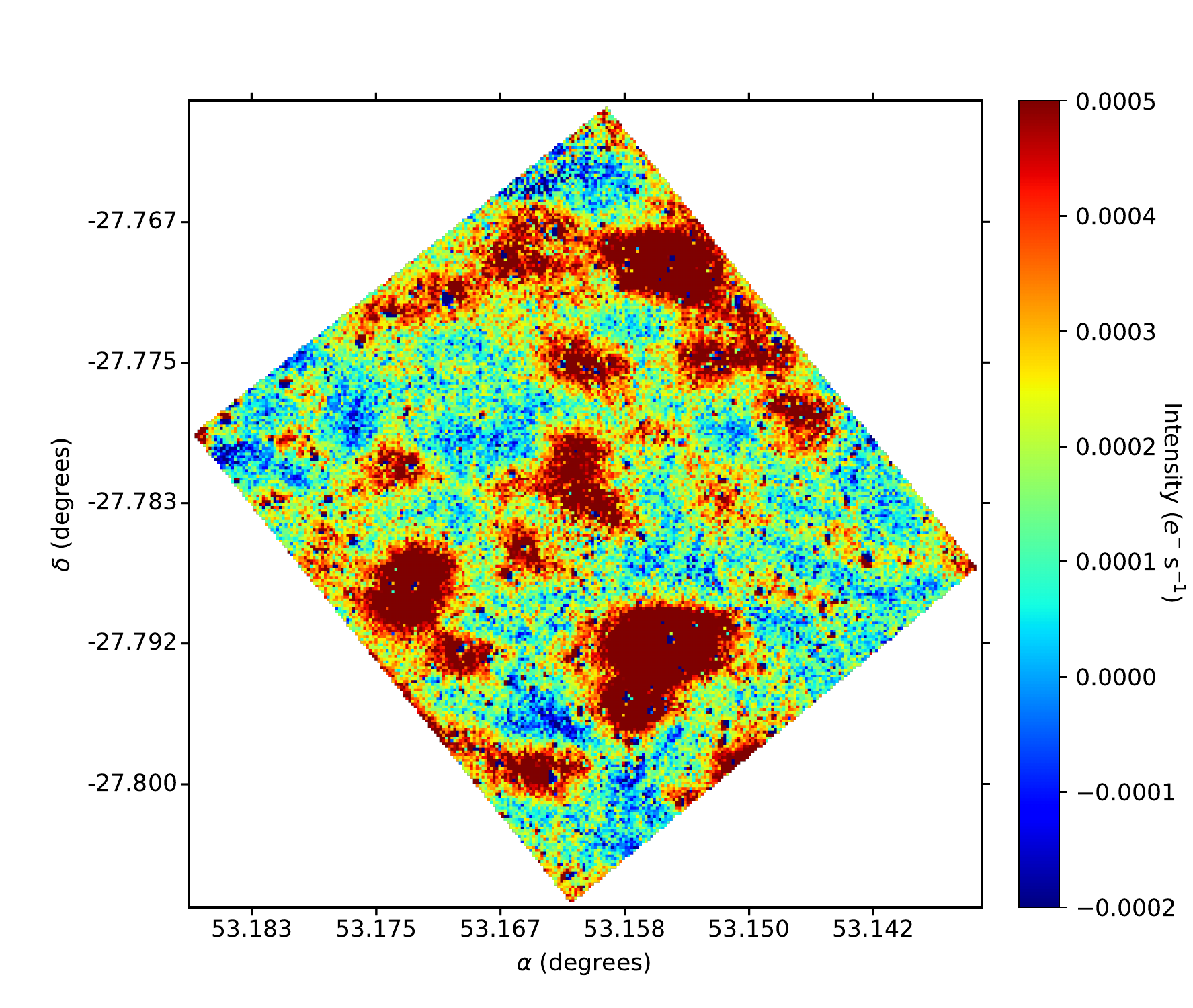}
\put(55,215){\color{black} \textbf{ABYSS-XDF}}
\put(55,205){\color{black} \textbf{F125W}}
\end{overpic}
\begin{overpic}[width=0.49\textwidth, clip, trim=80 0 0 40]{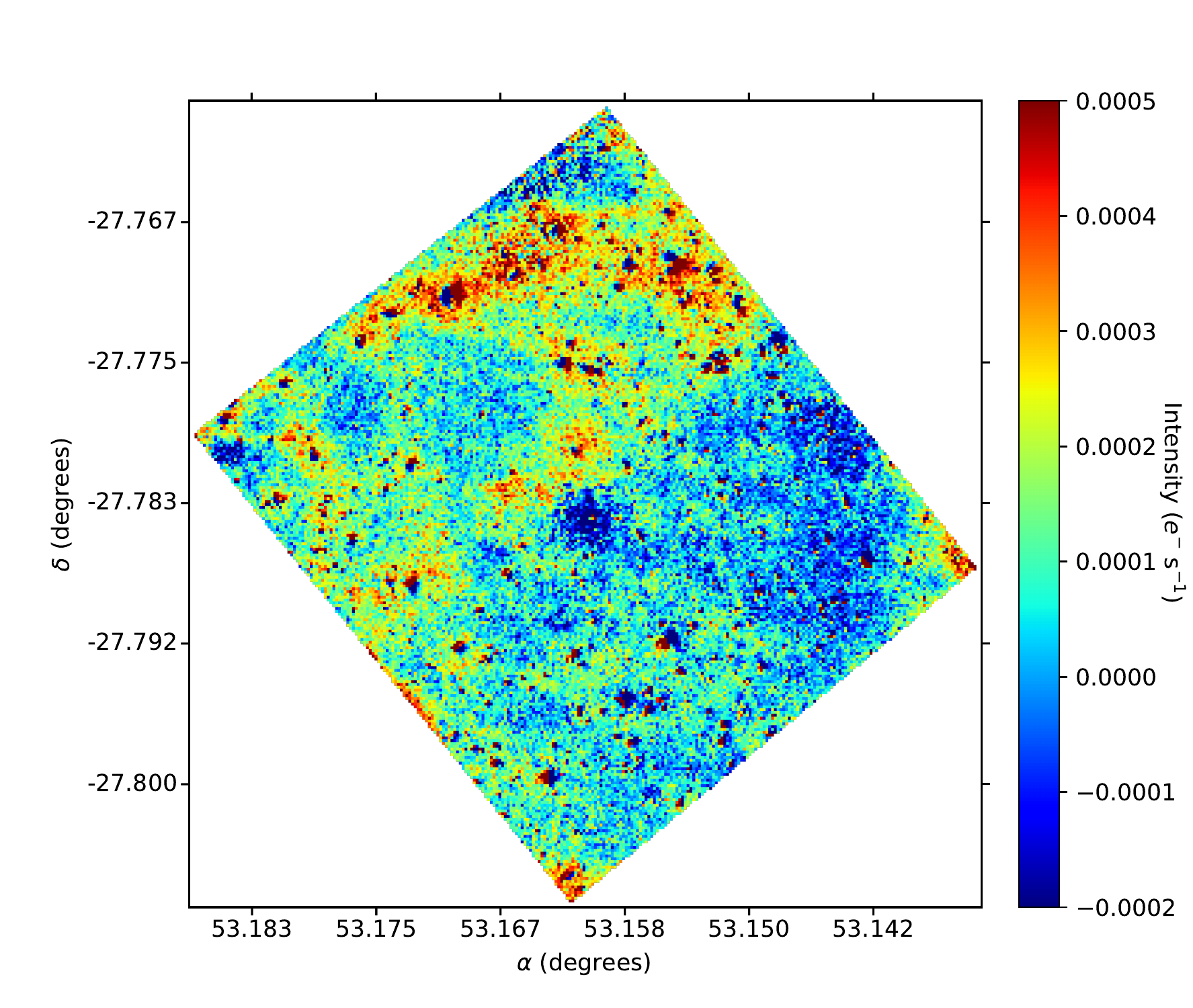}
\put(10,215){\color{black} \textbf{ABYSS-HUDF12}}
\put(10,205){\color{black} \textbf{F125W}}
\end{overpic}

\caption[]{Comparison of the \textsf{ABYSS} mosaics of the HUDF WFC3 IR and the previous releases. Each plot represents the intensity difference between our version of the mosaics (\textsf{ABYSS}) and the reference versions (left panels: XDF, right panels: HUDF12), for the F105W filter (top panels) and the F125W filter (bottom panels). See the colour bar for reference.} 
\vspace{-0.25cm}
\label{fig:HUDF_bin1}
\end{figure*}

\begin{figure*}[]
\centering
\vspace{-0.25cm}
\begin{overpic}[width=0.49\textwidth, clip, trim=0 0 80 40]{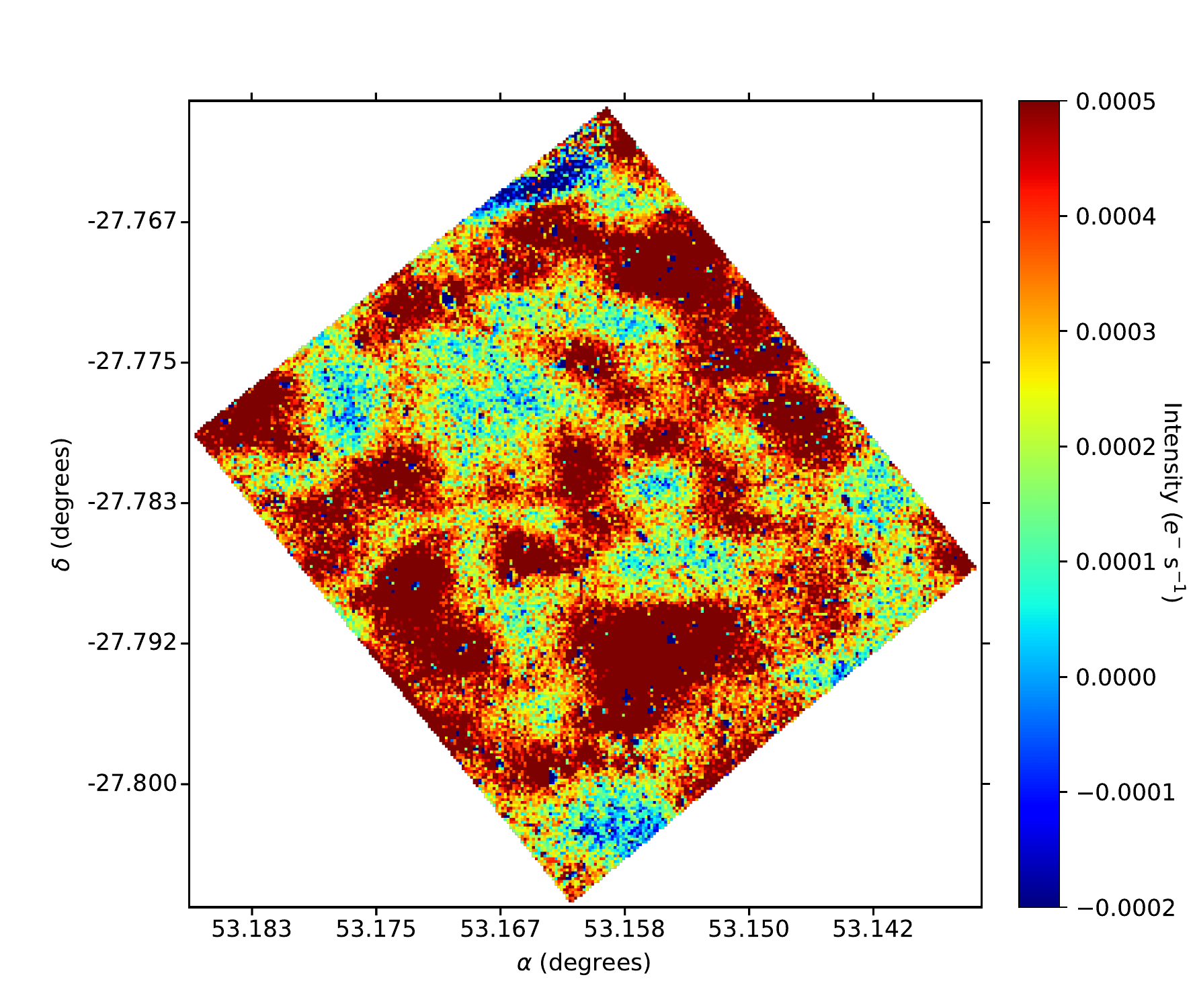}
\put(55,215){\color{black} \textbf{ABYSS-XDF}}
\put(55,205){\color{black} \textbf{F140W}}
\end{overpic}
\begin{overpic}[width=0.49\textwidth, clip, trim=80 0 0 40]{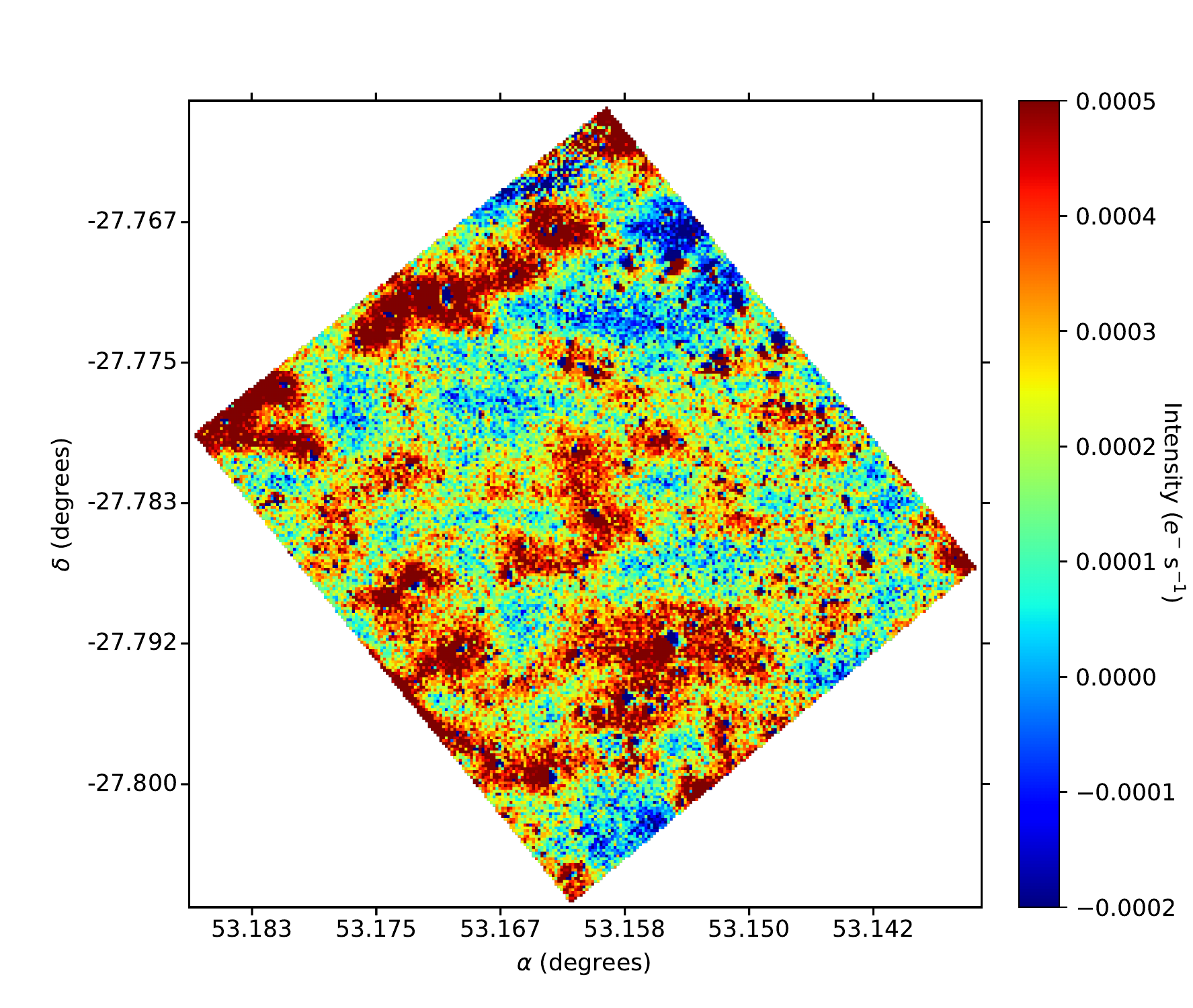}
\put(10,215){\color{black} \textbf{ABYSS-HUDF12}}
\put(10,205){\color{black} \textbf{F140W}}
\end{overpic}
\begin{overpic}[width=0.49\textwidth, clip, trim=0 0 80 40]{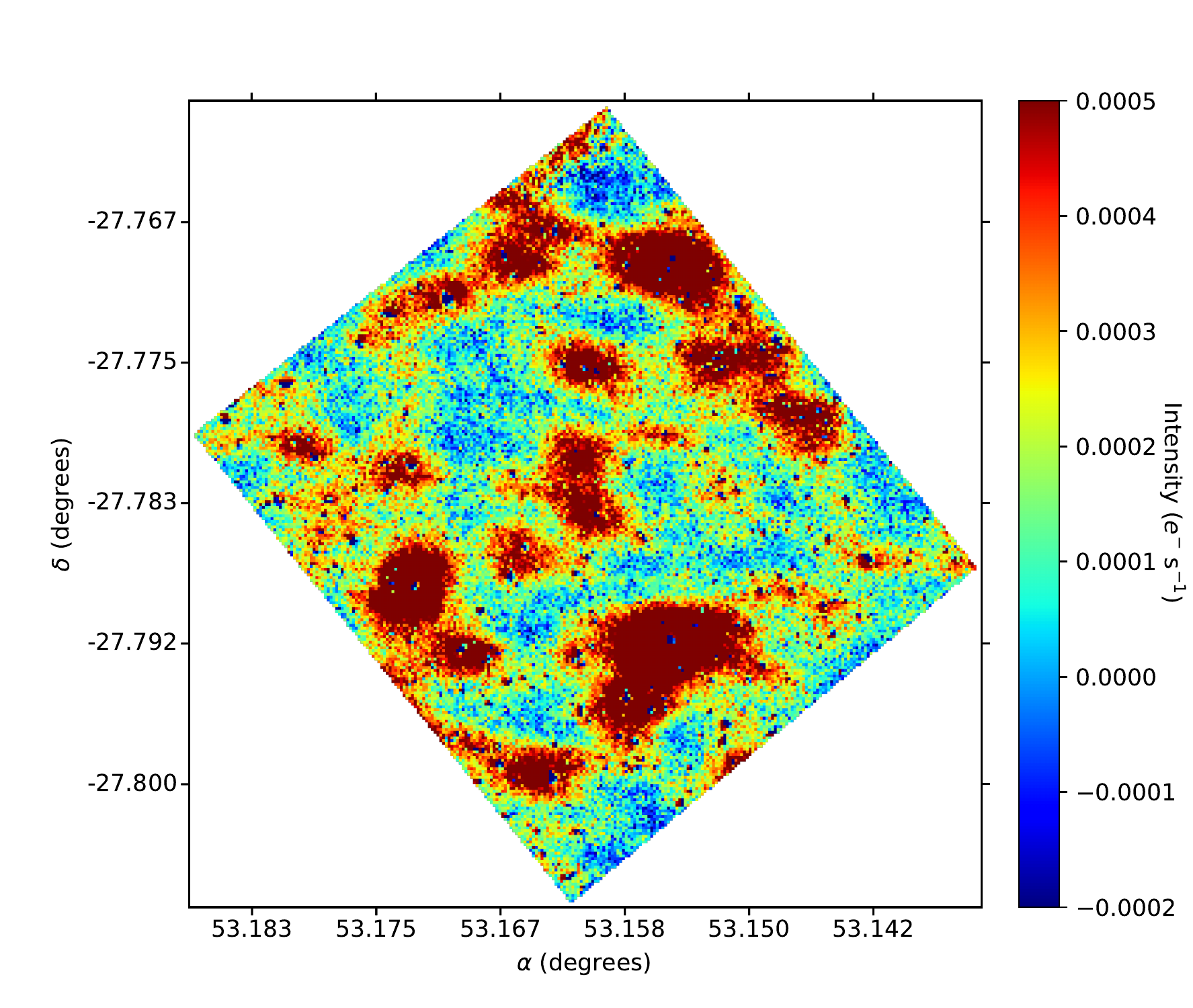}
\put(55,215){\color{black} \textbf{ABYSS-XDF}}
\put(55,205){\color{black} \textbf{F160W}}
\end{overpic}
\begin{overpic}[width=0.49\textwidth, clip, trim=80 0 0 40]{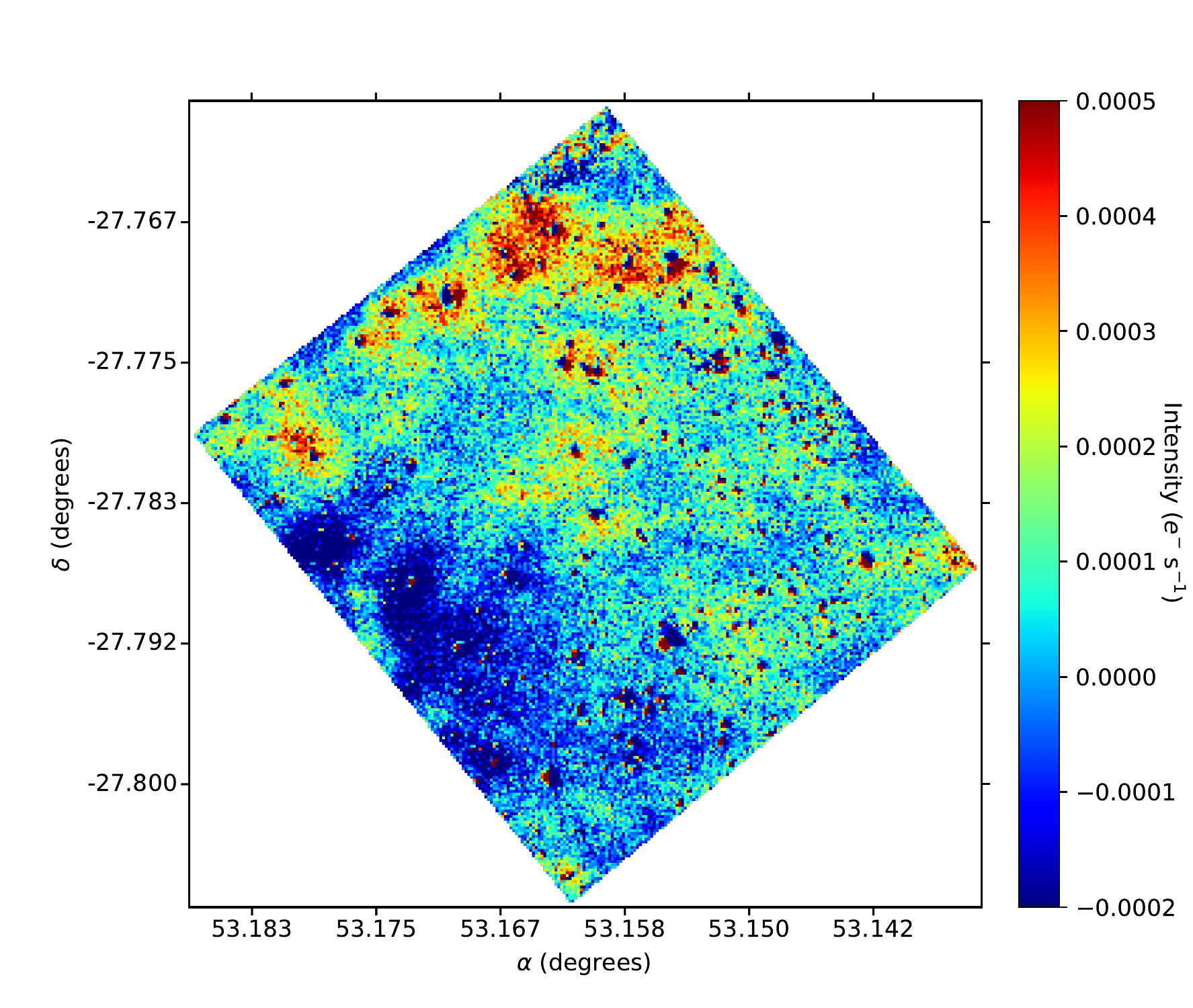}
\put(10,215){\color{black} \textbf{ABYSS-HUDF12}}
\put(10,205){\color{black} \textbf{F160W}}
\end{overpic}

\caption[]{Comparison of the \textsf{ABYSS} mosaics of the HUDF WFC3 IR and the previous releases. Each plot represents the intensity difference between our \textsf{ABYSS} mosaics of the HUDF WFC3 IR and the previous reductions (left panels: XDF, right panels: HUDF12), for the F140W filter (top panels) and the F160W filter (bottom panels). See the colour bar for reference.} 
\vspace{-0.25cm}
\label{fig:HUDF_bin2}
\end{figure*}

In Figs.\,\ref{fig:HUDF_bin1} and \ref{fig:HUDF_bin2} we perform a comparison of the low surface brightness properties of the three different versions of the mosaics compared in this paper: HUDF12, XDF and our dedicated version of the HUDF mosaics which we called \textsf{ABYSS}. We perform the same analysis for the four available filters. In order to reduce the noise and enhance the differences between our mosaics and the previous versions in the low-surface brightness regions, we have re-sampled the image binning the data to 0.6 arcsec boxes ($10\times10$ pixels). 

Visual inspection of the binned difference frames reveals that the XDF mosaics present large over-subtracted regions, which are centred over the regions where the most extended objects are. They are easily identifiable in the difference intensity images that we present on Figs.\,\ref{fig:HUDF_bin1} (for F105W and F125W) and \,\ref{fig:HUDF_bin2} (for F140W and F160W). Comparing the light distribution of the largest objects from the \textsf{ABYSS} mosaics and the morphology of the difference intensity images for the XDF we find that they create almost a mirror image of the intensity mosaics. Moreover, the distribution of the residuals for the XDF is strikingly similar in the four filters, being more clear in the F105W and more noisy in the F140W (this is an expected effect due to the relative depth of the mosaics). 

Although, in principle, this effect may be caused both by over-subtraction in the regions around the largest objects of the XDF or by under-subtraction in our \textsf{ABYSS} version, because we also applied a two-dimensional sky background correction to our mosaics, as \citet[][see Sect.\,\ref{Subsec:gradients}]{Illingworth2013}, it is highly unlikely that the correction that we applied generate under-subtraction around the largest objects. First, because the typical scale-length of our gradients is one third or half of the field-of-view, much larger than the size of the largest objects on the HUDF, while the variations that we detect on the binned difference frames are much smaller. And secondly and most importantly, two-dimensional sky-subtraction tends to fit the extended light of the galaxies, over subtracting these regions, not the opposite effect. 

Interestingly, we found that the HUDF12 presents a similar effect at a lower scale. The distribution of residuals for the F105W and the F140W filters are similar for the XDF and the HUDF12 mosaics when comparing with our mosaics, although the differences are less intense. Inspecting the difference intensity images corresponding to those mosaics (right panels of Figs.\,\ref{fig:HUDF_bin1} and Figs.\,\ref{fig:HUDF_bin2}), we can easily see how the extended light around one of the largest elliptical galaxies, HUDF-5 \citep[$\alpha=53.15545, \delta=-27.79150$,][]{Buitrago2017}, was over subtracted, similarly as in the case of the XDF mosaics. Nevertheless, the west region of the F125W of the HUDF12 is less over subtracted than the rest, presenting a lower level of residual intensity in Fig.\,\ref{fig:HUDF_bin1}. We found a similar effect in the south-east region of the HUDF12 F160W mosaic (see Fig.\,\ref{fig:HUDF_bin2}), where the HUDF12 predicts more flux than our own mosaics. Neither this effect nor the discussed on F125W was found when comparing to the XDF, suggesting that the issue is due to the HUDF12 mosaics.

\begin{figure}[]
\centering
\includegraphics[width=0.49\textwidth]{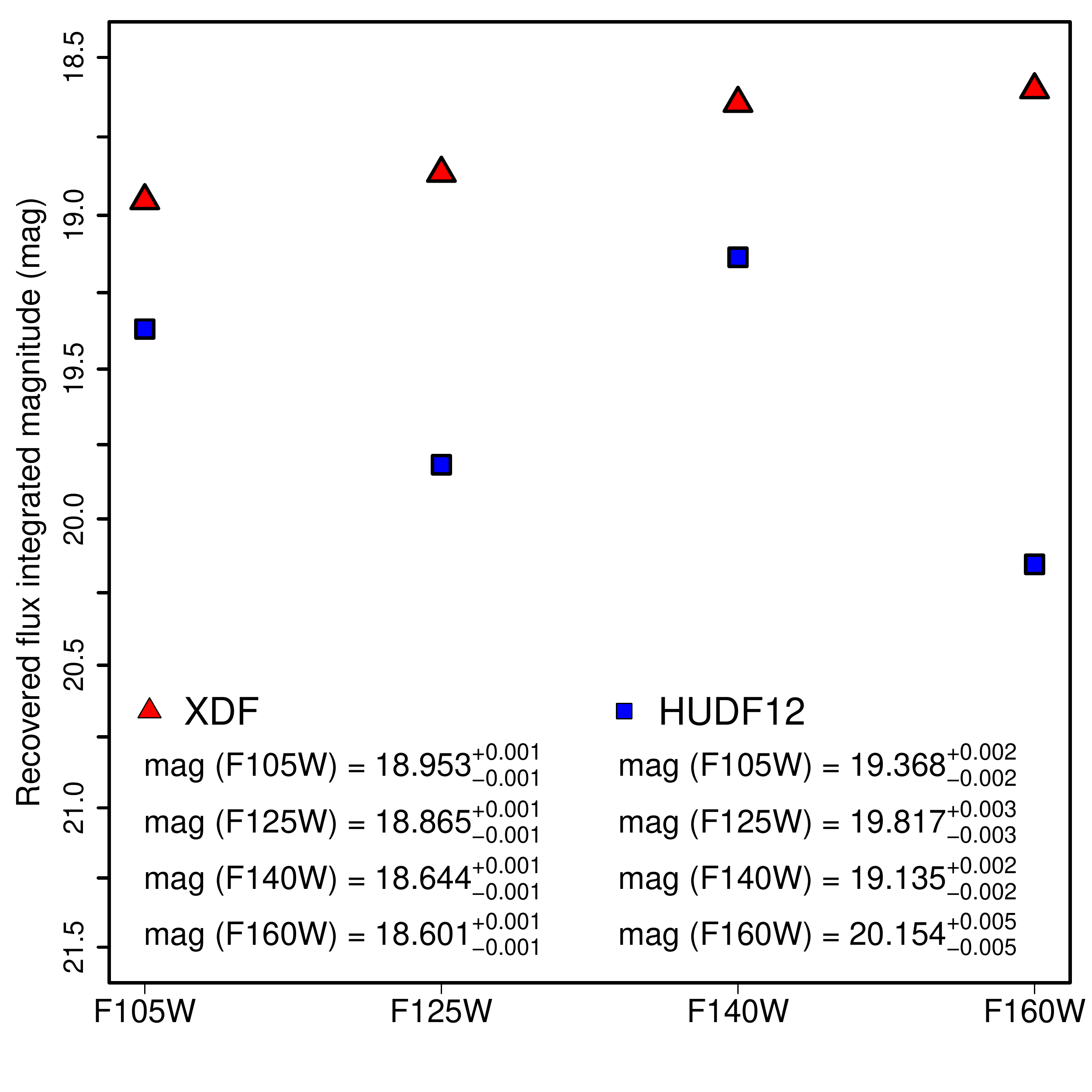}
\caption[]{The missing light of the HUDF measured as the integrated magnitude of the regions with $\overline{\mu} > 26$ \magarc. These are calculated with the difference images between the \textsf{ABYSS} mosaics and the XDF (red triangles) and the HUDF12 (blue squares). See the legend for the values and uncertainties of the different bands.}
\label{fig:lostlight}
\end{figure}

As an illustrative exercise, we calculated the equivalent integrated magnitude of the light recovered on the low surface brightness regions of the \textsf{ABYSS} HUDF mosaics, compared to the previous versions. In order to do that, first we calculate the mean surface brightness on 1.2 arcsec boxes (20 pixels). Due to the fact that small differences in magnitude at the brightest cores of the galaxies may dominate over the differences on the dimmest regions of the images, we select only the regions were the mean surface brightness is lower than $\overline{\mu}=26$ \magarc\ for each individual filter. Finally, we integrate the difference in flux between our mosaics and the reference images (XDF and HUDF12). We show the resulting equivalent magnitudes in Fig.\,\ref{fig:lostlight}. We found that the amount of recovered light is equivalent to a $\sim 19$ mag source when comparing to the XDF and to a $\sim 20$ mag for the HUDF12. This is comparable to the brightness of some of the largest objects in the HUDF. This result is nearly constant for all the filters, although we found that the integrated magnitude for the F125W and F160W images of the HUDF12 is dimmer, in agreement with the results found on the difference maps (see Figs.\,\ref{fig:HUDF_bin1} and \ref{fig:HUDF_bin2}). We must remark that this result highly depends on the size of the image. In a larger area, the total magnitude missing will be larger. In this sense, the values provided here should serve only to compare the equivalent effect of over-subtraction of low surface brightness features on brighter sources. 

We conclude that our images successfully recover most of the extended light around the largest objects from the HUDF WFC3 IR. The recovered light is easily identified as positive counts when subtracting the previous versions of the mosaics from the \textsf{ABYSS} HUDF WFC3 IR images around these objects. 

\subsubsection{Photometric consistency with previous HUDF WFC3 IR images}

\begin{figure*}[]
\centering
\vspace{0.25cm}
\includegraphics[width=0.49\textwidth]{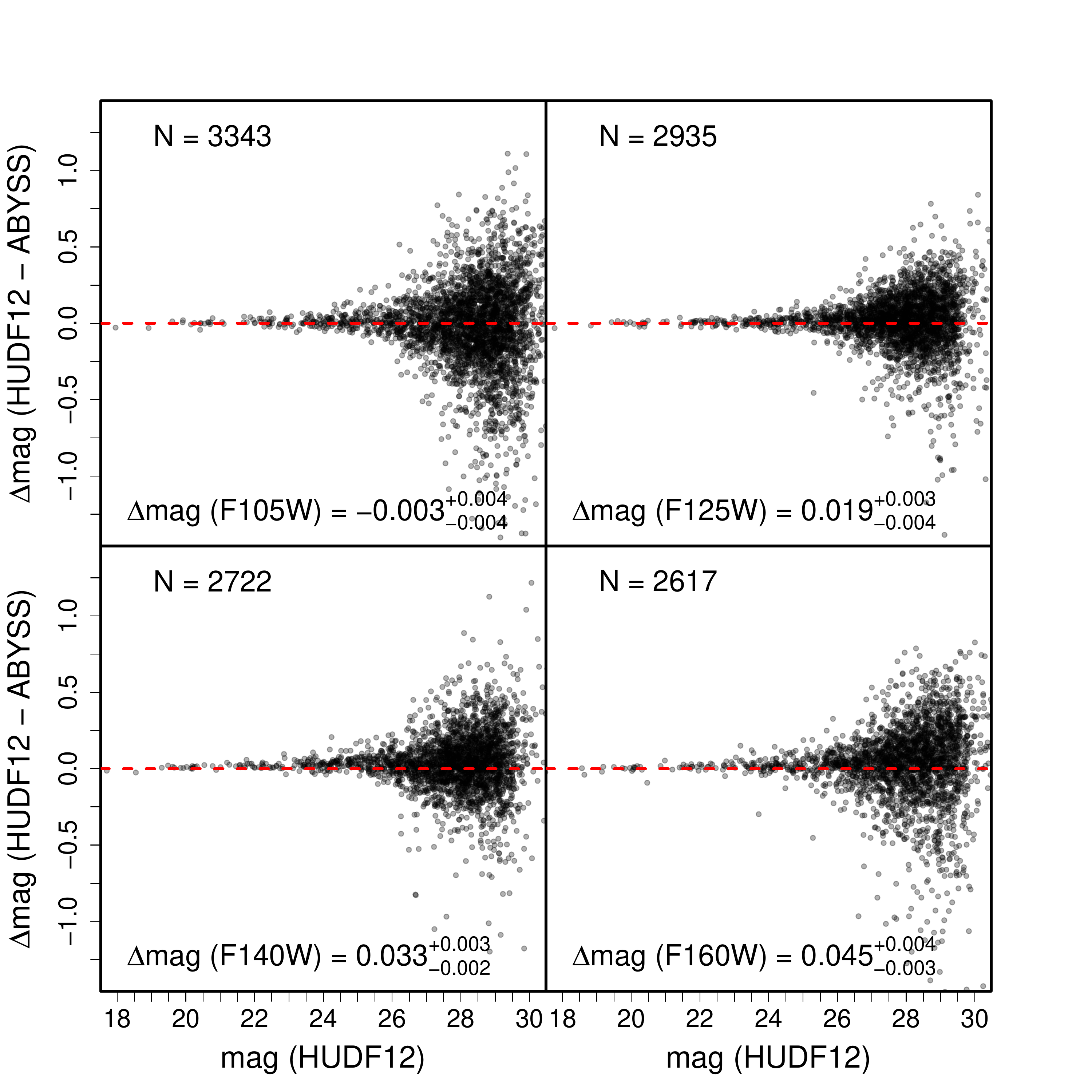}
\includegraphics[width=0.49\textwidth]{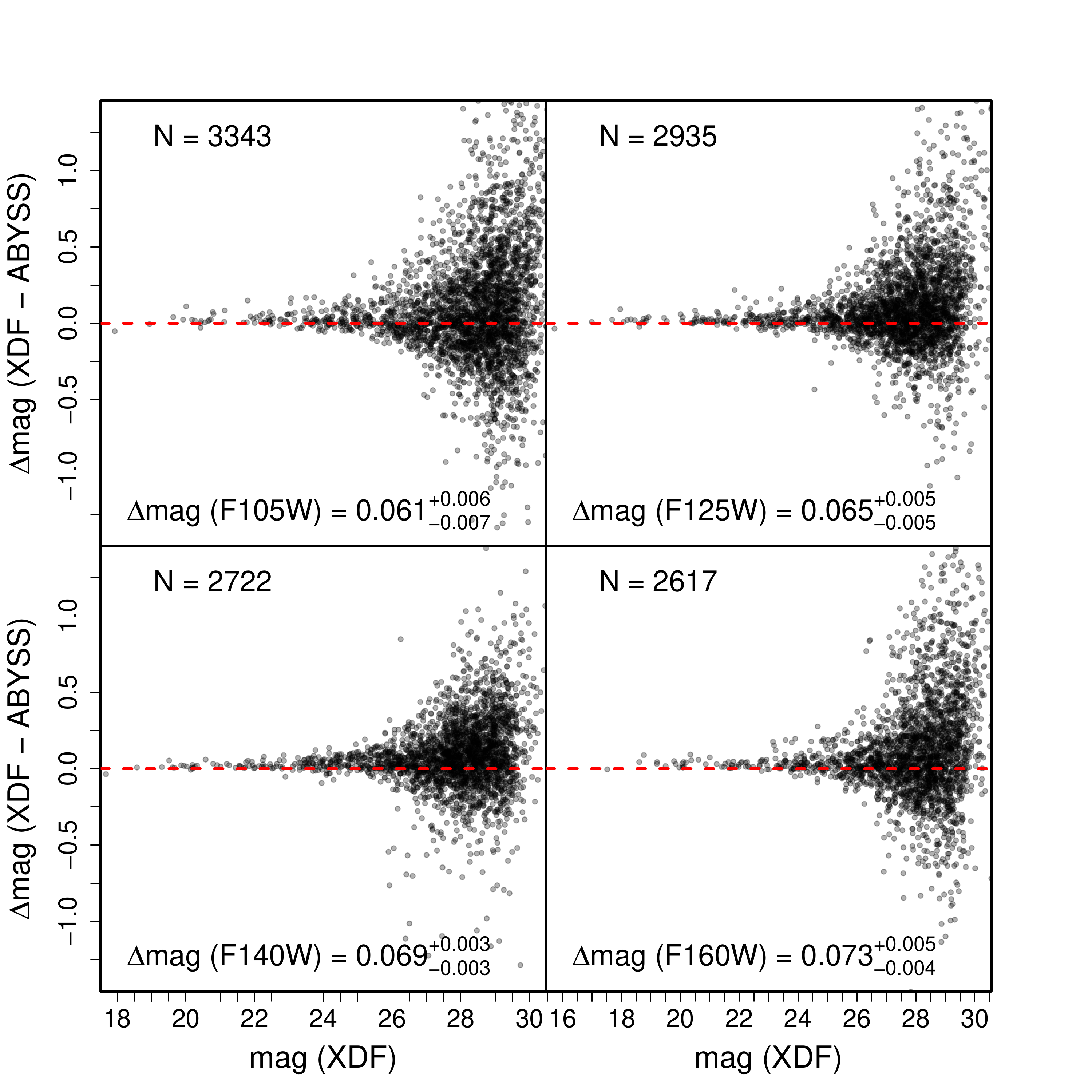}
\caption[]{Comparison of the HUDF photometry of several sources for our {\textsf{ABYSS}} reduction and the previous releases. Each panel shows the magnitude differences with their corresponding reference catalogue as a function of the object magnitude. \emph{Left panel:} \textsf{ABYSS} vs. HUDF12. \emph{Right panel:} \textsf{ABYSS} vs. XDF. The magnitudes are measured using fixed non-parametric apertures in each object and filter calculated with {\textsf{Gnuastro}} using the HUDF12 segmentation maps.} 
\vspace{0.25cm}
\label{fig:HUDF_photometry}
\end{figure*}

In this section we study the photometric properties of the final mosaics, comparing with the results from the previous reductions of the WFC3 IR mosaics for the HUDF. In order to do that, we measure and compare the magnitude of $\sim 2500$ objects on the HUDF and identified using {\textsf{NoiseChisel}} and measured using {\textsf{MakeCatalog}} (both part of {\textsf{Gnuastro}}, see \citealt{Akhlaghi2016} for a discussion about separate detection and catalogue production on astronomical surveys). We created a catalogue per filter and reduction set (12 final catalogues in total). We provide the number of objects identified on the mosaics created for this test on the legend of each panel in Fig.\,\ref{fig:HUDF_photometry}. 

The {\textsf{NoiseChisel}} segmentation maps are highly dependent on the shape and amount of extended light of the objects. Because of this reason, we used one mosaic per filter to calculate the segmentation maps. In this case, we used the mosaics from the HUDF12. Notice that the exact photometry of the objects depends on this choice, but not the conclusions from this test. Thus, the segmentation maps are fixed for the three versions used to compare and independent for each one of the four filters. Finally, {\textsf{MakeCatalog}} calculates the integrated magnitude, correcting for the local sky background for each object.  

In Fig.\,\ref{fig:HUDF_photometry} we compare the differences in magnitude between our reduction and the HUDF12 and the XDF mosaics. We found that the median differences between the magnitudes are $\Delta m_{HUDF12-ABYSS}=0.026^{+0.016}_{-0.013}$ mag when comparing to the HUDF12 versions of the mosaics, and slightly larger ($\Delta m_{XDF-ABYSS}=0.067^{+0.006}_{-0.005}$ mag) when comparing with the XDF mosaics. These differences are negligible and (most importantly) not systematic. The objects in our filters are not systematically brighter or dimmer than in the previous releases. In addition, we found that they are compatible with the differences that we found when comparing the XDF with the HUDF12 mosaics using the same method and apertures ($\Delta m_{F105W} = 0.071^{+0.006}_{-0.006}$, $\Delta m_{F125W} = 0.047^{+0.004}_{-0.005}$, $\Delta m_{F140W} = -0.030^{+0.003}_{-0.003}$, $\Delta m_{F160W} = 0.021^{+0.005}_{-0.004}$). We conclude that the photometric analysis in fixed apertures does not reveal any systematic bias or significant differences when comparing to the results and differences between the previous versions of the HUDF mosaics. 

\subsubsection{Surface brightness limiting magnitude maps}
\label{Subsec:maglim}
\begin{figure*}[]
\centering
\vspace{-0.25cm}
\begin{overpic}[width=0.49\textwidth, clip, trim=0 0 70 20]{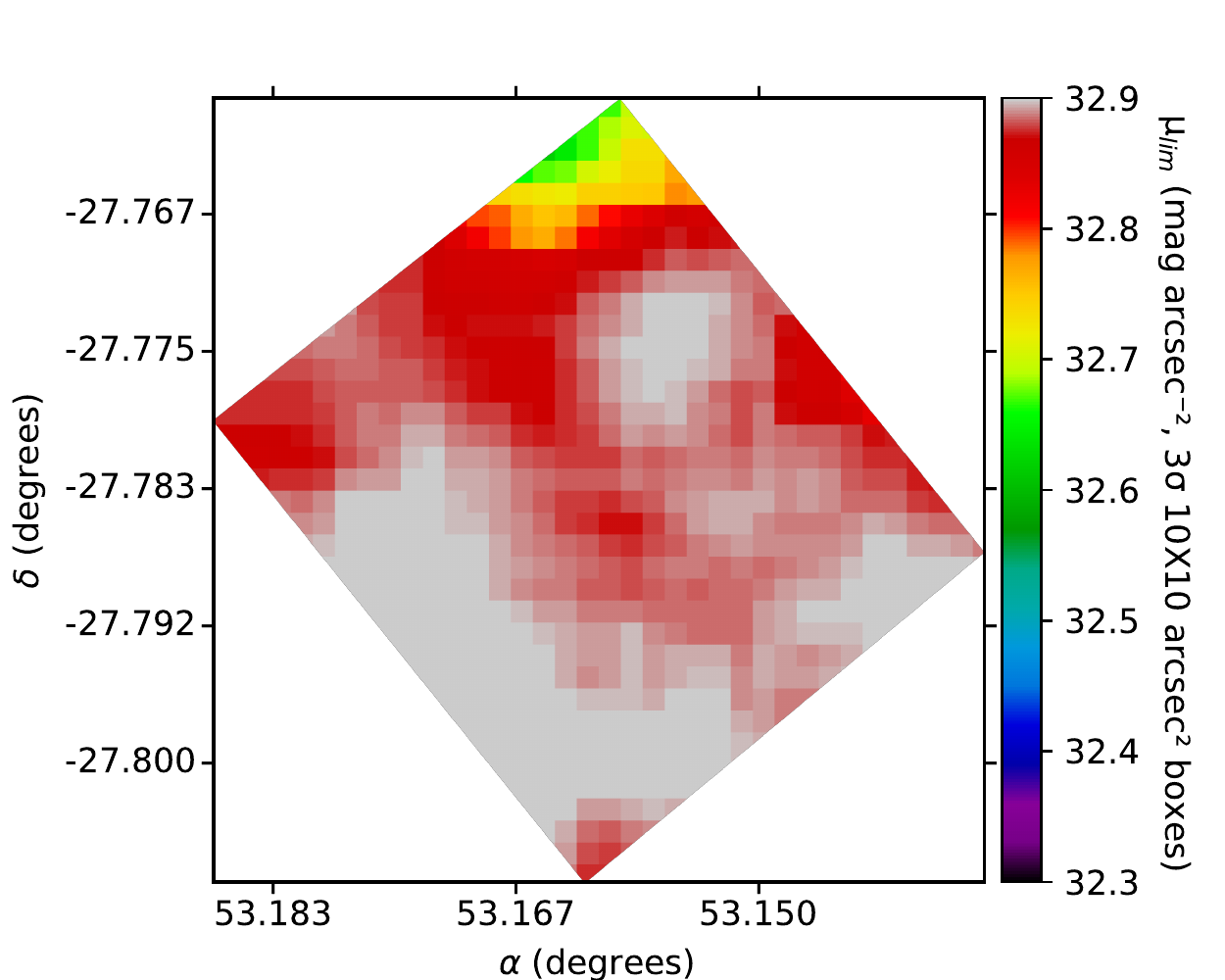}
\put(70,210){\color{black} \textbf{ABYSS}}
\put(70,200){\color{black} \textbf{F105W}}
\end{overpic}
\begin{overpic}[width=0.49\textwidth, clip, trim=62 0 8 20]{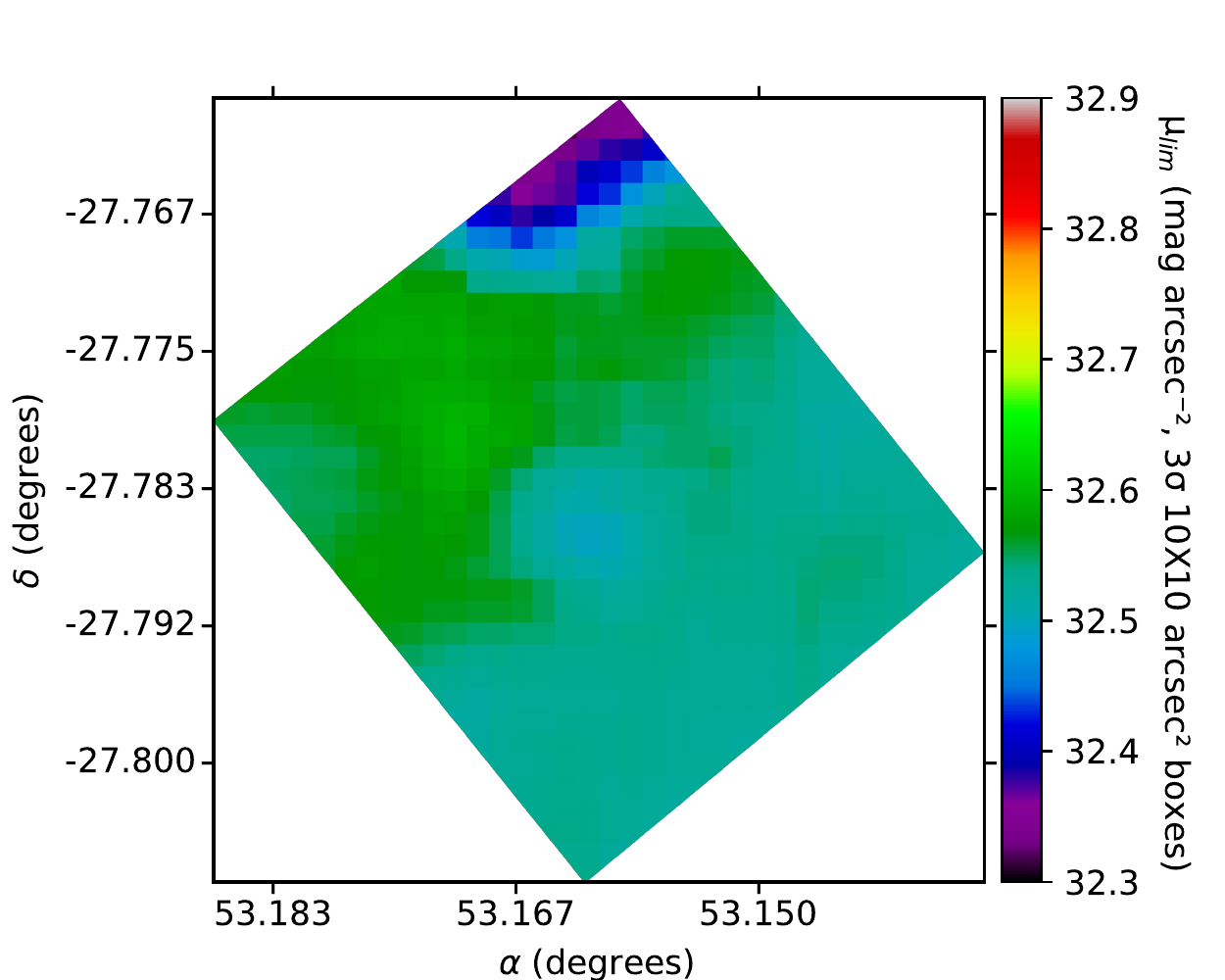}
\put(15,210){\color{black} \textbf{ABYSS}}
\put(15,200){\color{black} \textbf{F125W}}
\end{overpic}

\begin{overpic}[width=0.49\textwidth, clip, trim=0 0 70 20]{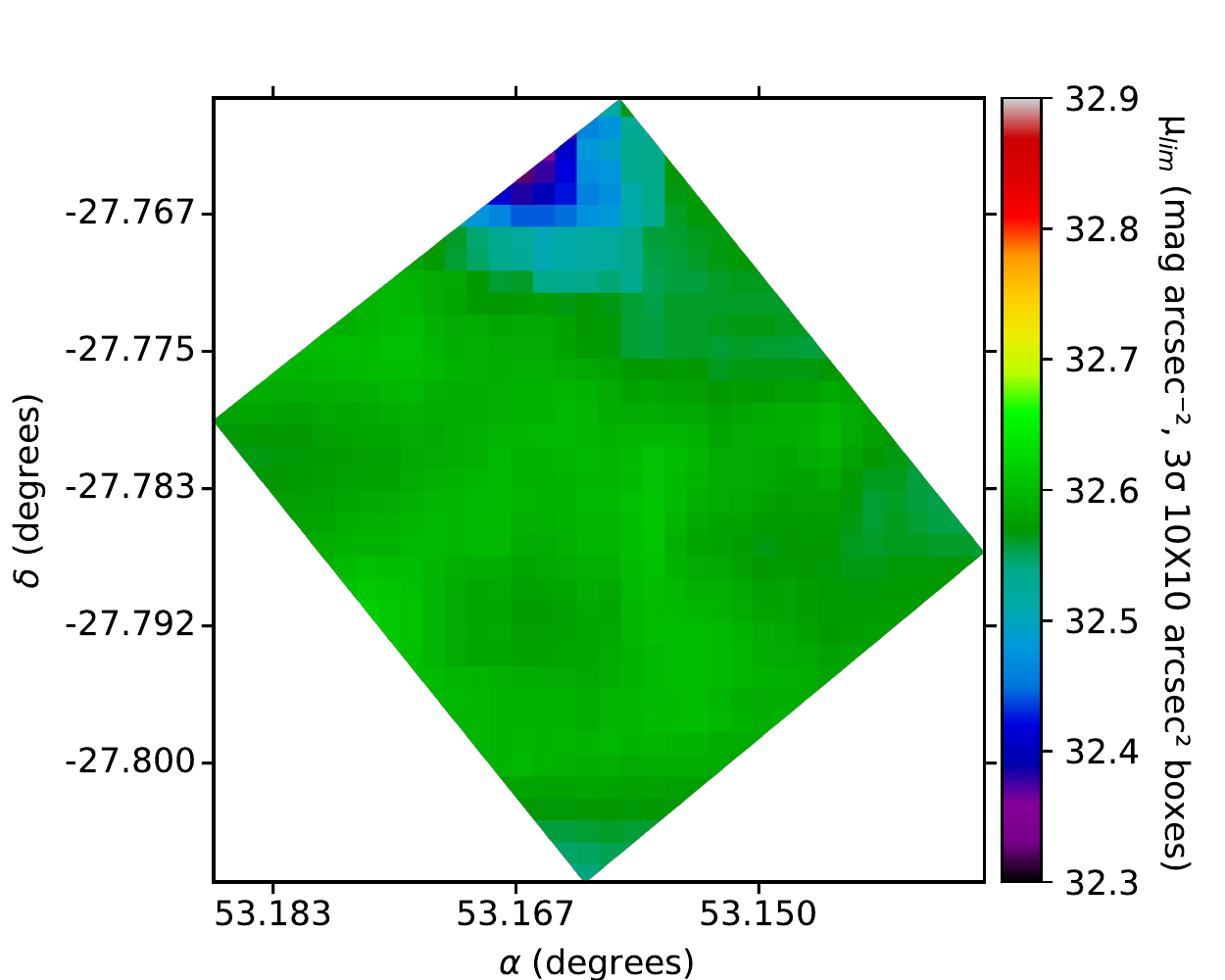}
\put(70,210){\color{black} \textbf{ABYSS}}
\put(70,200){\color{black} \textbf{F140W}}
\end{overpic}
\begin{overpic}[width=0.49\textwidth, clip, trim=62 0 8 20]{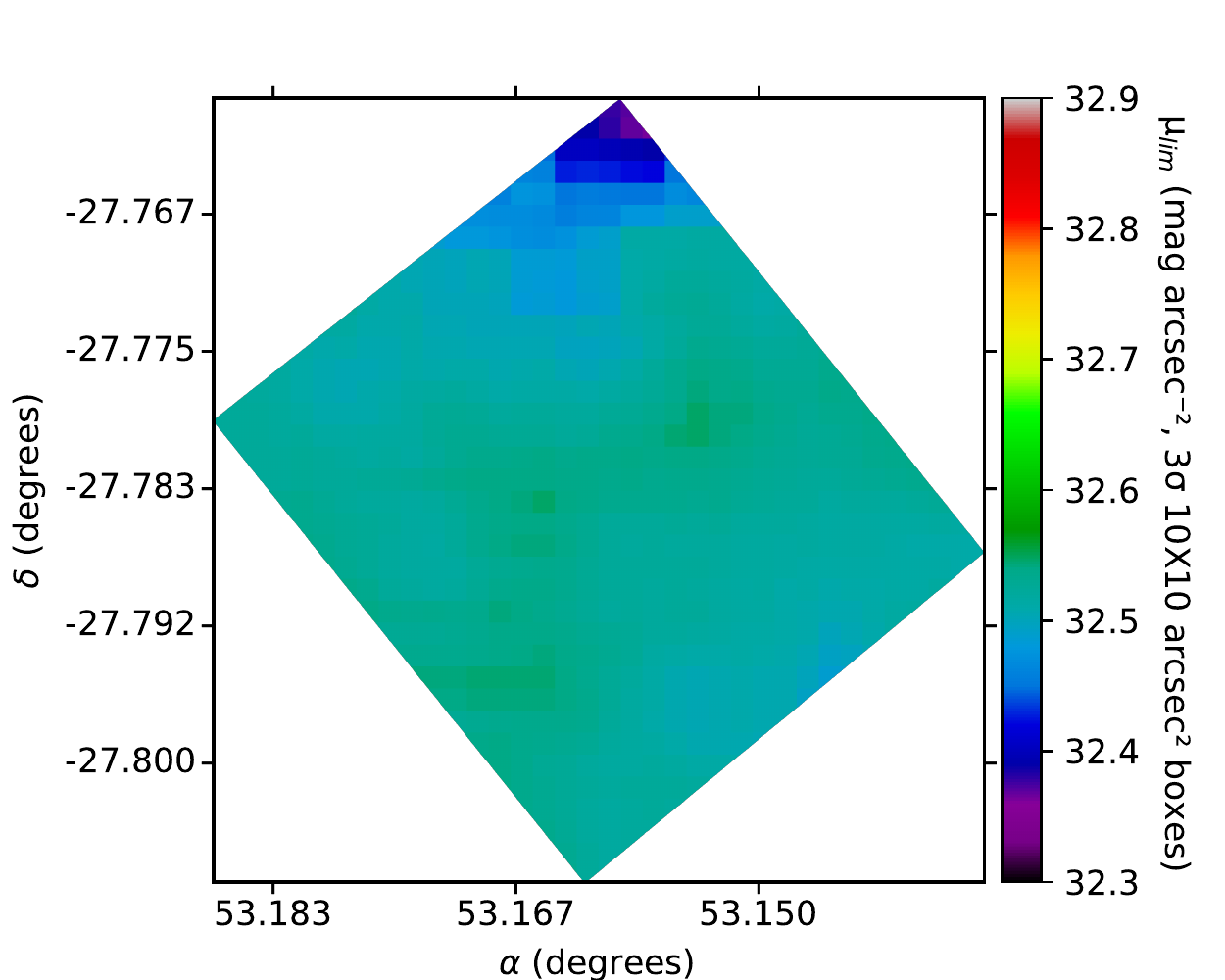}
\put(15,210){\color{black} \textbf{ABYSS}}
\put(15,200){\color{black} \textbf{F160W}}
\end{overpic}

\caption[]{Surface brightness limiting magnitude maps for the \textsf{ABYSS} mosaics of HUDF. From top to bottom and left to right, F105W, F125W, F140W, and F160W bands. Each panel represents in colour scale the surface brightness limiting magnitude for each mosaic, measured as the 3$\sigma$ upper limit of the sky level on $10\times 10$ arcsec$^{2}$ boxes \citep{Trujillo2016}, as a function of the position on the mosaics. All the panels are at the same scale (see the colour bar on the right panels for reference).} 
\vspace{0.25cm}
\label{fig:depth_maps}
\end{figure*}

In Fig.\,\ref{fig:depth_maps} we present the surface brightness limiting magnitude maps, calculated using the standard deviation estimation of {\textsf{NoiseChisel}} for the F105W, F125W, F140W, and F160W mosaics. We find that despite removing a large amount of pixels because of persistence and gradient corrections, the relative depth between filters has not notably changed. We summarise the results in Table \ref{table:HUDF_detail}. In our version, as well as in the previous releases, the deepest filter is the F105W band (\mulim $=32.89^{+0.01}_{-0.02}$ \magarc, 3$\sigma$ in $10\times10$ arcsec boxes), followed by F140W (\mulim$=32.54^{+0.03}_{-0.01}$ \magarc), F125W (\mulim$=32.58^{+0.01}_{-0.02}$ \magarc), and F160W (\mulim$=32.52^{+0.01}_{-0.02}$ \magarc). The relative differences between the surface brightness limiting magnitudes for the same filters of different versions are very small. We note that even after the extremely conservative selection criteria that we have carried out to select the images and the valid pixels (mainly due to persistence effects), our final mosaics present compatible limiting magnitudes than previous HUDF releases. The limiting magnitude for the \textsf{ABYSS} F105W image is $0.01$ \magarc\ deeper than in the XDF and $0.04$ \magarc\ deeper when comparing with the F105W HUDF12 image. The F160W image is the most affected presenting a limiting magnitude $0.15$ \magarc\ brighter than in the XDF and HUDF12. This is because the F160W images are more affected by persistence effects (see Sect.\ref{Subsec:persistence}). Finally, the the F125W presents a compatible depth with XDF and HUDF12, while our F140W mosaic is slightly deeper ($\sim0.06$ \magarc) than the previous reductions (see Fig.\,\ref{fig:depth_maps}). 

Nevertheless, the above limiting surface brightness magnitude should be understood as the formal limiting magnitude and not the effective ones from the images. The effective limiting magnitudes are affected by systematic effects (such as sky over-subtraction) not included in this measurement of the pixel noise.  
The surface brightness limiting magnitude is position dependent over the field of view, with differences of almost $0.5$ \magarc. It is notable the presence of a shallower region on our mosaics on the north corner, due to the effect of conservative masking of the WFC3 IR chip cosmetic defect called the "wagon wheel" (see Sect.\,\ref{Subsec:dataquality}). We find that this region can present a surface brightness limiting magnitude $\sim 1$ \magarc\ brighter than the rest of the mosaics. Any analysis including data from this region should be done carefully. 

The side effect of our reduction process to avoid systematic biases (in particular, removal of persistence contamination) is that the F160W mosaic present slightly brighter formal limiting surface brightness than the previous versions of the HUDF, if we only take into account the pixel noise. Nevertheless, as shown in Sect.\,\ref{Subsec:mosaic_difference}, those previous reductions were dominated by systematic biases, much larger than the relative differences found in surface brightness limiting magnitude. As a result of this, our mosaics contain much more information on the outskirts of the extended objects of the HUDF than any previous version of the data. We will comment on this in Sect.\,\ref{Subsect:profiles}. We find that the conservative persistence masking and removal of gradients that we have performed do not affect significantly the depth of our mosaics, which present surface brightness limiting magnitudes similar to those from the previous versions of the HUDF, while reducing the systematic biases.




\begin{table*}
{\small 
\begin{center}
\begin{tabular}{ccccccc}
\toprule
Filter & No. exposures & Exposure time & \mulim\ \textsf{ABYSS} &  \mulim\ XDF & \mulim\ HUDF12 \\
 &  & [s] & [\magarc] & [\magarc] & [\magarc] \\
(1)&(2)&(3)&(4)&(5)&(6)\\
\midrule


F105W & 264 & 312072 & $32.888^{+0.013}_{-0.018}$ & $32.877^{+0.017}_{-0.017}$ & $32.846^{+0.021}_{-0.017}$ \\[1.2ex]

F125W & 234 & 175837 &  $32.539^{+0.031}_{-0.014}$ & $32.517^{+0.026}_{-0.027}$ &  $32.503^{+0.026}_{-0.024}$ \\[1.2ex]

F140W & 103 & 86352  & $32.585^{+0.012}_{-0.024}$ & $32.517^{+0.014}_{-0.017}$ &  $32.521^{+0.013}_{-0.012}$ \\[1.2ex]

F160W & 271 & 234696 & $32.522^{+0.013}_{-0.016}$ & $32.677^{+0.018}_{-0.020}$ &  $32.660^{+0.018}_{-0.019}$ \\
\hline
\bottomrule
\end{tabular}
\caption{Summary of the data used to create the {\textsf{ABYSS}} HUDF WFC3 IR mosaics and surface brightness limiting magnitude comparison with previous mosaics. \emph{Columns:} 1) Filter identifier. 2) Number of exposures included in the final mosaic. 3) Total exposure time. 4) Median surface brightness limiting magnitude (3$\sigma$ measured on $10\times10$ arcsec$^2$ boxes) for the \textsf{ABYSS} mosaics. 5) Median surface brightness limiting magnitude for the XDF mosaics, measured in the same way. 6) Median surface brightness limiting magnitude for the HUDF12 mosaics, measured in the same way. The surface brightness limits shown here refer to the pixel noise of the images and do not account for systematic effects.}
\label{table:HUDF_detail}
\end{center}
}
\end{table*}

\subsection{Surface brightness profiles}
\label{Subsect:profiles}

\begin{figure*}[]
\centering
\vspace{0.25cm}

\begin{overpic}[clip, trim=0 0 0 0, width=0.33\textwidth]{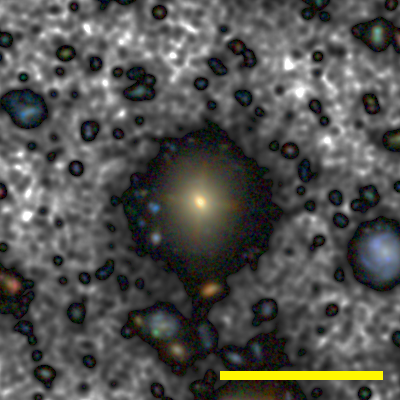}
\put(3,160){\color{black}
\colorbox{white}{\textbf{1) HUDF-1}}}
\put(115,17){\color{yellow}{\textbf{10 arcsec}}}
\end{overpic}
\begin{overpic}[clip, trim=0 0 0 0, width=0.33\textwidth]{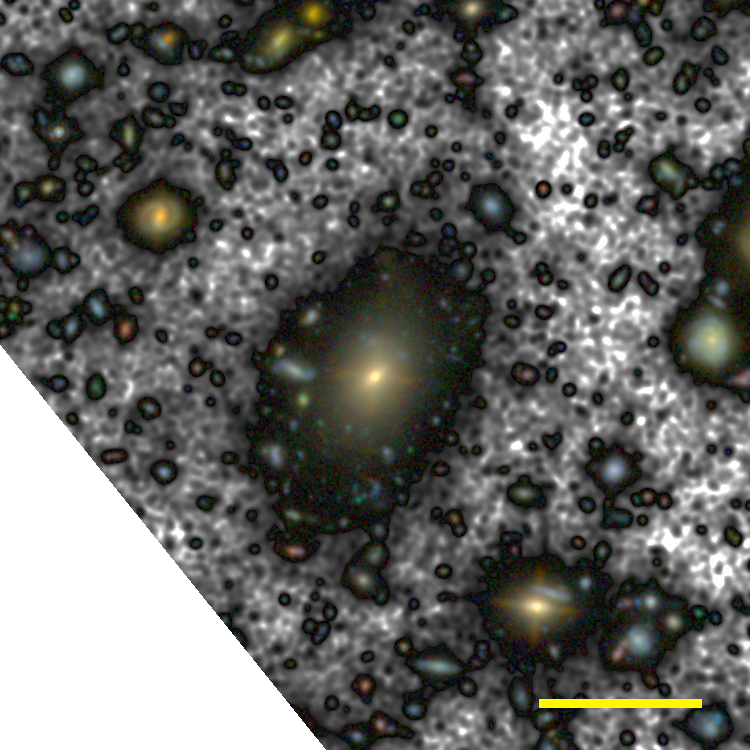}
\put(3,160){\color{black} \colorbox{white}{\textbf{2) HUDF-2}}}
\end{overpic}
\begin{overpic}[clip, trim=0 0 0 0, width=0.33\textwidth]{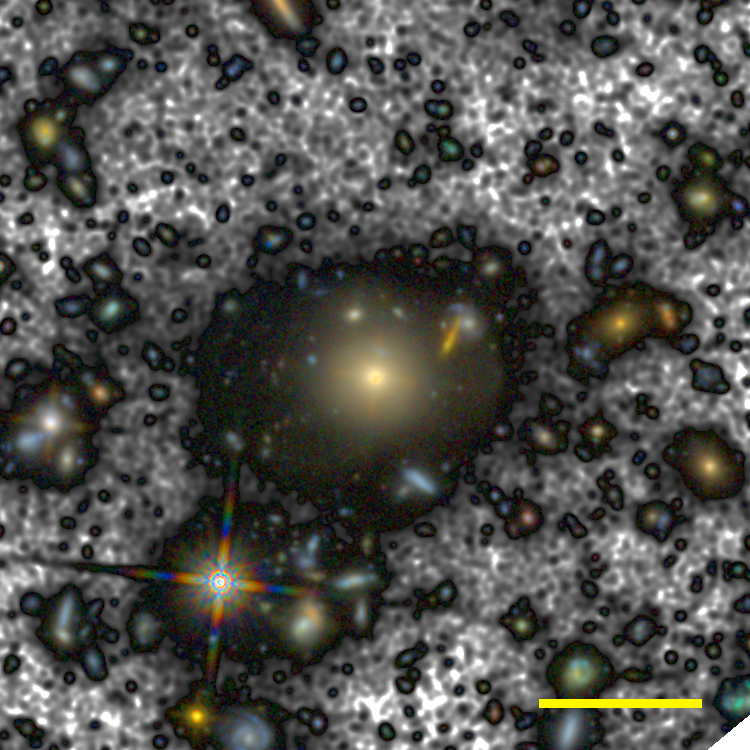}
\put(3,160){\color{black} \colorbox{white}{\textbf{3) HUDF-5}}}
\end{overpic}

\begin{overpic}[clip, trim=0 0 0 0, width=0.33\textwidth]{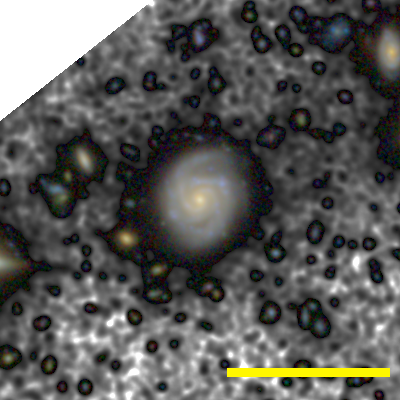}
\put(3,160){\color{black} \colorbox{white}{\textbf{4) ABYSS-1}}}
\end{overpic}
\begin{overpic}[clip, trim=0 0 0 0, width=0.33\textwidth]{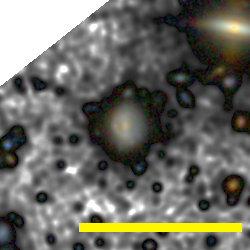}
\put(3,160){\color{black} \colorbox{white}{\textbf{5) ABYSS-2}}}
\end{overpic}
\begin{overpic}[clip, trim=0 0 0 0, width=0.33\textwidth]{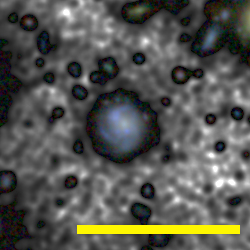}
\put(3,160){\color{black} \colorbox{white}{\textbf{6) ABYSS-3}}}
\end{overpic}

\caption[]{Luminance-RGB images of the selected targets for the surface brightness profile analysis: \emph{Top row, from left to right:} 1) HUDF-1 ($24\times 24$ arcsec), 2) HUDF-2 ($45\times 45$ arcsec), 3) HUDF-5 ($45\times 45$ arcsec). \emph{Bottom row, from left to right:} 1) ABYSS-1 ($24\times 24$ arcsec), 2) ABYSS-2 ($15\times 15$ arcsec), 3) ABYSS-3 ($15\times 15$ arcsec). The yellow segment represents 10 arcsec in all images. The high signal-to-noise parts of the mosaics are represented with colours (\emph{red:} F160W, \emph{green:} mean of F125W and F140W bands, \emph{blue:} F105W). The low signal-to-noise regions are represented as a black and white background (black regions are brighter than white regions) according to the mean image of the four mosaics (F105W, F125W, F140W, F160W).} 
\vspace{0.25cm}
\label{fig:targets}
\end{figure*}

\begin{figure*}[t!]
\centering
\vspace{0.25cm}
\begin{overpic}[width=0.33\textwidth, trim=0 60 0 0, clip]{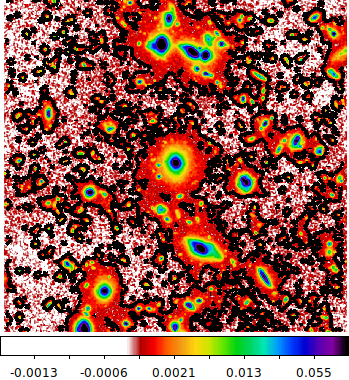}
\put(120,147){\color{black} \colorbox{white}{\textbf{ABYSS}}}
\put(7,147){\color{black} \colorbox{white}{\textbf{HUDF-1}}}
\end{overpic}
\begin{overpic}[width=0.33\textwidth, trim=0 60 0 0, clip]{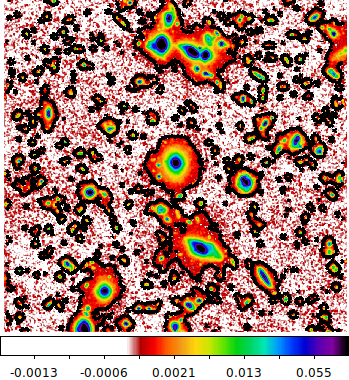}
\put(120,147){\color{black} \colorbox{white}{\textbf{HUDF12}}}
\end{overpic}
\begin{overpic}[width=0.33\textwidth, trim=0 60 0 0, clip]{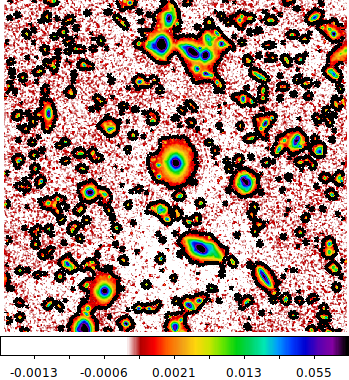}
\put(120,147){\color{black} \colorbox{white}{\textbf{XDF}}}
\end{overpic}

\begin{overpic}[width=0.33\textwidth, trim=0 60 0 0, clip]{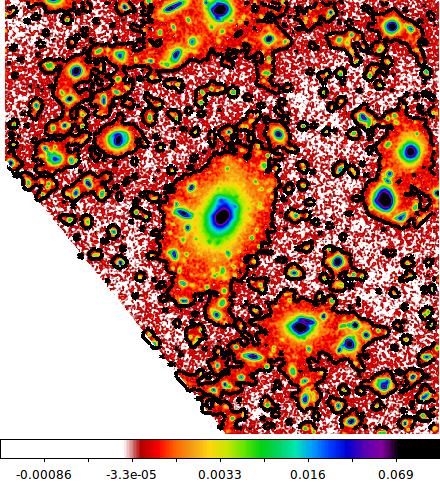}
\put(120,154){\color{black} \colorbox{white}{\textbf{ABYSS}}}
\put(7,154){\color{black} \colorbox{white}{\textbf{HUDF-2}}}
\end{overpic}
\begin{overpic}[width=0.33\textwidth, trim=0 60 0 0, clip]{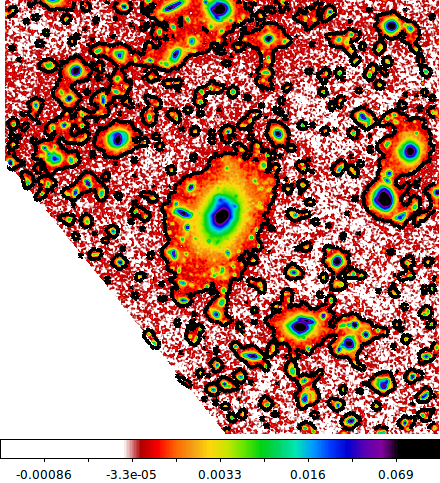}
\put(120,154){\color{black} \colorbox{white}{\textbf{HUDF12}}}
\end{overpic}
\begin{overpic}[width=0.33\textwidth, trim=0 60 0 0, clip]{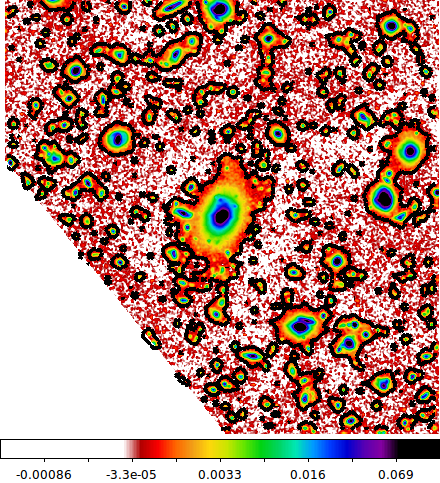}
\put(120,154){\color{black} \colorbox{white}{\textbf{XDF}}}
\end{overpic}

\begin{overpic}[width=0.33\textwidth, trim=0 0 0 0, clip]{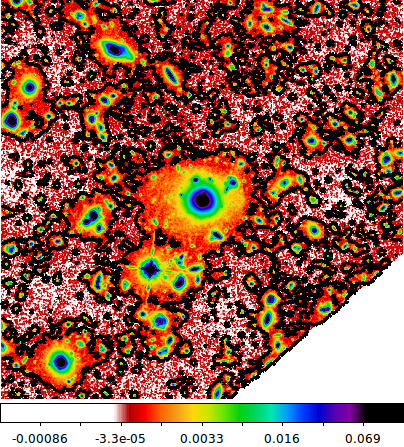}
\put(120,180){\color{black} \colorbox{white}{\textbf{ABYSS}}}
\put(7,180){\color{black} \colorbox{white}{\textbf{HUDF-5}}}
\end{overpic}
\begin{overpic}[width=0.33\textwidth, trim=0 0 0 0, clip]{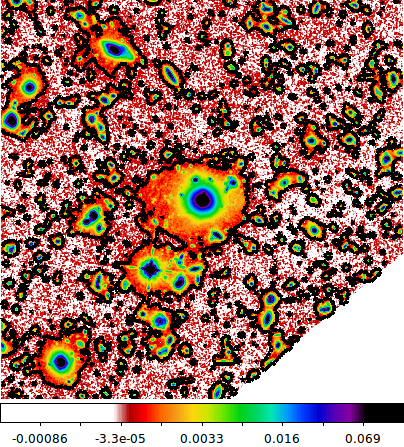}
\put(120,180){\color{black} \colorbox{white}{\textbf{HUDF12}}}
\end{overpic}
\begin{overpic}[width=0.33\textwidth, trim=0 0 0 0, clip]{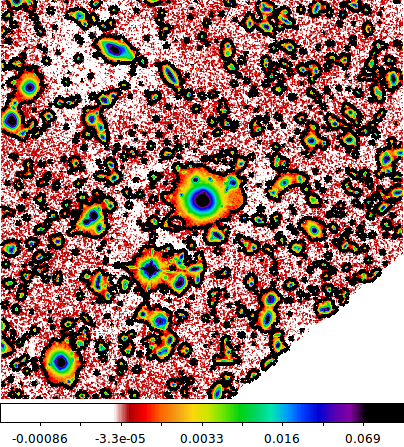}
\put(120,180){\color{black} \colorbox{white}{\textbf{XDF}}}
\end{overpic}

\caption[]{F105W intensity images of HUDF-1 (top row, $54\times 54$ arcsec), HUDF-2 (central row, $54\times 54$ arcsec), and HUDF-5 (bottom row, $72\times 72$ arcsec) \citep{Buitrago2017} for our version of the HUDF mosaics (\textsf{ABYSS}, left column), the HUDF12 mosaics \citep[][central column]{Koekemoer2012}, and the XDF mosaics \citep[][right column]{Illingworth2013}. The black contours represent the $\mu_{F105W} = 29$ \magarc\ isophote. All images are at the same colour scale.} 
\vspace{-0.25cm}
\label{fig:contours1}
\end{figure*}
\begin{figure*}[t!]
\centering
\vspace{0.25cm}
\begin{overpic}[width=0.33\textwidth, trim=0 60 0 0, clip]{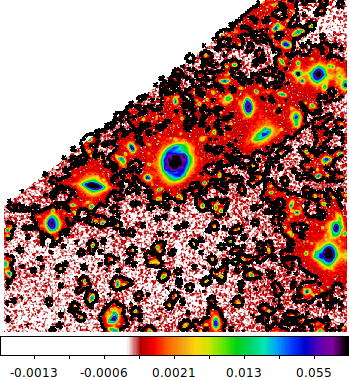}
\put(120,147){\color{black} \colorbox{white}{\textbf{ABYSS}}}
\put(7,147){\color{black} \colorbox{white}{\textbf{ABYSS-1}}}
\end{overpic}
\begin{overpic}[width=0.33\textwidth, trim=0 60 0 0, clip]{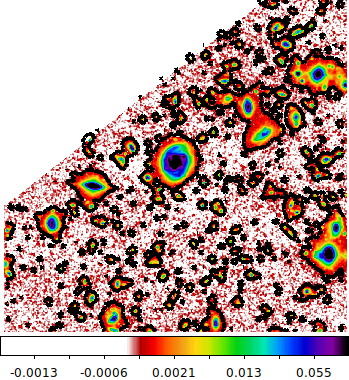}
\put(120,147){\color{black} \colorbox{white}{\textbf{HUDF12}}}
\end{overpic}
\begin{overpic}[width=0.33\textwidth, trim=0 60 0 0, clip]{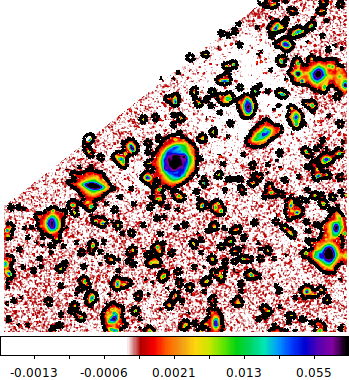}
\put(120,147){\color{black} \colorbox{white}{\textbf{XDF}}}
\end{overpic}

\begin{overpic}[width=0.33\textwidth, trim=0 60 0 0, clip]{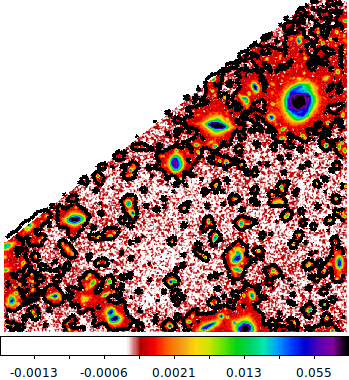}
\put(120,147){\color{black} \colorbox{white}{\textbf{ABYSS}}}
\put(7,147){\color{black} \colorbox{white}{\textbf{ABYSS-2}}}
\end{overpic}
\begin{overpic}[width=0.33\textwidth, trim=0 60 0 0, clip]{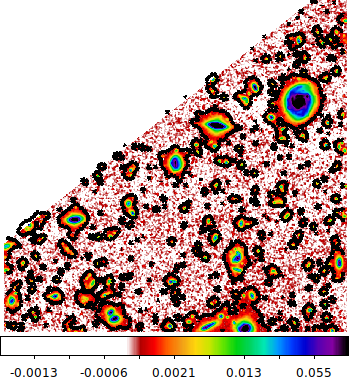}
\put(120,147){\color{black} \colorbox{white}{\textbf{HUDF12}}}
\end{overpic}
\begin{overpic}[width=0.33\textwidth, trim=0 60 0 0, clip]{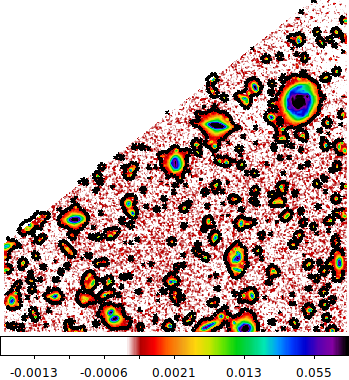}
\put(120,147){\color{black} \colorbox{white}{\textbf{XDF}}}
\end{overpic}

\begin{overpic}[width=0.33\textwidth, trim=0 0 0 0, clip]{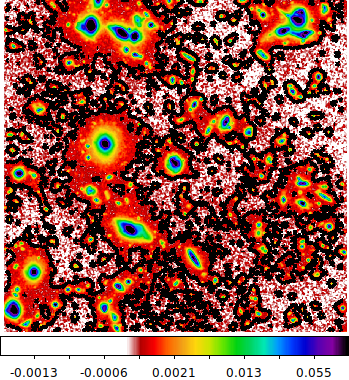}
\put(120,177){\color{black} \colorbox{white}{\textbf{ABYSS}}}
\put(7,177){\color{black} \colorbox{white}{\textbf{ABYSS-3}}}
\end{overpic}
\begin{overpic}[width=0.33\textwidth, trim=0 0 0 0, clip]{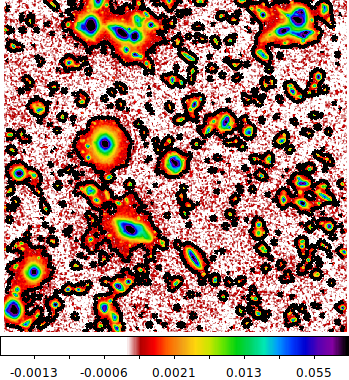}
\put(120,177){\color{black} \colorbox{white}{\textbf{HUDF12}}}
\end{overpic}
\begin{overpic}[width=0.33\textwidth, trim=0 0 0 0, clip]{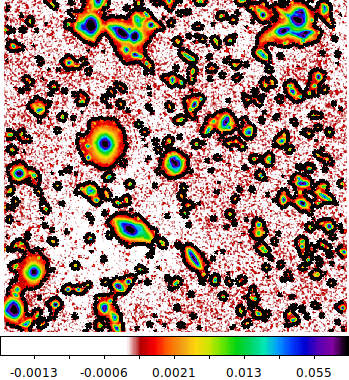}
\put(120,177){\color{black} \colorbox{white}{\textbf{XDF}}}
\end{overpic}

\caption[]{F105W intensity images of ABYSS-1 (top row), ABYSS-2 (central row), and ABYSS-3 (bottom row) for our version of the HUDF mosaics (\textsf{ABYSS}, left column), the HUDF12 mosaics \citep[][central column]{Koekemoer2012}, and the XDF mosaics \citep[][right column]{Illingworth2013}. The field of view is $54\times 54$ arcsec on all the images. The black contours represent the $\mu_{F105W} = 29$ \magarc\ isophote. All images are at the same colour scale.}
\vspace{-0.25cm}
\label{fig:contours2}
\end{figure*}

In this section we study the surface brightness profiles of several objects (see Table \ref{table:targets}) from the HUDF, comparing the results from our reduction and the previous releases of the mosaics. First, in Sect.\,\ref{Subsec:LargeObjects} we focus our attention on those objects with large angular sizes, which are more sensitive to over-subtraction and the primary target of the reduction process applied on this work. We start with HUDF-5, the largest one from our sample \citep[$z=0.607$, following the nomenclature from][]{Buitrago2017} and one of the objects most affected by the sky over-subtraction of the previous versions, continuing with HUDF-1 ($z=0.618$), HUDF-2 ($z=0.619$), and ABYSS-1 ($z=0.622$), a spiral galaxy from \citet{Elmegreen2005}. Finally, we will study whether there are significant changes on the structure of the objects with smaller ($R_{lim}<10$ arcsec) angular sizes in Sect.\,\ref{Subsec:Smallobjects}. For this, we have selected two additional spiral galaxies from \citet{Elmegreen2005}: ABYSS-2 ($z=0.607$) and ABYSS-3 ($z=1.28$). We show the cutouts of the targets in Fig.\,\ref{fig:targets}. The main selection criteria for the spiral galaxies was the relative isolation to avoid light contamination from nearby sources. We refer to Table \,\ref{table:targets} for the sky coordinates, main properties, and IAU ID. 

In order to properly analyse the surface brightness profiles of the target objects we first performed a careful manual masking of all the nearby objects. These masks were created using the sum of all the four filters. For consistency, we use the same masks for every filter and version of the HUDF. Secondly, we analysed the surface brightness profiles of each galaxy using concentric elliptical apertures. We take into account both the uncertainties of the intensity dispersion along the elliptical aperture and the intrinsic uncertainty of each pixel (sky noise) using bootstrapping and Monte Carlo simulations. The sky noise is measured on each mosaic independently, using the standard deviation maps used for the limiting magnitude analysis (see Sect.\,\ref{Subsec:maglim}). Finally, to improve the sky estimation we subtract the local sky level using the median value of all the valid appertures that are beyond a certain limit \citep{Pohlen2006}, which is set at 140 kpc for HUDF-2 and HUDF-5, following the method described on \citet{Buitrago2017}. For the three spiral galaxies (ABYSS-1, ABYSS-2 and ABYSS-3) and HUDF-1 (notably smaller than HUDF-2 and HUDF-5), we measure the local sky level beyond 80 kpc, which is well beyond the visual limit of the objects. The geometric parameters for the profiles (position angle and axis ratio) were calculated using {\textsf{SEXtractor}}. We detail the position angles, axis ratios, and the local sky region limit for each object in Table \ref{table:targets}. We remark that in order to do a fair comparison we apply the same masks, geometric parameters, elliptical aperture algorithm, and local sky level region to all the mosaics, regardless of their wavelength and version. For the sake of simplicity, we do not apply any type of PSF correction. The main objective of the surface brightness profile analysis that we present here is to illustrate the differences on the low surface brightness regions of the different mosaics. The differences of the PSF between the three versions studied are minimal, as they are created from similar datasets. Thus, any attempt to correct the individual images from PSF effect would introduce additional uncertainties. We will study the shape of the surface brightness profiles of the galaxies on the HUDF according to the new mosaics in a forthcoming paper (Borlaff et al. in prep). 

\begin{table*}
{\small 
\begin{center}
\begin{tabular}{ccccccccc}
\toprule
ID & IAU ID & MUSE ID & $\alpha$ & $\delta$ & $z$ & PA & $b/a$ & Sky background limiting radius \\
(1)&(2)&(3)&(4)&(5)&(6)&(7)&(8)&(9)\\
& & & (degrees) & (degrees) & & (degrees) & & (kpc)\\

\midrule
HUDF-1 & J033237.30-274729.3 & 5 & 53.16164 & -27.78025 & 0.618 & -14.18 & 0.878 & 80 \\[1.2ex] 
HUDF-2 & J033241.40-274717.1 & 870 & 53.17254 & -27.78812 & 0.619 & -30.61 & 0.599 & 140 \\[1.2ex] 
HUDF-5 & J033237.30-274729.3 & 862 & 53.15545 & -27.79150 & 0.667 & 75.18 & 0.900 & 140 \\[1.2ex] 
ABYSS-1 & J033240.78-274615.6 & 1 & 53.16993 & -27.77106 & 0.622 & -15.82 & 0.884 & 80\\[1.2ex] 
ABYSS-2 & J033242.25-274625.3 & 916 & 53.17606 & -27.77371 & 1.288 & -3.31 & 0.890 & 80\\[1.2ex] 
ABYSS-3 & J033237.96-274651.9 & 7 & 53.15815 & -27.78109 & 0.620 & 64.69 & 0.860 & 80\\[1.2ex] 
\hline
\bottomrule
\end{tabular}
\caption{Selected targets for the surface brightness profile comparison of the HUDF WFC3 IR mosaics. \emph{Columns:} 1) ID. 2) IAU ID. 3) MUSE catalog ID. 4) Right ascension (degrees). 5) Declination (degrees). 6) Spectroscopic $z$ from the MUSE HUDF catalog \citep{Bacon2017}. 7) Position angle (degrees, anti-clockwise from north). 8) Axis ratio of the minor over the major axis. 9) Minimum galactocentric radius used to calculate the local sky background (kpc).}
\label{table:targets}
\end{center}
}
\vspace{-0.5cm}
\end{table*}

\subsubsection{Surface brightness profiles of large objects}
\label{Subsec:LargeObjects}

HUDF-5 \citep[][]{Buitrago2017} is an elliptical galaxy ($z=0.667$, $\log_{10}(M/M_{\odot}) = 11.19^{+0.09}_{-0.05}$) and one of the brightest objects from the HUDF. It presents a noticeable shell envelope, and is one of the objects most affected by the aggressive sky subtraction from the previous releases. In Figs.\,\ref{fig:contours1} and \ref{fig:contours2} we represent the intensity images of the six targets in the F105W filter (which is the deepest mosaic in our reduction) in three different panels, for the \textsf{ABYSS}, HUDF12, and XDF version of the mosaics. On top of each panel we represent with black contours their respective $\mu=29$ \magarc\ isophotes. We observe that: 1) the $\mu=29$ \magarc\ isophotal contours are more extended in our reduction than in the previous versions of the mosaics, specially when compared to the XDF, and 2) the new diffuse light tends to appear around the largest objects on the field-of-view, while the objects with small angular size almost do not change their respective contours. This is a expected result, as aggressive sky-subtraction tends to affect mostly to the envelopes of the largest objects. 
 
\begin{figure*}[h!]
\centering
\includegraphics[page=1, width=0.49\textwidth, trim=0 92 40 0, clip]{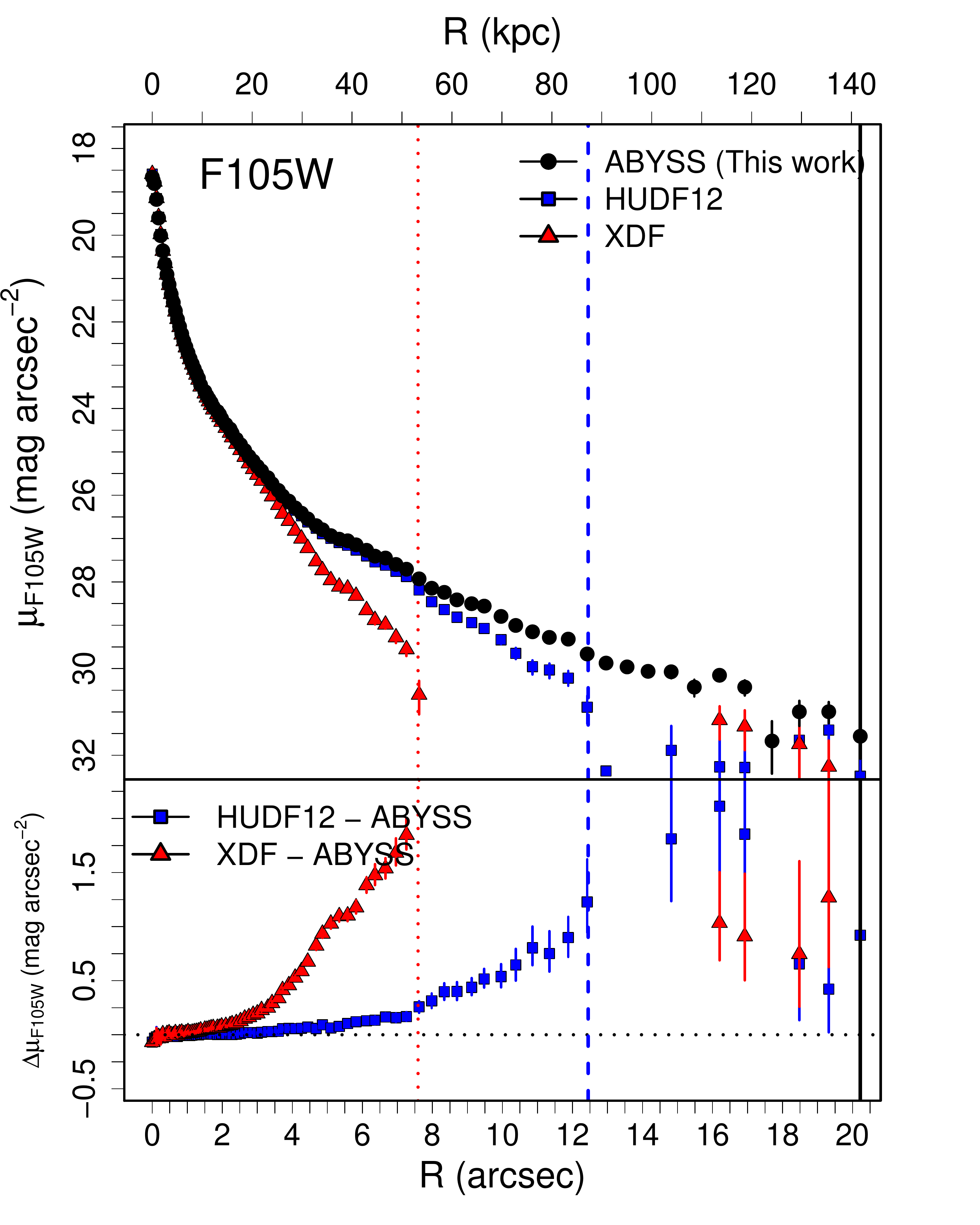}
\includegraphics[page=2, width=0.49\textwidth, trim=0 92 40 0, clip]{PROFILES/ewest_profile.pdf}

\includegraphics[page=3, width=0.49\textwidth, trim=0 0 40 85, clip]{PROFILES/ewest_profile.pdf}
\includegraphics[page=4, width=0.49\textwidth, trim=0 0 40 85, clip]{PROFILES/ewest_profile.pdf}

\caption[]{Comparison of the surface brightness profiles of the elliptical galaxy HUDF-5 \citep[$\alpha=53.15545, \delta=-27.79150$,][]{Buitrago2017} for the F105W (top left), F125W (top right), F140W (bottom left) and F160W filters (bottom right), using our own reduction of the HUDF WFC3 mosaics (\textsf{ABYSS}, black dots), the HUDF12 \citep[][blue squares]{Koekemoer2012} and the XDF \citep[][red triangles]{Illingworth2013}. The top plot of each panel shows the surface brightness profile for each reduction. Black solid, blue dashed and red dotted lines represent the elliptic aperture with largest semi-major axis that presents a signal-to-noise ratio higher than 3 over the sky-level. The bottom plot represents the difference in magnitude of each previous reduction with the \textsf{ABYSS} version of the mosaics as a function of galactocentric radius. Consult the legend on the figure.}  
\label{fig:HUDF5}
\end{figure*}

Nevertheless, we note that the intensity images are not a direct proxy of the shape of the surface brightness profile, as they are not corrected by the small sky residuals that appear on the surface brightness profiles (local sky background). The same effect shown on Figs.\,\ref{fig:contours1} and \ref{fig:contours2} could be caused by sky background under-subtraction on our mosaics. In order to confirm the validity of these findings, we must analyse the surface brightness profiles corrected by the local sky background of the target galaxy. In Fig.\,\ref{fig:HUDF5} we show the results of the surface brightness analysis for HUDF-5. We found several interesting results: 

\begin{enumerate}
\item Even after applying a local sky level correction, the XDF mosaics present a significant over-subtraction when comparing to the HUDF12 or our dedicated mosaics. This effect is noticeable at surface magnitudes fainter than $\mu \sim 26-27$ \magarc.
\item Our dedicated mosaics recover a significant amount of light when compared to the HUDF12 mosaics and the XDF. The maximum differences in surface brightness range from $\Delta \mu=0.5-1.25$ \magarc, being higher for the deeper images (F105W and F160W). 
\item Interestingly, the XDF version of the F140W mosaic appears to be less over-subtracted than the rest of the filters, obtaining a limiting radius at $R\sim10$ arcsec, while the rest of the filters reaches only $R\sim8$ arcsec.  
\item For the F105W, F125W, and F160W images, the surface brightness profiles of {\textsf{ABYSS}} extend much further out (up to $R \sim 25$ arcsec in the F105W and F160W bands) than in the HUDF12 and specially than the XDF mosaics, where the limit is almost one third smaller than the limiting radius of the \textsf{ABYSS} mosaics. 
\end{enumerate}

The general agreement up to a certain radius of the HUDF12 and our reduction, contrary to the results obtained with the XDF, is a quantitative proof of that the latter mosaics present a high and systematic over-subtraction of the outskirts around the most extended objects, such as HUDF-5. The over-subtraction can be as high as $\Delta \mu \sim 1.5$ \magarc\ at a surface brightness magnitude of $\mu \sim 28 $ \magarc\ for the XDF mosaics even after applying local sky correction. We obtain a similar result for the HUDF12 mosaics at $\mu \sim 29 $ \magarc. This result demonstrates that the depth of previous reductions was dominated by systematic biases rather than the sky noise. We also found interesting that the surface brightness profile of the shallower mosaic (F140W) of XDF presents $S/N>3$ to larger galactocentric radius than the rest of the filters, which are much deeper. In addition, visual inspection of the surface brightness profiles from our images reveals that they tend to reach the sky level smoothly, following the general shape of the surface brightness profile, rather than showing a sharp down-bending profile at the outskirts as observed on the XDF profiles. 

In Fig.\,\ref{fig:HUDF1} we represent the surface brightness profiles of HUDF-1, another one of the elliptical galaxies studied in \citet{Buitrago2017}. We found a very similar result as in the case of HUDF-5. The surface brightness profiles of the \textsf{ABYSS} mosaics recover a significant extension that has been completely removed in the case of XDF and HUDF12. Again, we find that the XDF is the most affected by over subtraction, with $\Delta \mu \sim 1.5$ \magarc\ with respect to the \textsf{ABYSS} profile at a surface brightness magnitude of $\mu \sim 29 $ \magarc. We recover also a significant amount of light compared to the HUDF12, reaching $\Delta \mu \sim 1.5$ \magarc\ at the limiting radius in the F105W and $\Delta \mu \sim 1$ \magarc\ in the rest of the images. The XDF profiles show a down-bending break that is not that strong in the HUDF12, and it is completely removed in our version of the mosaics, where the surface brightness profile reaches the sky level smoothly, following a nearly exponential shape.

We show the surface brightness profile analysis of HUDF-2 in Fig.\,\ref{fig:HUDF2}. For this object we detect a similar over subtraction of the XDF profiles. The XDF F160W surface brightness profile reaches $S/N=3$ at half the extension of the \textsf{ABYSS} and HUDF12 profiles. Interestingly, we found that the \textsf{ABYSS} and HUDF12 profile agree remarkably well for this object, reaching the limiting $S/N$ at very similar radius ($R_{lim}\sim 13$ arcsec, with the exception of F125W and F160W which extends to $R\sim15.5-16$ arcsec). In F160W, we observe that the HUDF12 surface brightness profile is $\Delta \mu \sim 0.3$ \magarc\ brighter than the \textsf{ABYSS} between $R=4$ and $R=13$ arcsec. For the rest of the filters, the \textsf{ABYSS} surface brightness profiles are $\Delta \mu \sim 0.3$ brighter than the HUDF12 ones beyond $R>11$ arcsec. This result agrees with the observed difference maps (see Sect.\,\ref{Subsec:mosaic_difference}), where we detected more flux on the southern section of the F160W HUDF12 image than in our F160W \textsf{ABYSS} mosaic.

We continue the analysis with the spiral galaxy ABYSS-1 (see Table \ref{table:targets}, and bottom left panel of Fig.\,\ref{fig:targets}). In Fig.\,\ref{fig:ABYSS1} we compare the surface brightness profiles based on our own reduction of the HUDF mosaics (\textsf{ABYSS}), the HUDF12, and the XDF, for the four different filters (F105W, F125W, F140W and F160W). We summarise the results in several points:
\begin{enumerate}
    \item As in the case of HUDF-5, the surface brightness profiles from our dedicated mosaics present higher intensity and signal-to-noise ratio on the outskirts, up to radius of $R\sim7-7.5$ arcsec. Similarly to the previous case, the maximum differences in surface brightness range from $\Delta \mu=0.5-1.5$~\magarc.
    \item The F140W band surface brightness profiles for the \textsf{ABYSS} and XDF version of the mosaics present a more similar shape than in the previous cases. 
    \item The surface brightness profiles of the four filters of the \textsf{ABYSS} mosaics and the F140W from XDF suggest the presence of an extended component clearly detectable at $R>5.5$~arcsec, which could be due to PSF effects or an extended stellar halo. The effect is less clear in the F125W image, which reaches $S/N=3$ at $R\sim6$ arcsec. 
    \item There is a clear discrepancy between different filters on the shapes of the profiles for the XDF version of the mosaics, being the F140W the most extended of all XDF surface brightness profiles and the only approaching our own reduction.     
\end{enumerate}

The presence of an extended component detectable beyond $R=5$ arcsec in these surface brightness profile does not necessarily imply that it is associated with a physical component (that is, a stellar halo). In order to identify the true nature of such extended light it would be necessary to estimate the amount of PSF scattered light from the nearby objects and the ABYSS-1 galaxy itself, in a similar way as done in \citet{Trujillo2016}. As demonstrated in that work, the scattered light of the sources creates a background of light that contaminates the images and the surface brightness profiles. Assuming that the observer knows the behaviour of the PSF, it is possible to model and correct it, recovering the true shape of the objects to a certain degree \citep[see][for a similar work with HST ACS data on GOODS-N]{Borlaff2017}. Nevertheless, such field of scattered light can be easily mistaken with the sky background. Removing it from the final mosaics in the reduction process neglects the possibility of recovering the information at the very low surface brightness limits.  


A detailed analysis of the new low surface brightness structures is well beyond the scope of the present paper and will be addressed in a forthcoming paper (Borlaff et al. in prep). We conclude that the surface brightness profiles that we present are a valid benchmark between the different reduction processes, as they use the same masks and surface brightness profile analysis and local sky background correction procedure. In addition, we found that our reduction process allows us to recover up to $\Delta \mu \sim 1-1.5$~\magarc\ on the surface brightness profiles when compared to the previous versions of the HUDF WFC3 IR mosaics and almost twice the radial size when compared to the XDF surface brightness profiles, even after local sky background correction. 

\subsubsection{Surface brightness profiles of small objects}
\label{Subsec:Smallobjects}

In this section we will analyse the properties of the surface brightness profiles of ABYSS-2 and ABYSS-3 (see Table \ref{table:targets}), two spiral galaxies with smaller angular sizes compared to the previous targets. The main objective is to do a consistency test and determine if there is any significant differences on the surface brightness profiles of such objects when comparing to the previous versions of the mosaics. 

In Fig.\,\ref{fig:ABYSS2} we show the surface brightness profiles of the spiral galaxy ABYSS-2. Given that the aggressive sky background correction applied on the previous versions of the HUDF affected specially to the most extended objects, the surface brightness profiles of more compact objects should be more similar than the large objects, regardless of the version of the mosaics. Indeed, we find that the structure of the surface brightness profiles is more similar in the case of ABYSS-2 than in the previous ones, especially in comparison with HUDF-5. The surface brightness differences are reduced to $\Delta \mu \sim 0.5$ \magarc\  at the limiting radius, and the relative differences between the HUDF12 and the XDF are notably smaller. Moreover, we still detect that the F140W profile of XDF has a more similar shape to our surface brightness profiles of {\textsf{ABYSS}} mosaics, suggesting that the over-subtraction could be less aggressive in this filter than in the rest. In addition to these results, we detect a tail of extended light in the {\textsf{ABYSS}} mosaics (possibly due to PSF scattered light) which dominates the surface brightness profile from $R \sim 2-2.5$ arcsec and it is not visible on the previous versions on the HUDF. This result is compatible with the analysis of ABYSS-1.

Finally, we analyse the surface brightness profile of ABYSS-3 (see Fig.\,\ref{fig:ABYSS3}). This is the smallest object of all the six selected galaxies. The surface brightness profile shows a clear exponential decline without deviations (with small signs of a down-bending at $R\sim1.5$ arcsec) until the limiting radius. We find no signs of a bulge at the inner regions. For this object we find a remarkably good agreement between the three versions of the mosaics, obtaining very similar limiting radii for all of them on each band ($R_{lim}\sim2.8-3.2$ arcsec) and no significant differences on the surface brightness profiles along all the visible radius. We do not find signs of extended scattered light, in contrast to the profiles of ABYSS-2. 

Therefore, the photometry and structure of objects with relatively small angular sizes is similar and compatible to that observed in the previous versions of the HUDF. This analysis provides a consistency test, demonstrating that our mosaics preserve the properties of the small objects while recovering the extended light from the largest sources in the field of view. The results provided here demonstrate that the cause of the differences in flux between the XDF and HUDF12 is a systematic bias caused by over-subtraction of the sky background. We show that our reduction pipeline provides a viable process to reduce systematic biases prior to the co-adding of the final mosaics. This process reduces the need for sky background subtraction, while preserving the structure of the brightest galaxies and their extended envelopes, maintaining the limiting magnitude of the faintest objects at the same time.


\section{Conclusions}
\label{Sec:Conclusions}

The low surface brightness Universe is the next frontier for many studies in galaxy evolution and cosmology. Many observational and theoretical works demonstrate that there are very extended and complex structures larger than the visible size of their host galaxies below the limiting magnitude of most current surveys. Moreover, the cosmological dimming substantially limits our capabilities to study the structure of extended objects at high redshift, most of which are only accessible through space-based observations. It is then mandatory to improve the reduction techniques of the cosmological deep fields from HST and other space telescopes. In this paper we test a number of corrections to improve the low surface brightness limits of the HUDF WFC3 IR mosaics. We have obtained a dedicated version of the images which we named {\textsf{ABYSS}} which we made publicly available for the benefit and use of all the astronomical community. We found several interesting results: 
\begin{enumerate}

    \item The XDF version of the HUDF WFC3 IR mosaics is dominated by a systematic bias in the form of a significant over-subtraction of the sky background around the objects with large angular size. A similar result (to a lesser extent) is obtained for the HUDF12. We successfully recover a significant amount of over-subtracted diffuse light around the largest objects of the HUDF, not detected by the previous versions of the mosaics. The integrated magnitude of the recovered light is equivalent to a $m\sim19$ mag object for the XDF and $m\sim20$ mag for the HUDF12 mosaics, comparable to the brightest galaxies on the image. 
    
    \item A significant fraction of the images of the HUDF are (at least partially) affected by persistence effects at the very low surface brightness regime, biasing the sky background estimation. The cause of this is the observation of bright sources (astronomical or calibration runs) in the previous hours to the scheduled HUDF observations.
    
    \item We propose and test a sky background correction method, based on careful masking using noise-based, non-parametric methods as {\textsf{Gnuastro NoiseChisel}} to detect and flag the extended envelopes of the sources on the field of view. We demonstrate that this method can improve the sky background determination more than one order of magnitude.

    \item We studied the surface brightness profiles of six objects in the HUDF, demonstrating that our reduction pipeline can preserve the properties of the smallest sources in the HUDF, while recovering the low surface brightness structures of the outskirts of the largest galaxies.  
\end{enumerate}

Systematic biases can dominate over the sky noise. It is for this reason that the noise level of the mosaics is not always a good proxy of the real depth of astronomical images. The measured value of the intensity dispersion of the sky dominated pixels it is not very sensitive to the effects of sky background over-subtraction. Hypothetically, an image reduced with a sky-subtraction process that completely fits and subtracts the sources on the field of view might obtain a similar surface brightness limiting magnitude (measured as the standard deviation of sky-noise level) as a reduction that preserves the outskirts of the largest sources. The latter image clearly contains more information and therefore is deeper than the over subtracted version, despite of having the same formal surface brightness limiting magnitude. We have shown that, despite the small loss of exposure time and the corresponding slight increase of surface brightness limiting magnitudes, our mosaics contain more signal and information about the outskirts of galaxies than previous versions of the WFC3 IR HUDF.  

Most, if not all of these problems will be found on the deep observations to be performed by HST near IR successor, the JWST. In particular, persistence will be present in its three image detectors (NIRCam, NIRISS, and MIRI). Mitigation of persistence effects on JWST deep cosmological surveys \citep{Leisenring2016} have to be based mainly on the following observational strategy: 1) careful scheduling of the previous observations and 2) large dithering patterns to avoid observing regions of the sky with those regions that are affected by persistence on all the exposures. In addition, deep and wide survey dedicated missions, (i.e., EUCLID, MESSIER) can improve their results using these techniques, which would enable a new generation of low surface brightness studies based on their legacy data. 

\pagebreak

In this paper we have reviewed many of these systematic effects and proposed solutions to them, applying the methods on the deepest image of the Universe ever taken, the HUDF, creating our own version called {\textsf{ABYSS}}: a low surface brightness dedicated reduction for the HUDF WFC3 IR mosaics. We make the results and the calibration files publicly available to the community - as well as the {\textsf{ABYSS pipeline}}, \footnote{{\textsf{ABYSS}} is a low surface brightness dedicated reduction for the HUDF WFC3 IR mosaics. . It is freely and publicly available at http://www.iac.es/proyecto/abyss/}, hoping to promote further analysis and improvements to the proposed reduction methods. 

\begin{acknowledgements}
The authors would like to thank the XDF and HUDF12 teams for their extraordinary work in which this contribution is based on, truly standing on the shoulders of giants. This work would not have been possible without the kind assistance of all the members of the STScI Help Desk. We specially thank Knox Long by its work on persistence effects and its extraordinary support during this work. A.B. acknowledges support from the International Space Science Institute (ISSI) as part of the Exploring the Ultra-Low Surface Brightness Universe International team. Some/all of the data presented in this paper were obtained from the Mikulski Archive for Space Telescopes (MAST). STScI is operated by the Association of Universities for Research in Astronomy, Inc., under NASA contract NAS5-26555. I.T. and C.M.L acknowledges the support from the SUNDIAL EU Network and from the European Union's Horizon 2020 research and innovation programme under Marie Sklodowska-Curie grant agreement No 721463 and the SUNDIAL ITN network. C.M.L. acknowledges financial support from the Spanish Ministry of Economy and Competitiveness (MINECO) under grant number AYA2016-76219-P. N.C acknowledged support from the Spanish Programa Nacional de Astronim{\'{i}}a y Astrof{\'{i}}sica under grant AYA2016-75808-R. C.G.G. acknowledges support from the European
Research Council (ERC) Consolidator Grant funding scheme (project ConTExt, grant number 648179). Support for MAST for non-HST data is provided by the NASA Office of Space Science via grant NNX09AF08G and by other grants and contracts. This work was partly done using GNU Astronomy Utilities {\textsf{Gnuastro}} version 0.5. {\textsf{Gnuastro}} is a generic package for astronomical data manipulation and analysis which was primarily created and developed for research funded by the Monbukagakusho (Japanese government) scholarship and ERC advanced grant 339659-MUSICOS. Support for this work was provided by the Spanish Ministerio de Econom\'{i}a y Competitividad (MINECO; grant AYA 2016-77237-C3-I-P). This research made use of Astropy,\footnote{http://www.astropy.org} a community-developed core Python package for Astronomy \citep{astropy:2013, astropy:2018}. 
\end{acknowledgements}

\pagebreak
\normalsize

\begin{appendix}
\onecolumn

\section{Surface brightness profiles of the objects commented in Section \ref{Subsect:profiles}}
\label{Appendix:profiles}

In this appendix we show the surface brightness profiles for the five objects (HUDF-1, HUDF-2, ABYSS-1, ABYSS-2, and ABYSS-3, see Table \ref{table:targets}) analysed on Sect.\,\ref{Subsect:profiles}, comparing the results for the \textsf{ABYSS}, XDF, and HUDF12 versions on the mosaics.

\vspace{-0.2cm}
\begin{figure*}[b!]
\centering
\vspace{-0.2cm}
\includegraphics[page=1, width=0.45\textwidth, trim=0 92 40 0, clip]{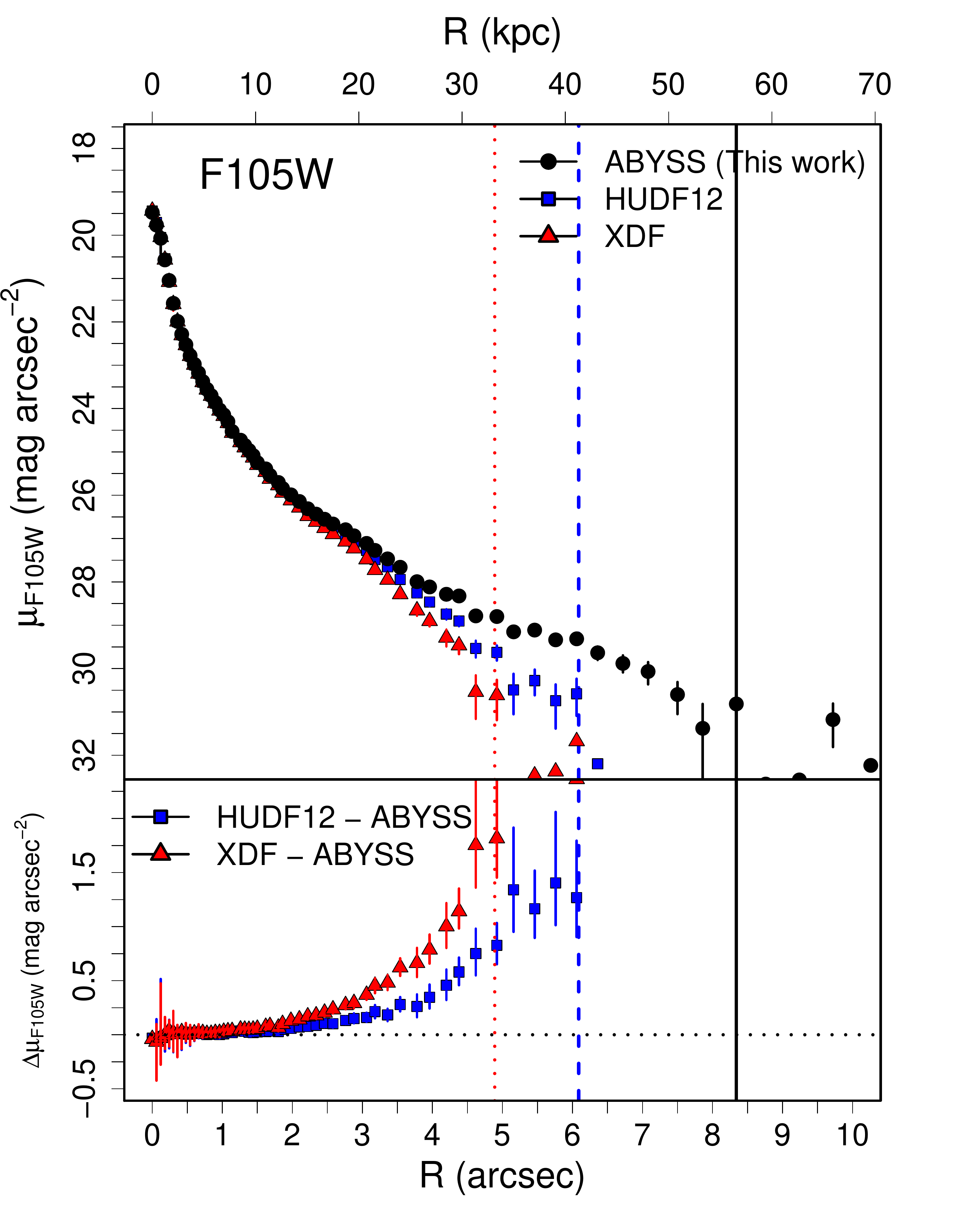}
\includegraphics[page=2, width=0.45\textwidth, trim=0 92 40 0, clip]{PROFILES/hudf1_profile.pdf}

\includegraphics[page=3, width=0.45\textwidth, trim=0 0 40 85, clip]{PROFILES/hudf1_profile.pdf}
\includegraphics[page=4, width=0.45\textwidth, trim=0 0 40 85, clip]{PROFILES/hudf1_profile.pdf}
\vspace{-0.5cm}
\caption[]{Comparison of the surface brightness profiles of the elliptical galaxy HUDF-1 \citep[$\alpha=53.16164, \delta=-27.78025$][]{Buitrago2017} for the F105W (top left), F125W (top right), F140W (bottom left), and F160W filters (bottom right), using our own reduction of the HUDF WFC3 mosaics (\textsf{ABYSS}, black dots), the HUDF12 \citep[][blue squares]{Koekemoer2012}, and the XDF \citep[][red triangles]{Illingworth2013}. The top plot of each panel shows the surface brightness profile for each reduction. Black solid, blue dashed, and red dotted lines represent the elliptic aperture with largest semi-major axis that presents a $S/N$ ratio higher than 3 over the sky-level. The bottom plot represents the difference in magnitude of each previous reduction with the \textsf{ABYSS} version of the mosaics as a function of galactocentric radius. Consult the legend on the figure.}
\label{fig:HUDF1}
\end{figure*}


\begin{figure*}[t!]
\centering
\includegraphics[page=1, width=0.49\textwidth, trim=0 92 40 0, clip]{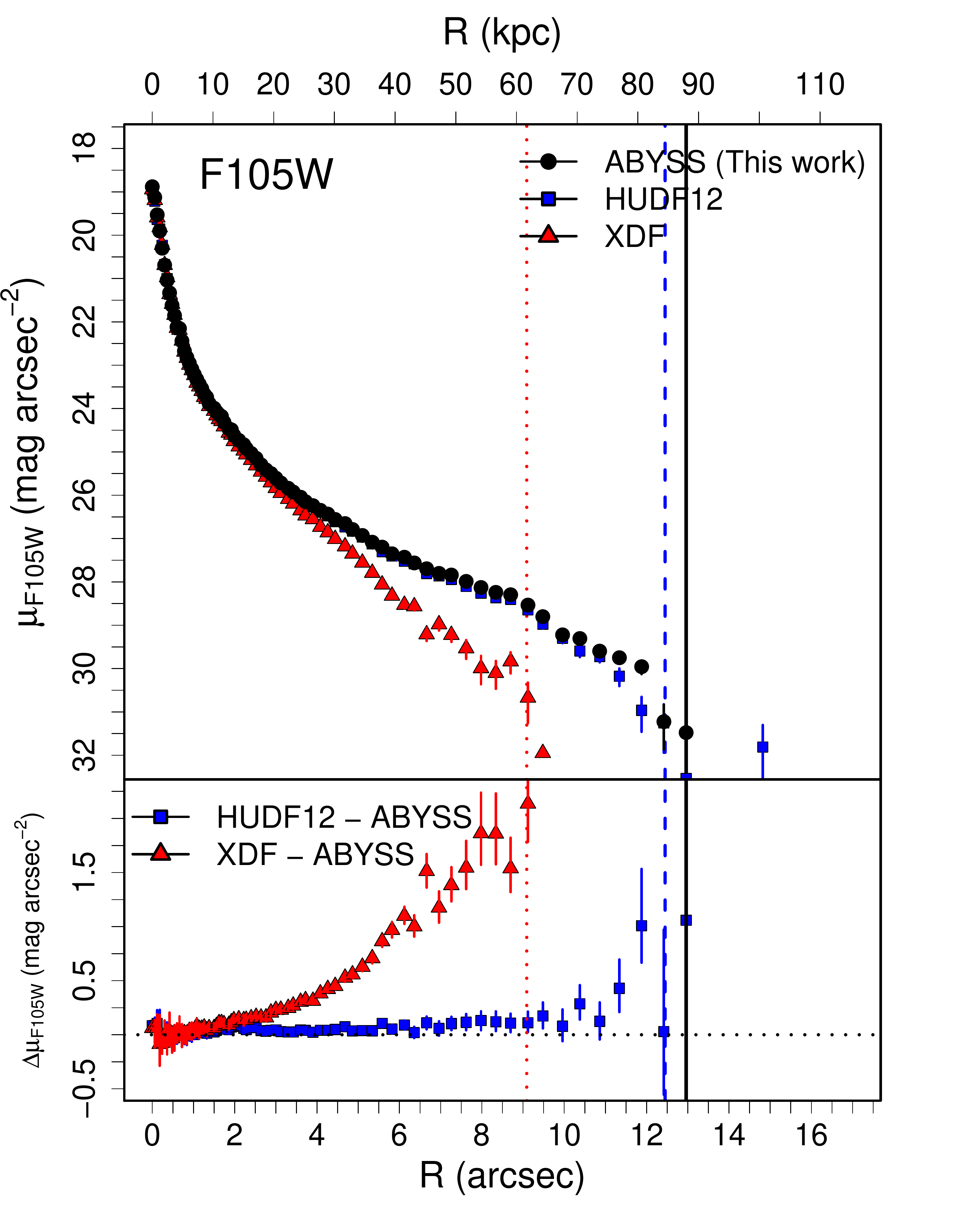}
\includegraphics[page=2, width=0.49\textwidth, trim=0 92 40 0, clip]{PROFILES/esouth_profile.pdf}

\includegraphics[page=3, width=0.49\textwidth, trim=0 0 40 85, clip]{PROFILES/esouth_profile.pdf}
\includegraphics[page=4, width=0.49\textwidth, trim=0 0 40 85, clip]{PROFILES/esouth_profile.pdf}

\caption[]{Comparison of the surface brightness profiles of the elliptical galaxy HUDF-2 \citep[$\alpha=53.17254, \delta=-27.78812$][]{Buitrago2017} for the F105W (top left), F125W (top right), F140W (bottom left), and F160W filters (bottom right), using our own reduction of the HUDF WFC3 mosaics (\textsf{ABYSS}, black dots), the HUDF12 \citep[][blue squares]{Koekemoer2012} and the XDF \citep[][red triangles]{Illingworth2013}. The top plot of each panel shows the surface brightness profile for each reduction. Black solid, blue dashed and red dotted lines represent the elliptic aperture with largest semi-major axis that presents a signal-to-noise ratio higher than 3 over the sky-level. The bottom plot represents the difference in magnitude of each previous reduction with the \textsf{ABYSS} version of the mosaics as a function of galactocentric radius. Consult the legend on the figure.}  
\vspace{0.25cm}
\label{fig:HUDF2}
\end{figure*}

\begin{figure*}[t!]
\centering
\includegraphics[page=1, width=0.49\textwidth, trim=0 92 40 0, clip]{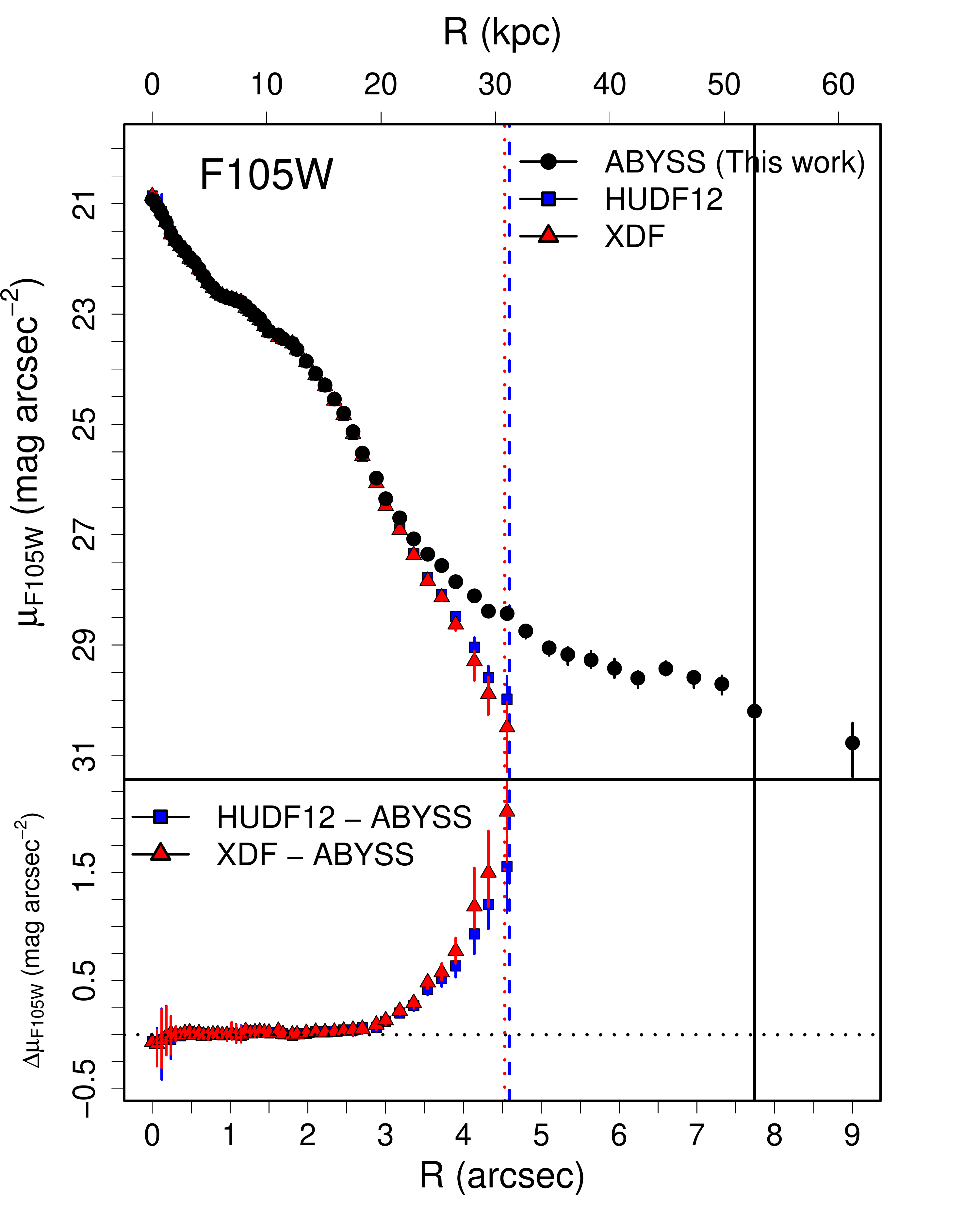}
\includegraphics[page=2, width=0.49\textwidth, trim=0 92 40 0, clip]{PROFILES/nsp_profile.pdf}

\includegraphics[page=3, width=0.49\textwidth, trim=0 0 40 85, clip]{PROFILES/nsp_profile.pdf}
\includegraphics[page=4, width=0.49\textwidth, trim=0 0 40 85, clip]{PROFILES/nsp_profile.pdf}

\caption[]{Comparison of the surface brightness profiles of the spiral galaxy ABYSS-1 \citep[$\alpha=53.16993, \delta=-27.77106$,][]{Elmegreen2005} for the F105W (top left), F125W (top right), F140W (bottom left), and F160W filters (bottom right), using our own reduction of the HUDF WFC3 mosaics (\textsf{ABYSS}, black dots), the HUDF12 \citep[][blue squares]{Koekemoer2012}, and the XDF \citep[][red triangles]{Illingworth2013}. The top plot of each panel shows the surface brightness profile for each reduction. Black solid, blue dashed, and red dotted lines represent the elliptic aperture with largest semi-major axis that presents a $S/N$ ratio higher than 3 over the sky-level. The bottom plot represents the difference in magnitude of each previous reduction with the \textsf{ABYSS} version of the mosaics as a function of galactocentric radius. Consult the legend on the figure.}  
\vspace{0.25cm}
\label{fig:ABYSS1}
\end{figure*}


\begin{figure*}[t!]
\centering
\includegraphics[page=1, width=0.49\textwidth, trim=0 92 40 0, clip]{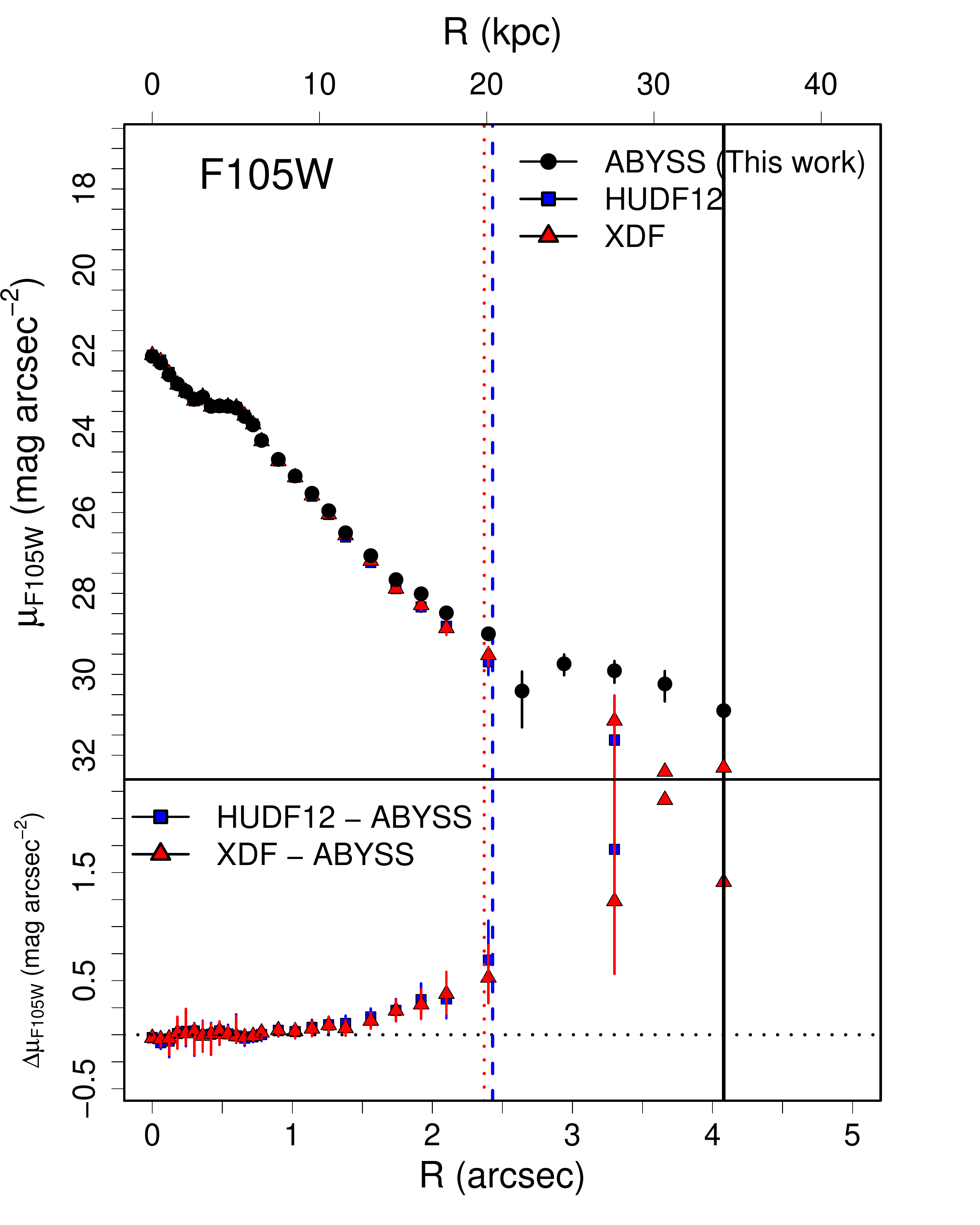}
\includegraphics[page=2, width=0.49\textwidth, trim=0 92 40 0, clip]{PROFILES/nsp2_profile.pdf}

\includegraphics[page=3, width=0.49\textwidth, trim=0 0 40 85, clip]{PROFILES/nsp2_profile.pdf}
\includegraphics[page=4, width=0.49\textwidth, trim=0 0 40 85, clip]{PROFILES/nsp2_profile.pdf}

\caption[]{Comparison of the surface brightness profiles of the spiral galaxy ABYSS-2 \citep[$\alpha=53.17606, \delta=-27.77371$,][]{Elmegreen2005} for the F105W (top left), F125W (top right), F140W (bottom left), and F160W filters (bottom right), using our own reduction of the HUDF WFC3 mosaics (\textsf{ABYSS}, black dots), the HUDF12 \citep[][blue squares]{Koekemoer2012}, and the XDF \citep[][red triangles]{Illingworth2013}. The top plot of each panel shows the surface brightness profile for each reduction. Black solid, blue dashed, and red dotted lines represent the elliptic aperture with largest semi-major axis that presents a $S/N$ ratio higher than 3 over the sky-level. The bottom plot represents the difference in magnitude of each previous reduction with the \textsf{ABYSS} version of the mosaics as a function of galactocentric radius. Consult the legend on the figure.}  
\vspace{0.25cm}
\label{fig:ABYSS2}
\end{figure*}


\begin{figure*}[t!]
\centering
\includegraphics[page=1, width=0.49\textwidth, trim=0 92 40 0, clip]{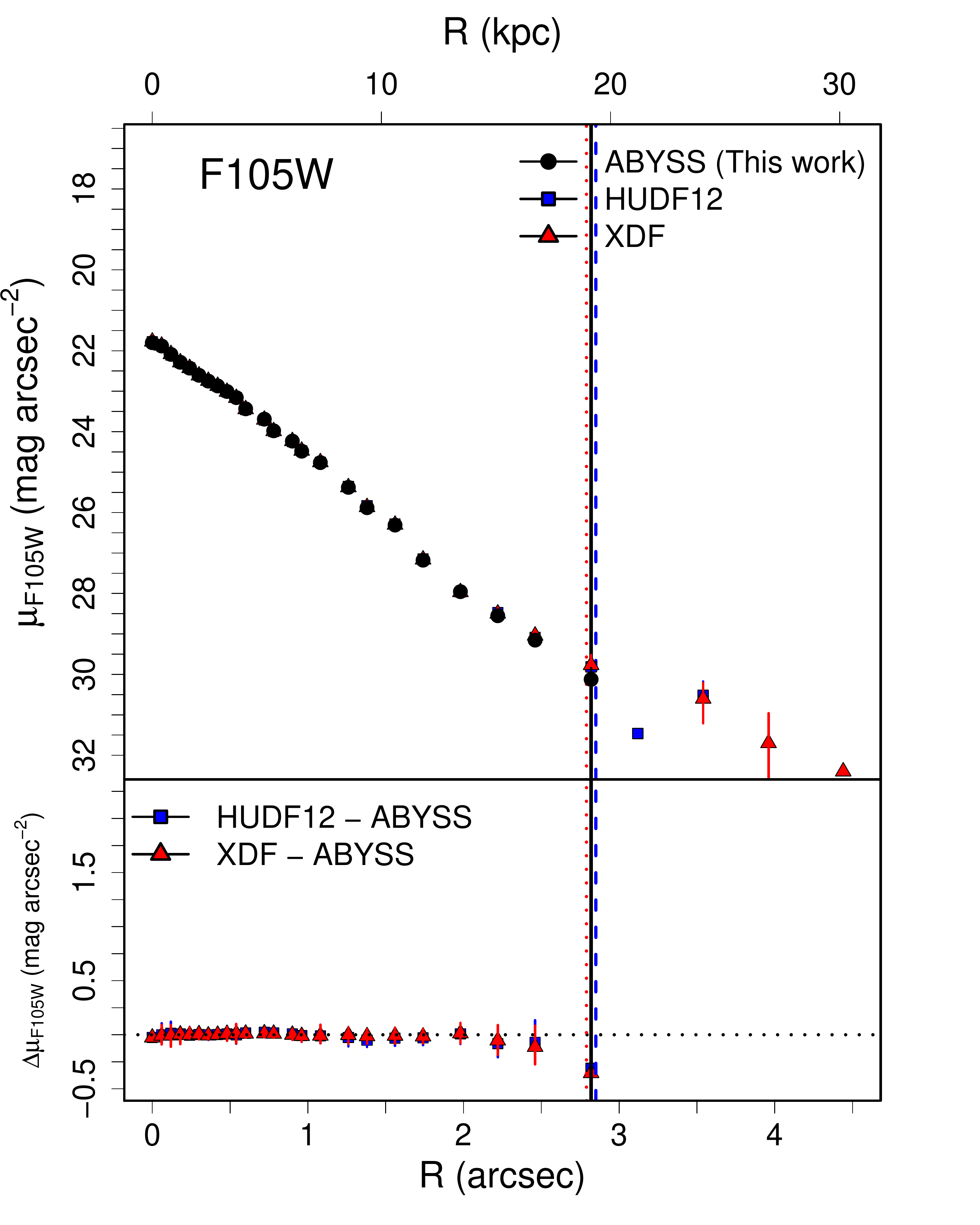}
\includegraphics[page=2, width=0.49\textwidth, trim=0 92 40 0, clip]{PROFILES/nsp3_profile.pdf}

\includegraphics[page=3, width=0.49\textwidth, trim=0 0 40 85, clip]{PROFILES/nsp3_profile.pdf}
\includegraphics[page=4, width=0.49\textwidth, trim=0 0 40 85, clip]{PROFILES/nsp3_profile.pdf}

\caption[]{Comparison of the surface brightness profiles of the spiral galaxy ABYSS-3 \citep[$\alpha=53.15815, \delta=-27.78109$,][]{Elmegreen2005} for the F105W (top left), F125W (top right), F140W (bottom left), and F160W filters (bottom right), using our own reduction of the HUDF WFC3 mosaics (\textsf{ABYSS}, black dots), the HUDF12 \citep[][blue squares]{Koekemoer2012}, and the XDF \citep[][red triangles]{Illingworth2013}. The top plot of each panel shows the surface brightness profile for each reduction. Black solid, blue dashed, and red dotted lines represent the elliptic aperture with largest semi-major axis that presents a $S/N$ ratio higher than 3 over the sky-level. The bottom plot represents the difference in magnitude of each previous reduction with the \textsf{ABYSS} version of the mosaics as a function of galactocentric radius. Consult the legend on the figure.}  
\vspace{0.25cm}
\label{fig:ABYSS3}
\end{figure*}

\end{appendix}
\clearpage
\newpage
\twocolumn
\small  
\bibliographystyle{aa}
\bibliography{borlaff_HUDF.bib}{}

\end{document}